\documentclass{pinchcrNumRef}
\usepackage{graphicx,amsmath,amssymb,subfigure,color,verbatim,array,lscape}
\usepackage{yllmath}

\providecommand{\todo}[1]{}             
\providecommand{\todo}[1]{\red{#1}}     

\providecommand{\mutot}{\tilde{\mu}}
\providecommand{\htot}{\tilde{h}}

\providecommand{\gammadag}{\gamma^\dag}

\providecommand{\chn}{l}

\providecommand{\onlinecite}[1]{\cite{#1}}

\title{Theoretical Studies of Superconductor-Insulator Transitions}
\author{Yen Lee Loh and Nandini Trivedi}
\affiliation{Department of Physics, The Ohio State University, 191 W
Woodruff Avenue, Columbus, OH  43210}
\date{Started 2009-7-20, touched 2009-7-20, 2010-5-10, 2010-7-28,
2010-8-27, compiled \today}

\begin{document}
\maketitle
\tableofcontents
\maintext

\chapter{Introduction} \slabel{Introduction}




The superconductor-insulator transition (SIT) is an intriguing example of a
quantum phase transition in a fermionic system.  It is a theme that runs across boundaries between disciplines -- strongly correlated electrons, helium, cold atoms, and nuclear physics -- and is highly relevant to some of the biggest unsolved problems in superconductivity.
For a broad perspective, see the overview chapter ``Superconductor-Insulator Transitions: Present Status and Open Questions'' by N.~Trivedi and the review by Gantmakher and Dolgopolov \cite{gantmakher2010}.

In this article we study superconductor-insulator transitions within the general framework of an attractive Hubbard model.  This is a well-defined model of $s$-wave superconductivity which permits different tuning parameters (disorder and field).  Furthermore, it allows a comparison of various analytical and computational approaches in order to gain a complete understanding of the various effects of amplitude and phase fluctuations.
We present a systematic pedagogical approach, 
aiming to equip the ``lay'' reader 
with enough apparatus to be able to understand the numerical calculations,
reproduce some of the simpler results, 
and be able to tackle future problems related to inhomogeneous phases.
We go into considerable detail on mean-field theory (MFT) and the
Bogoliubov-de Gennes (BdG) approach, as these are a first line of attack
which can capture much of the physics, but we also outline cases where
this fails to capture phase fluctuations and more sophisticated
Quantum Monte Carlo (QMC) calculations are necessary.
We discuss the behavior of many observables, including densities of
states, superfluid stiffness, and dynamical conductivity, for the
disorder-tuned superconductor-insulator transition.

The general Hamiltonian is 
	\begin{align}
	H
	&=
	-	\sum_{ij\sigma} t_{ij} \cdag_{i\sigma} \cccc_{j\sigma}
	-	\sum_{i}
			U n_{i\up} n_{i\dn}
	-	\sum_{i\sigma}
		\mu_{i\sigma} n_{i\sigma}
	\end{align}
where $i$ and $j$ are site indices, $\sigma=\pm 1$ are fermion spin indices, $t_{ij}$ is the hopping from site $i$ to site $j$, $\cdag_{i\sigma}$ and $\cccc_{i\sigma}$ are fermion creation and annihilation operators, $U$ is the on-site attractive interaction, $\mu_{i\sigma} = \mu - v_i - h\sigma$ is the site-dependent chemical potential, $\mu$ is the average chemical potential, $v_i$ are disorder potentials on each site drawn independently from a uniform distribution $[-V,+V]$, $h$ is a uniform Zeeman field, and $n_{i\sigma}=\cdag_{i\sigma} \cccc_{i\sigma}$ is the local density for spin $\sigma$.  In particular, we will consider various combinations of the parameters $t$, $U$, $\mu$, $h$, disorder strength $V$, and temperature $T$.  
If necessary, the Hubbard model parameters $t$, $V$, $U$ can be connected to physical observables such as $\Delta_0$, $k_F$, $E_F$, the mean free path $l$, and the disorder energy scale $\hbar/\tau$.
We shall focus on transport and spectra, which are the properties most directly affected by the SIT (rather than thermodynamic properties). 

This model covers many situations of interest.
We begin with the tight-binding model (with hopping $t$) and introduce the other ingredients one by one (disorder $V$, attraction $U$, and Zeeman field $h$).
Adding $V$ alone 
leads to the Anderson model of localization
and the metal-insulator transition in a non-interacting system.
We use this as a convenient starting point for introducing various concepts (\sref{AndersonChapter}).
Conversely, with $U$ alone, 
the model describes a superconductor in which the size of the pairs can be controlled by the
strength of the attraction and allows one to access the beautiful physics of
the crossover from the BCS regime of large overlapping Cooper pairs to
the BEC regime of tightly bound bosons (\sref{CleanSuperconductor}).

The interplay between $U$ and $V$
produces a disorder-tuned SIT.
With analytical and numerical treatments of increasing sophistication, culminating in QMC simulations, we elucidate the bosonic nature of this transition and its consequences
(\sref{DisorderTunedSIT}).

Magnetic-field-tuned SITs are quite different due to the pairbreaking nature of magnetism.
The competition between $U$ and $h$ leads to a parallel-field-tuned SIT
from a fully paired superconductor
to a fully polarized Fermi liquid 
via a Fulde-Ferrell-Larkin-Ovchinnikov (FFLO) state,
in which the pairing and magnetism form a self-organized layered microstructure
-- an example of microscale phase separation (\sref{CleanSCInZeemanField}). 
Finally, when $U$, $h$, and $V$ are all present, a disordered version of the FFLO state (``dLO'') appears to survive at weak disorder (\sref{DirtySCInZeemanField}).
In dLO phases, Andreev bound states in domain walls contribute mid-gap weight in the density of states, which may be an explanation for the anomalous zero-bias tunneling conductance observed in purely parallel-field-tuned SITs.

\chapter{Free Fermions in a Random Potential} \slabel{AndersonChapter}

In this chapter we introduce the tight-binding model and obtain the dispersion relation and density of states.  We then discuss the properties of the Anderson Hamiltonian for on-site random potentials.  We calculate single-particle properties (density of states, localization length, and participation ratio) and two-particle properties (conductivity), which serve as a basis for comparing with later results on disordered superconductors.

\section{Tight-Binding Model on Square Lattice}
\begin{figure}
	\subfigure[
		Constant-energy contours of dispersion relation from \eref{square-lattice-dispersion}.
		The thick diamond indicates the contour at zero energy,
		which would be the Fermi surface for a half-filled system.
		]{
		\includegraphics[width=0.38\textwidth]{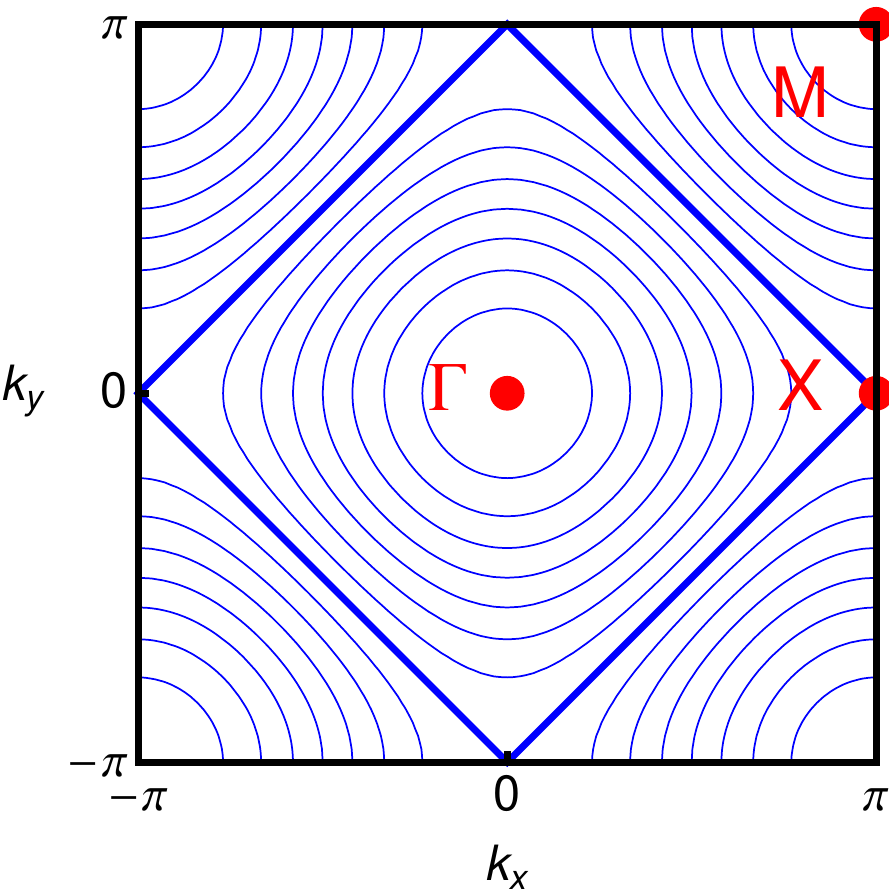}
		\label{tb-contours}
	}
	\hspace{0.04\textwidth}
	\subfigure[
		Density of states.
    The solid line is the analytic formula \eref{square-lattice-dos}.
    The histogram is for a $36\times 36$ square lattice with periodic boundary conditions.
		]{
		\includegraphics[width=0.58\textwidth]{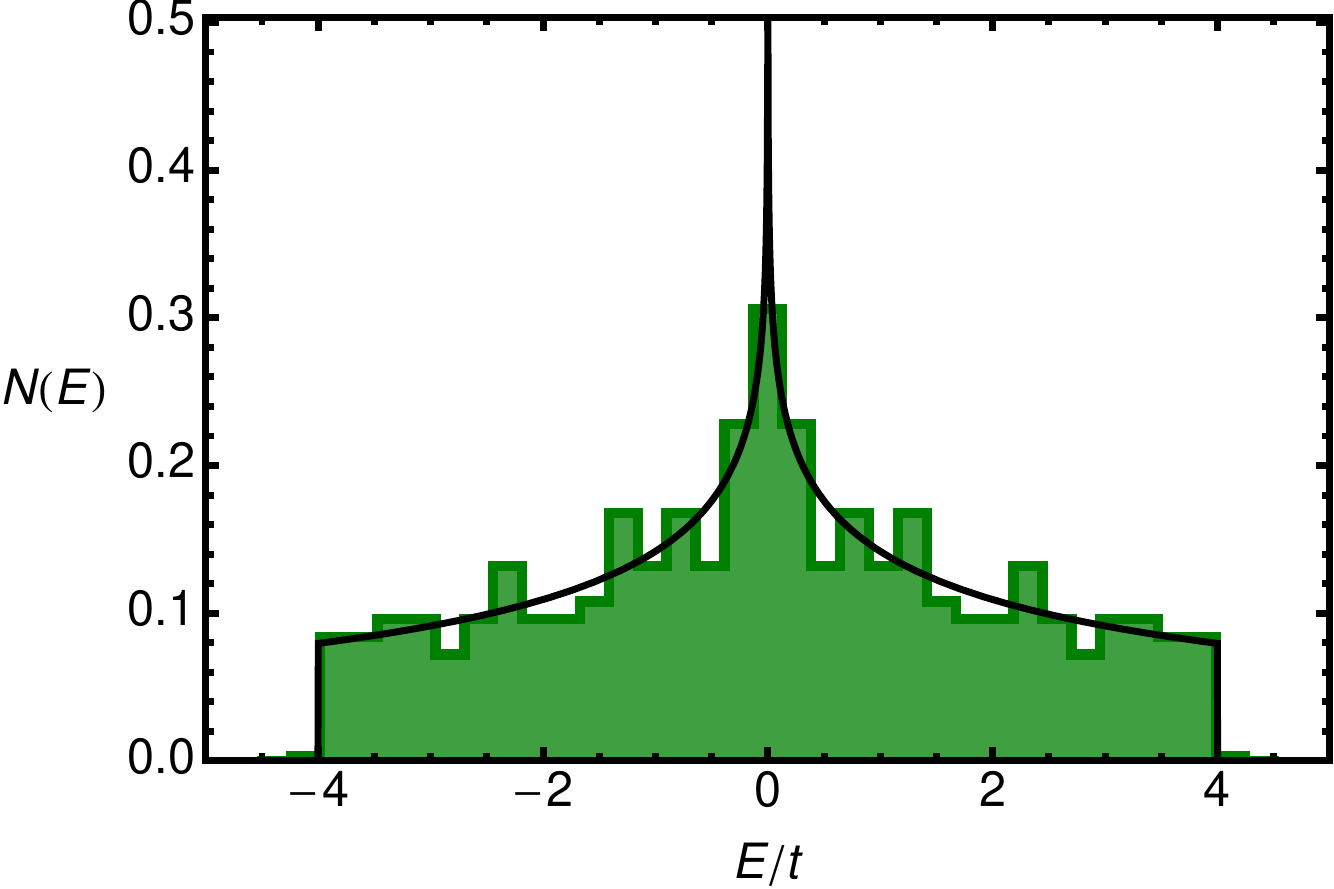}
		\label{tb-dos}
	}
	\caption{
		Properties of the square lattice tight-binding model.
	}
	\label{tb}
	\end{figure}

In this article we will focus on two-dimensional systems.  The starting point for our study is the tight-binding model Hamiltonian for spinless fermions on a square lattice:
	\begin{align}
	H
	&=
	-	\sum_{ij} t_{ij} \cdag_{i} \cccc_{j}
	-	\mu \sum_{i} n_{i}
	,
	\end{align}
where $i$ and $j$ are site indices, $t_{ij}$ are hopping amplitudes (such that $t_{ij}=t$ for nearest neighbors and $0$ otherwise), $\cdag_i$ and $\cccc_i$ are fermion creation and annihilation operators, $\mu$ is the chemical potential, and $n_i = \cdag_i \cccc_i$ are number operators. 
Since $H$ is translationally invariant, it can be diagonalized by Fourier transforming into momentum space:
	\begin{align}
	H
	&=
	\sum_{\kkk} (\vare_\kkk-\mu) \cdag_{\kkk} \cccc_{\kkk}
	\end{align}
where the dispersion relation is
	\begin{align}
	\vare_\kkk &= 
	 	-2t(\cos k_x + \cos k_y)
	.
	\elabel{square-lattice-dispersion}
	\end{align}
This dispersion relation has a bandwidth of $8t$, and it has stationary points at the $\Gamma$, $X$, and $M$ symmetry points of the Brillouin zone.
The density of states (per site) is
	\begin{align}
	N_\text{TB} (E) &= \frac{1}{N} \sum_\kkk \delta(E - \vare_\kkk) 
	= \int_{BZ} \frac{d^2 k}{(2\pi)^2}  \delta(E - \vare_\kkk)
	\end{align}
where $N$ is the number of sites in the system.
This can be written in terms of the complete elliptic integral $K$:
\footnote{There are several definitions for elliptic integrals; ours is the $\mathtt{EllipticK}$ function in Mathematica.}
	\begin{align}
	N_\text{TB} (E) &= 
	  \left|   \frac{2}{\pi^2 E} \Im K\left( \frac{16}{E^2} \right)  \right|
	.
	\elabel{square-lattice-dos}
	\end{align}
The DOS has step singularities at $E=-4t$ and $E=4t$ and a logarithmic singularity at $E=0$, as illustrated in Fig.~\ref{tb-dos}.  These van Hove singularities correspond to the  $\Gamma$, M, and X symmetry points in Fig.~\ref{tb-contours}.  For the finite systems treated in this article, the van Hove singularities are quite severely smeared out.

When the chemical potential lies within the band, this model is a Fermi liquid.  Its thermodynamic properties correspond to those of a Fermi liquid.  For example, the specific heat is linear in temperature.  However, the transport properties of this model are pathological: it has zero superfluid stiffness but infinite DC conductivity (with finite charge stiffness).  These concepts will be discussed in \sref{DisorderTunedSIT}.  In order to model real metals, one has to introduce some amount of disorder, as we do in the next section.

\section{Anderson Hamiltonian} \slabel{anderson}
Consider a tight-binding model with a disorder potential $v_i$ at each site, picked independently from a uniform distribution on $[-V, +V]$, where $V$ is the disorder strength.
\footnote{In the literature, the disorder bandwidth $W=2V$ is often used as the disorder strength parameter.}
This is known as the ``Anderson model'' of localization:
	\begin{align}
	H
	&=
	-	\sum_{ij} t_{ij} \cdag_{i} \cccc_{j}
	+	\sum_{i} (v_i - \mu) n_{i}
	.
	\end{align}
The hopping alone would produce plane-wave eigenstates with a bandwidth of $8t$,
whereas the disorder potential alone would produce site-localized eigenstates with a bandwidth of $2V$.  The competition between hopping and disorder makes this a non-trivial problem, and in some situations it can give rise to a metal-insulator transition.

We wish to calculate thermodynamic and transport properties averaged over disorder configurations.  Many methods have been developed for studying such problems, each with their own advantages: exact diagonalization (ED), Lanczos methods, Chebyshev methods, transfer matrices, mapping to nonlinear sigma models, and so on.
In this article we will focus on exact diagonalization,
\footnote{The word ``exact'' is misleading, because the diagonalization is usually performed numerically; for example, LAPACK 2003 uses Householder tridiagonalization and the ``Relatively Robust Representation'' algorithm.  Some authors prefer to write ``direct diagonalization,'' although this is also a misnomer, as most diagonalization methods involve an ``indirect'' stage (iteration to convergence).}  
 to set the stage for the formalism of the disordered superconductor in \sref{DisorderTunedSIT}.  

\begin{figure}
	\includegraphics[width=\textwidth]{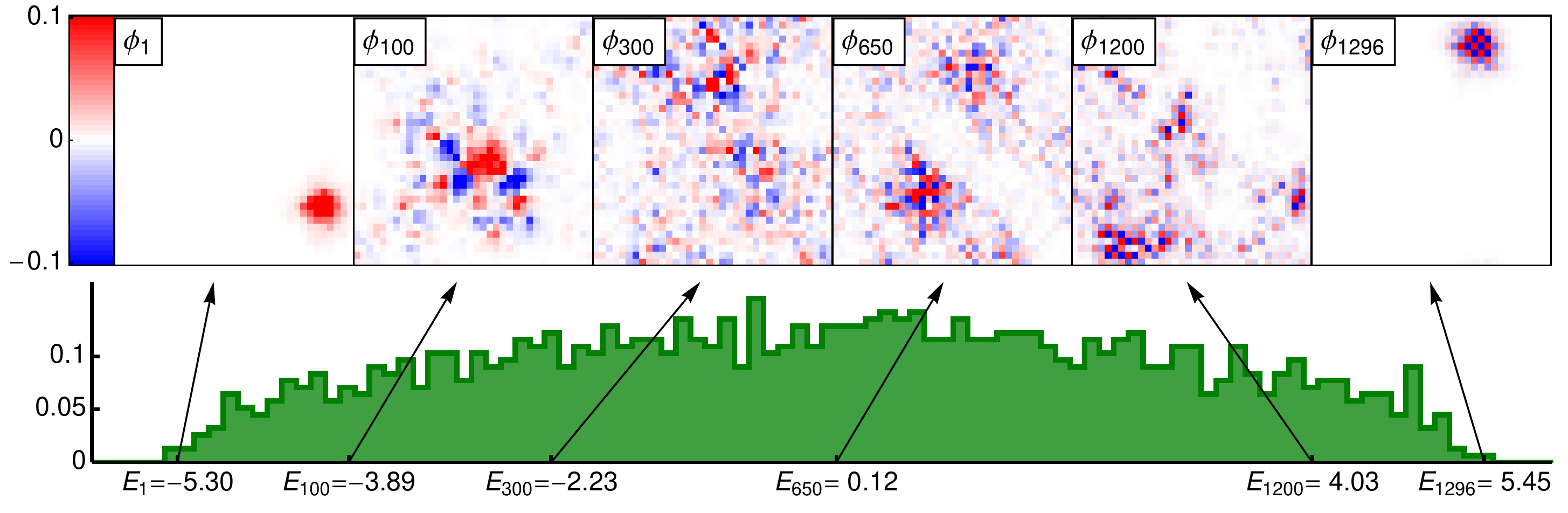}
	\caption{
		Six eigenstates of the Anderson model on a $36\times 36$ square lattice
		for a single disorder realization ($\mu=0, V=3t$).
		Red and blue colors indicate signs of eigenfunctions $\psi_{i\alpha}$.
		The states in the band tail are localized, 
		whereas the states in the band center are quasi-extended
		over the size of the system.
	}
	\label{anderson-evals-evecs}
	\end{figure}

Although it is convenient to write the Hamiltonian in operator form, for computational purposes it is necessary to construct the Hamiltonian matrix elements explicitly, such that $\hat{H} = \sum_{ij} H_{ij} \cdag_i \cccc_j$:
	\begin{align}
	H_{ij}
	&=
	- t_{ij} +  (v_i - \mu) \delta_{ij}
	.
	\end{align}
For example, for a $3\times 3$ square lattice with open boundaries, the Hamiltonian matrix is a sparse matrix with only five non-zero diagonals:
	\begin{align}
	\mathbf{H} 
	&=
	\psmat{
		v_{11} & -t     &        & -t     &        &        & \\
		-t     & v_{12} & -t     &        & -t     &        & \\
		       &  -t    & v_{13} &        &        & -t     & \\
		-t     &        &        & v_{21} & -t     &        & -t \\
		       & -t     &        & -t     & v_{22} & -t     &        & -t \\
		       &        & -t     &        & -t     & v_{23} &        &        & -t \\
		       &        &        & -t     &        &        & v_{31} & -t     & \\
		       &        &        &        & -t     &        & -t     & v_{32} & -t \\
		       &        &        &        &        & -t     &        & -t     & v_{33} \\
	}	
	.
	\end{align}
The next step is to diagonalize the Hamiltonian matrix to find the eigenvalues $E_\alpha$ and unitary eigenvector matrix $\phi_{i\alpha}$ such that $\sum_j H_{ij} \phi_{j\alpha} = E_\alpha \phi_{i\alpha}$, where $\alpha$ labels the eigenmodes.  Then, $\gamma^\dag$ and $\gamma$ are operators that create and annihilate fermions in eigenmodes, where
$\cccc_i = \sum_\alpha \phi_{i\alpha} \gamma_\alpha$
and
$\gamma_\alpha = \sum_i \phi^*_{i\alpha} \cccc_i$.
Figure~\ref{anderson-evals-evecs} shows localized and ``extended'' eigenstates for a single disorder realization.

	\begin{figure}
	\subfigure[
		Mobility edge trajectories (solid),
			for the cubic lattice (3D) Anderson model with box disorder, $v_i \in {[}-V,+V{]}$,
			as well as band edge trajectores (dashed) for comparison.
		The blue crosses indicate the parameters for the data points 
			in Fig.~\ref{KramerLocalizationLength}.
	]{
		\includegraphics[width=0.48\textwidth]{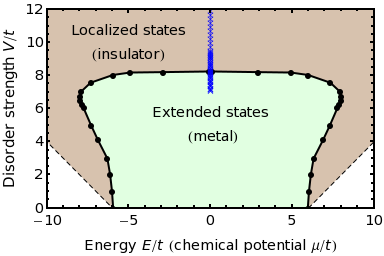}
		\label{KramerMobilityEdges}
	}
	\hspace{0.04\textwidth}
	\subfigure[
		Divergence of the localization length $\xi$ at the critical disorder strength $V_c$.
		For $V<V_c$, $\xi$ is to be interpreted as a scaling length instead.
	]{
		\includegraphics[width=0.48\textwidth]{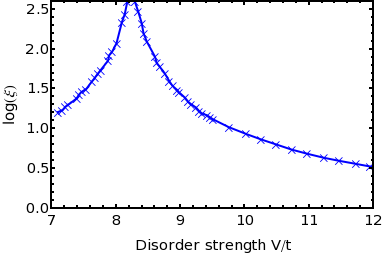}				
		\label{KramerLocalizationLength}
	}
	\subfigure[
		Phase diagram of the square lattice (2D) Anderson model with box disorder.
	]{
		\includegraphics[width=0.5\textwidth]{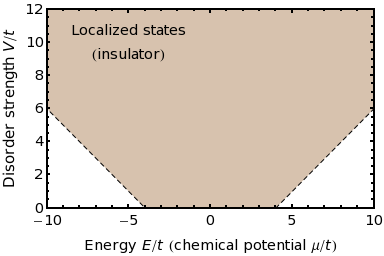}	\label{SquareLatticeAndersonPhaseDiagram}
	}	
	\subfigure[
		Localization length $\xi_\text{loc}$
			on strips of width $M$ with cylindrical boundary conditions,
			calculated using the Green function method, \eref{GreenFunctionMethod}.  
		The thick black curve is the infinite-size limit, 
			extracted using a rudimentary form of finite-size scaling.
		$\xi_\text{loc}$ is finite for all $V>0$, but it diverges strongly as $V\rightarrow 0$.
		]{
		\includegraphics[width=0.5\textwidth]{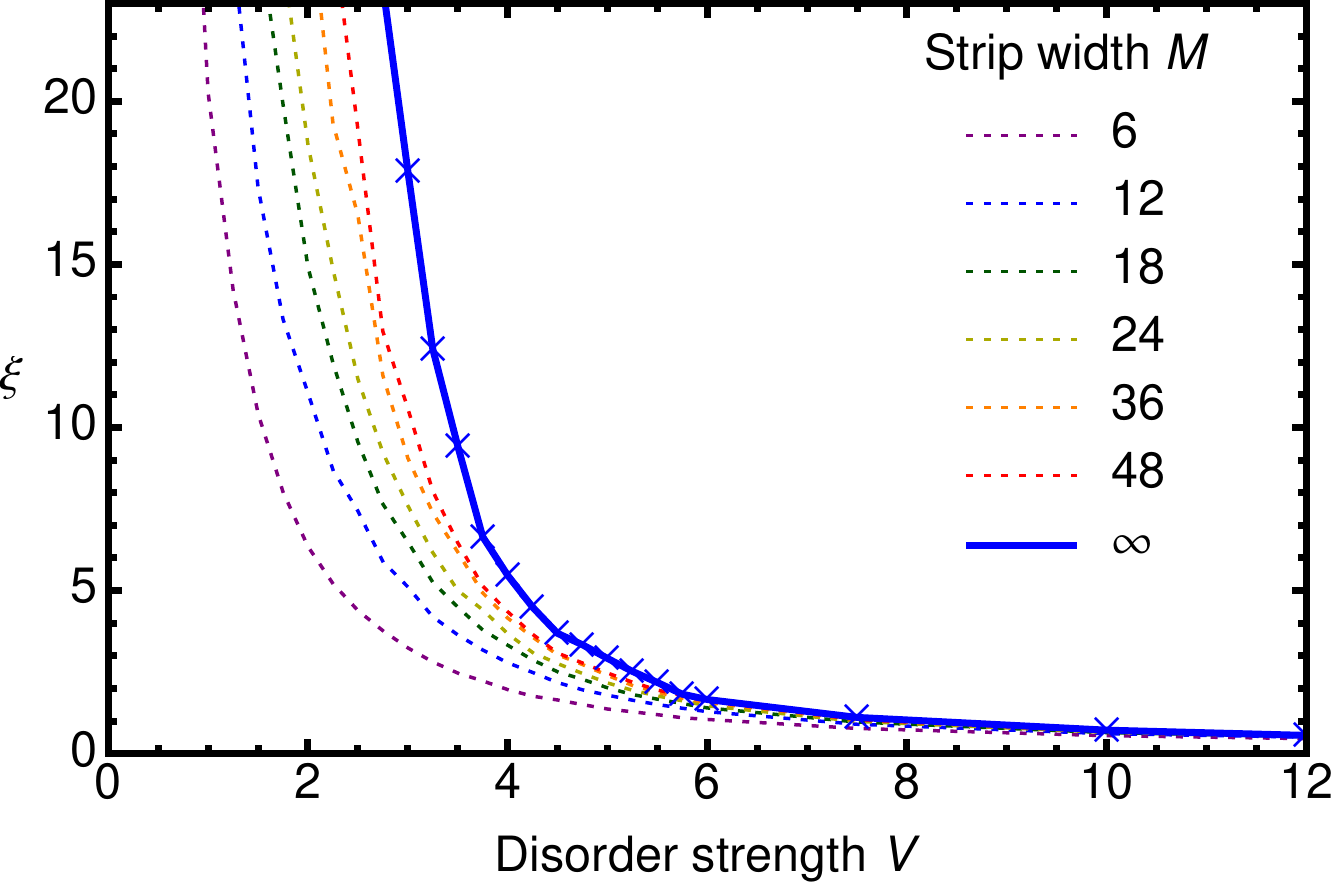}	\label{andersonLocLen}
	}	
	\caption{
		Properties of the Anderson model.  3D figures are adapted from the review by Kramer and MacKinnon.
	}
	\label{CubicLatticeAndersonProperties}
	\label{SquareLatticeAndersonProperties}
	\end{figure}

\section{Participation Ratio and Localization Length}
The Anderson model of localization has been extensively studied in the context of metal-insulator transitions.  
The nature of the eigenstates is now well understood in terms of the scaling theory of localization \cite{PhysRevLett.42.673} and the theory of weak localization \cite{PhysRevB.22.5142}.  (See the chapter by K. Slevin and T. Ohtsuki for more on the topic of Anderson localization.)

For 3D systems, eigenstates can be extended or localized.
Extended states occupy the center of each band, whereas localized states exist near the edges of the band (the tails in the density of states).
The boundaries between regions of extended and localized states are called mobility edges.
The loci of the mobility edges form a curve in the $(E,V)$ plane\cite{bulka1985}, as illustrated in Fig.~\ref{CubicLatticeAndersonProperties}.
Sweeping the chemical potential across a mobility edge produces a metal-insulator transition (MIT), which is an example of a quantum phase transition.
Figure~\ref{CubicLatticeAndersonProperties} also serves as a phase diagram in the $(\mu,V)$ plane.

In 1D and 2D systems, an infinitesimal amount of disorder is sufficient to make all states localized.
However, localized states appear to be extended if their localization length is greater than the box size,
as is evident in Fig.~\ref{anderson-evals-evecs}.

One way to quantify the spatial extent of an eigenstate $\phi_{i\alpha}$ is via its participation number
	\begin{align}
	p_\alpha
	&= 1 \bigg/ \sum_i \left| \phi_{i\alpha} \right|^4
	,
	\elabel{ParticipationNumber}
	\end{align}
which describes the number of sites over which the wavefunction has an appreciable magnitude.
\footnote{In the literature one often encounters the participation ratio $p_\alpha/N$, which is the fraction of sites over which the wavefunction has an appreciable magnitude.}
Sometimes $R_\alpha=(p_\alpha)^{1/d}$ is identified as an ``eigenmode radius,'' an indicator of the spatial extent of eigenmode $\alpha$.

\begin{figure}
	\subfigure[
		Response of a $200\times 20$ strip of the Anderson model
			to a source at one edge, calculated by brute force linear system solution.
		Periodic boundary conditions were used in the short direction.
		Parameters were $E=-3$, $V=1.5$, $t=1$.	
		]{
		\includegraphics[width=\textwidth]{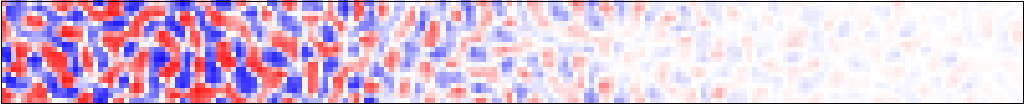}	\label{AndersonStrip}
	}	
	\subfigure[
		Eigenvalues of $\GGG_l$ as a function of strip length $l$.
		The largest eigenvalue of $\GGG_l$ decays exponentially with $l$.
		The length scale for this decay gives the localization length.
		]{
		\includegraphics[width=\textwidth]{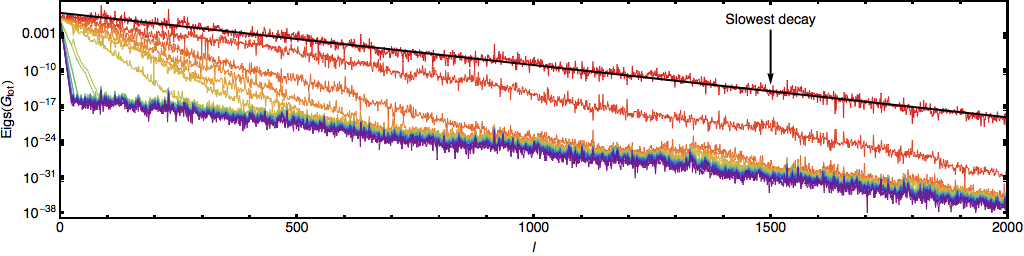}	\label{AndersonGFMethod2}
	}	
	\caption{
		Illustration of the Green function and localization length on a strip.
	}
	\label{AndersonGFMethod}
	\end{figure}

Another important concept is the localization length $\xi_\text{loc}$.  Qualitatively, this is the length scale for the exponential decay of an eigenfunction far from its center of mass, but for practical calculations, $\xi_\text{loc}$ is usually defined as the decay length of the transmission coefficient along a long strip, and can be calculated using transfer matrix methods or Green function methods\cite{mackinnon1981}.  The Green function $\GGG_l$ is the result of evolving the Schr\"odinger equation $l$ layers away from a source at the 0th layer (see Fig.~\ref{AndersonGFMethod} for an illustration).
It can be calculated using the recursion relation given by 
	\begin{align}
	\mathbf{A}_{-1} &= \mathbf{0} 
	,\nonumber\\	
	\mathbf{A}_0 &= \mathbf{1} 
	,\nonumber\\
	\mathbf{A}_l &= (E\mathbf{1} - \mathbf{H}_l) \mathbf{A}_{l-1} - \mathbf{A}_{l-2} 
			,\qquad{l=1,\dotsc,L}
	,\nonumber\\
	\mathbf{G}_l &= (\mathbf{A}_l) ^{-1}
	,\nonumber\\
	\xi_\text{loc} {}^{-1}  &= \lim_{l\rightarrow\infty}	\frac{1}{2(l-1)} \ln \tr (\mathbf{G}_l)^2
	,
	\elabel{GreenFunctionMethod}
	\end{align}
where $\mathbf{H}_L$ is the $M\times M$ Hamiltonian for the $L$th slice of the strip.  One must take special precautions against numerical instabilities, for example, by periodically restarting the recursion.
Typically, one calculates $\xi_\text{loc}$ as a function of strip width $M$ and then uses finite-size scaling to extrapolate to $M\rightarrow \infty$.  See Refs.~\cite{mackinnon1981,bulka1985,kramerReview} for details.

Figure~\ref{KramerLocalizationLength}, adapted from Ref.~\cite{kramerReview}, shows $\xi_\text{loc}$ as a function of box disorder strength for the cubic lattice Anderson model, clearly showing a metal-insulator transition.  Similar results have been obtained for other disorder distributions, indicating universal behavior for the critical exponents at the MIT.  

Figure~\ref{andersonLocLen} shows $\xi_\text{loc} (V)$ for the square lattice.  Because of weak localization in two dimensions, $\xi_\text{loc}$ is finite for all $V > 0$.  Nevertheless, $\xi_\text{loc}$ diverges strongly as $V\rightarrow 0$, and for $V\lesssim 2t$, it greatly exceeds the dimensions of the systems we will be simulating.

In principle, the average ``eigenmode radius'' at a given energy, $R(E)$, and the localization length, $\xi_\text{loc} (E)$, are distinct quantities, as the former refers to the bulk of the eigenmodes whereas the latter refers to the tails.  Nevertheless, both quantities diverge at the mobility edge, and for the purposes of this article we will loosely refer to both as $\xi_\text{loc}$.

\section{Single-Particle Properties}
\begin{figure*}
	\subfigure[Density of states]{
		\includegraphics[width=0.48\textwidth]{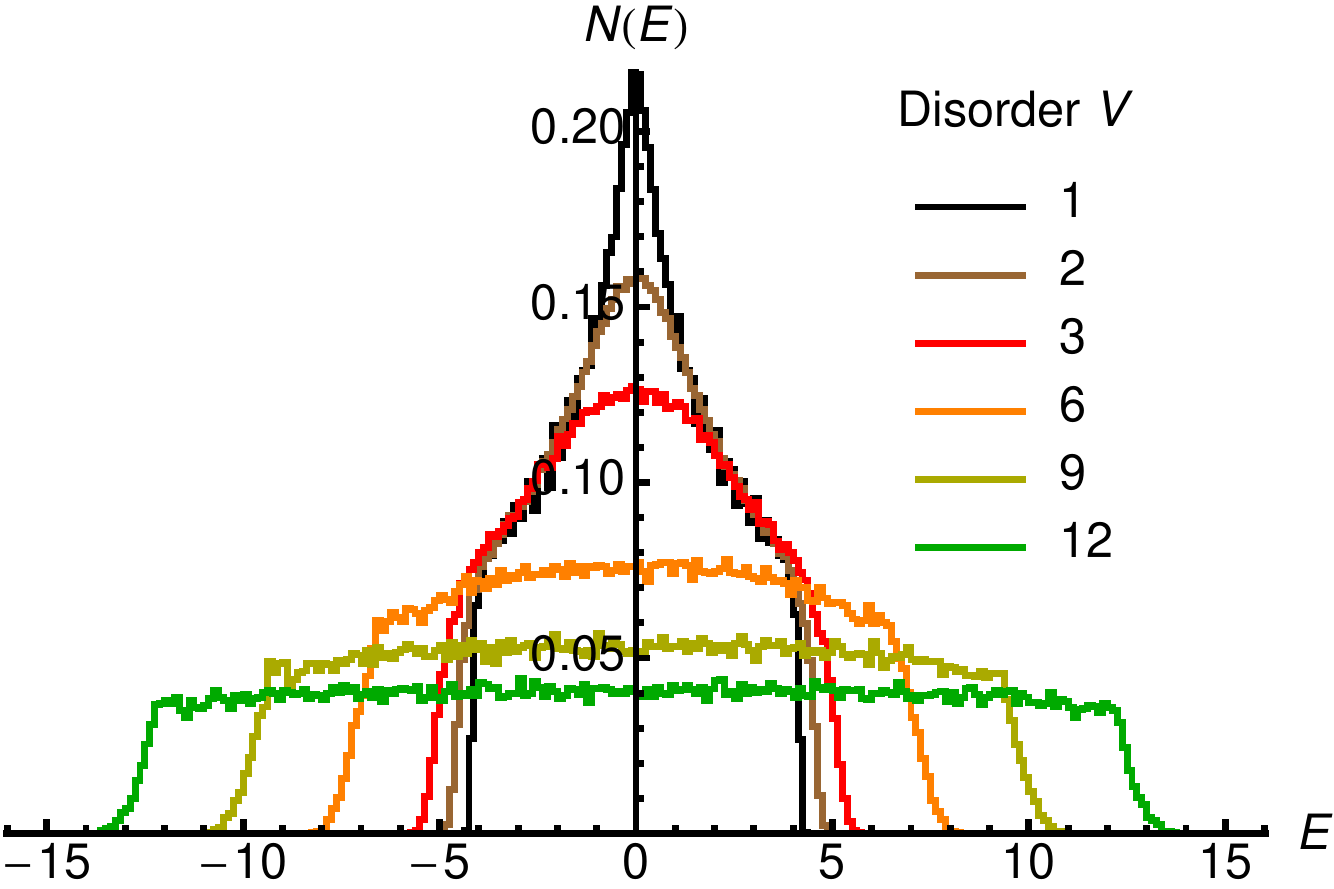}		\label{andersonAE}
	}
	\subfigure[Dynamical conductivity]{
		\includegraphics[width=0.48\textwidth]{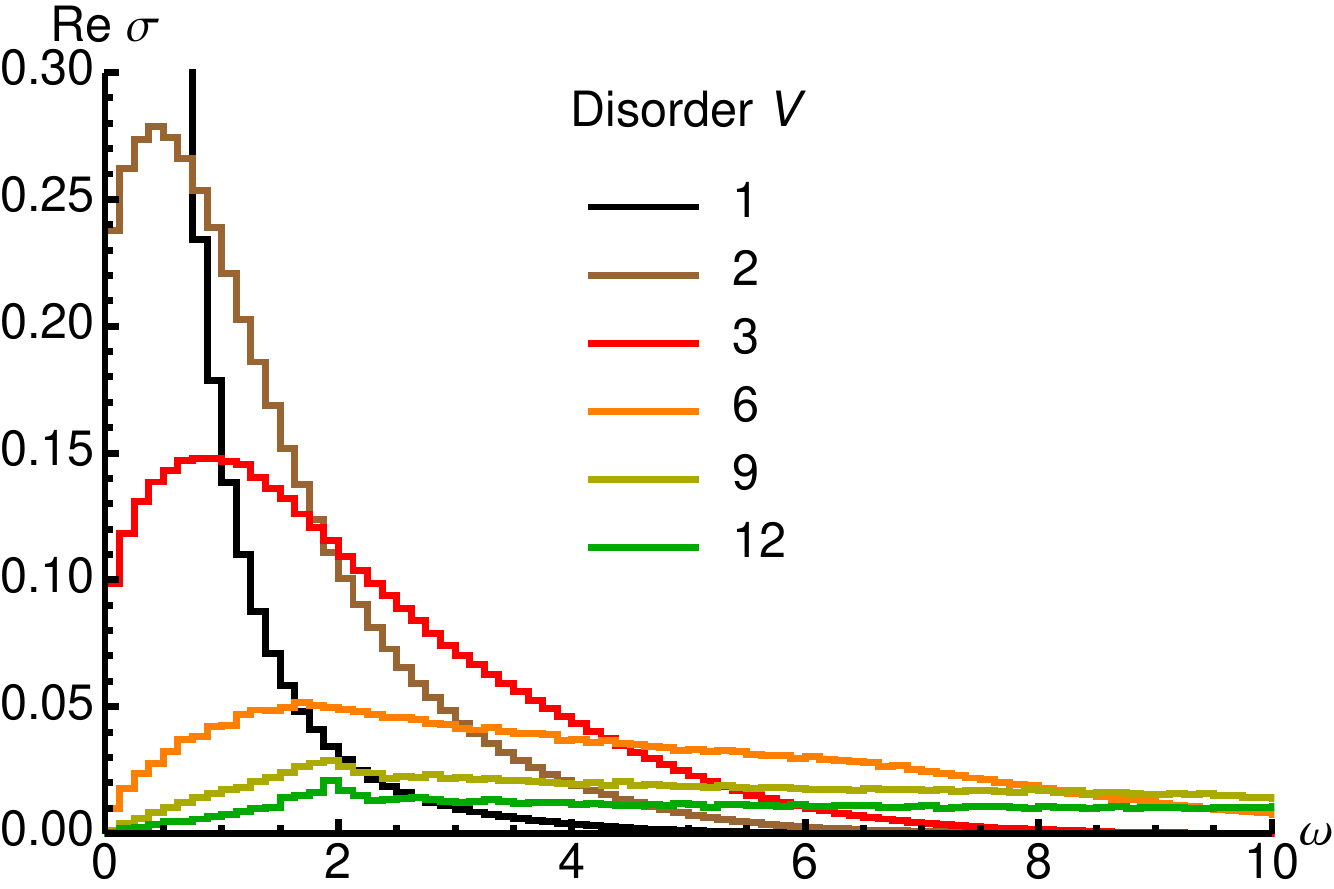}		\label{andersonReSigmaE}
	}
	\caption{
		\label{anderson}
		Properties of the Anderson model with box disorder on a $36\times 36$ square lattice.
		The first column shows the density of states $N(E)$.
		As disorder increases, the DOS gets flatter and more box-like,
			but no gaps appear.
		The second column shows the real part of the conductivity, $\Re \sigma(\omega)$,
		for chemical potential $\mu=0$.     
		At weak disorder, for the finite systems studied here,
			the dynamical conductivity has a large Drude-like peak.
		At strong disorder, the conductivity is suppressed,
			and it develops a ``soft gap'' (presumably due to Mott variable-range hopping),
			even though the density of states remains gapless.
		Note the suppression of $\sigma(\omega)$ at low frequency $\omega$.
		For an infinite 2D system, 
			the DC conductivity would be suppressed all the way to zero.
	}
	\end{figure*}

Many quantities can be computed by using the eigenmode representation.  
The (retarded) single-particle Green function is
	\begin{align}
	G_{ij} (E) 
	&=	\sum_\alpha \phi_{i\alpha}  \phi^*_{j\alpha} \frac{1}{E - E_\alpha + i0^+}
	\end{align}
where the infinitesimal shift in the denominator imposes causality.
From the Green function one can obtain single-particle spectral properties,
such as the local density of states (LDOS) $N_i(E)$, spectral function $A(\kkk,E)$, 
and total density of states (DOS) $N(E)$,
\footnote{The random potential breaks translational invariance, so that momentum is not a good quantum number; particles can scatter from $\kkk$ to $\kkk+\qqq$, and hence the spectral function and momentum distribution are also functions of the total momentum $\qqq$.  Averaging over many disorder realizations recovers translational invariance, such that only the $\qqq=\mathbf{0}$ term remains.}
	\begin{align}
	N_i(E) &= -\tfrac{1}{\pi} \Im G_{ii} (E)
	=  \sum_\alpha \delta(E - E_\alpha) \left| \phi_{i\alpha} \right|^2 ,
\\
	A(\kkk,E) &= -\tfrac{1}{\pi} \Im G_{\kkk\kkk} (E)
	=  \sum_\alpha \delta(E - E_\alpha) \left| \phi_{\kkk\alpha} \right|^2 ,
\\
	N(E) 
	&= \sum_i N_i(E) 
	= \sum_\kkk A(\kkk,E)
	=  \sum_\alpha \delta(E - E_\alpha)
,
	\end{align}
where $\phi_{\kkk\alpha}=\frac{1}{N} \sum_i e^{-i\kkk\cdot\rrr_i} \phi_{i\alpha}$ are the eigenfunctions in the momentum representation.
Note that sums of Dirac delta functions are most efficiently calculated by accumulating weights in bins.  
Furthermore, one can obtain static properties such as the local number density $\mean{n_i}$ and momentum distribution $\mean{n(\kkk)}$,
	\begin{align}
	\mean{n_i} 
	= \mean{\cdag_i \cccc_i}
	&= \int dE~ f(E) N_i(E)
	=  	\sum_\alpha f_\alpha \left| \phi_{i\alpha} \right|^2 ,
\\
	\mean{n(\kkk)} 
	= \mean{\cdag_\kkk \cccc_\kkk}
	&= \int dE~ f(E) A(\kkk,E)
	=  	\sum_\alpha f_\alpha \left| \phi_{\kkk\alpha} \right|^2 .
	\end{align}
where $f_\alpha$ are the Fermi occupation factors of the eigenmodes,
	\begin{align}
	f_\alpha &= f(E_\alpha) = \frac{1}{e^{\beta (E_\alpha - \mu)} + 1} 
	= \half - \half \tanh \tfrac{\beta (E_\alpha - \mu)}{2}
	.
	\end{align}
Figure \ref{andersonAE} shows the DOS for different disorder strengths.
The single-particle DOS is gapless.  For weak disorder ($V=1t$) it resembles the tight-binding DOS, whereas for strong disorder ($V=12t$) it approaches the uniform distribution of the disorder potential.

\section{Electromagnetic Response (DC and AC Conductivity)}
For the purposes of this article the most important two-particle property is the electromagnetic response tensor,
$\Upsilon_{\mu\nu} (\qqq,\omega) = -\frac{dj_{\nu} (\qqq,\omega)}{dA_\mu (\qqq,\omega)}$, which is the current response in direction $\nu$ to an applied vector potential with wavevector $\qqq$, frequency $\omega$, and polarization $\mu$.  
This can be calculated in the eigenbasis using the Kubo formula (see Sec.~\ref{AppendixAndersonKubo} for derivation),
	\begin{align}
	\frac{ \Im \Upsilon_{\mu\nu\qqq\omega}}{\pi}
	&= 
		\sum_{\alpha\beta} 
		\Gamma_{\alpha\beta\mu\qqq} \Gamma_{\beta\alpha\nu\bar\qqq}
		(f_\beta - f_\alpha)~
		 \delta(E_\alpha - E_\beta - \omega)
,\\
	\Re \Upsilon_{\mu\nu\qqq\omega}
	&=
		\mean{-k_{\mu\nu}}
+
	\calP \int_{-\infty}^{\infty} \frac{d\omega'}{\pi}~ 
  \frac{ \Im \Upsilon_{\mu\nu\qqq\omega'} }{ \omega - \omega' }
	,
	\end{align}
where
	\begin{align}
	\mean{-k_{\mu\nu}}
	&= \sum_{\alpha ij} r_{ij\mu} r_{ij\nu} t_{ij} 
		\phi^*_{i\alpha} \phi_{j\alpha} f_\alpha
,\nonumber\\
	\Gamma_{\alpha\beta\mu \qqq}
	&=\sum_{ij}  
		e^{-i\qqq\cdot\rrr_i} 
		r_{ij\mu}
		it_{ij}
		\phi^*_{i\alpha} \phi_{j\beta}
	.
	\end{align}
Here, $\mean{-k_{\mu\nu}}$ is the diamagnetic response, which is related to the mean kinetic energy,
and $r_{ij\mu}$ is the $\mu$th Cartesian component of the displacement vector from site $i$ to site $j$.

The dynamical conductivity (typically the optical or microwave conductivity) is 
$\sigma(\omega) = \frac{\Upsilon(\omega,\qqq=0)}{i\omega}$.  
Figure \ref{andersonReSigmaE} shows the behavior of $\Re \sigma(\omega)$ for the Anderson model at various disorder strengths. 
In an infinite 2D system, the DC conductivity $\sigma(\omega=0)$ is zero for all $V>0$.
However, because the calculations were done on finite 2D systems, there is a small finite DC conductivity.
Nevertheless, it can be seen that the AC conductivity is strongly suppressed at low frequencies.  
At weak disorder, this can be interpreted in terms of weak localization.
At strong disorder, this can be understood according to the theory of Mott variable-range hopping (VRH), which predicts the following temperature and frequency dependence of the conductivity:
	\begin{align}
	\sigma(\omega=0,T) &\propto \exp \left[-(T_0/T) ^ {1/(d+1)} \right]
,\\
	\Re \sigma(\omega,T=0) &\propto \omega^2 \left[\ln (\text{const}/\omega) \right]^4
	\end{align}
(where $d=2$ is the dimensionality).

\section{Summary}
In this section we have seen that the 2D Anderson model is generically an insulator.
For Anderson insulators, the single-particle spectrum is gapless, but the conductivity has a ``soft gap'' due to weak localization or variable-range hopping.  Due to the presence of low-lying fermionic excitations, an Anderson insulator can be classified as a Fermi insulator.

In the next section we will see that including attraction leads to a different type of insulator, a Bose insulator, in which the single-particle spectrum has a hard gap.

\chapter{Disorder-Tuned Superconductor-Insulator Transition} \slabel{DisorderTunedSIT}

\section{Clean Superconductor} \slabel{CleanSuperconductor}
We now proceed to the case of a clean superconductor modeled by the attractive Hubbard model on a square lattice.  The Hamiltonian is
	\begin{align}
	H
	&=
	-	\sum_{ij\sigma} t_{ij} \cdag_{i\sigma} \cccc_{j\sigma}
	-	\mu \sum_{i\sigma}		n_{i\sigma}
	-	\sum_{i}			U \cdag_{i\up} \cdag_{i\dn} \cccc_{i\dn} \cccc_{i\up}
	.
	\end{align}

\subsection{BCS mean-field theory (MFT)}
Most of the properties of this model can be understood within BCS mean-field theory.\cite{degennes,tinkham,bcs-long}
Here, in order to obtain quantitative results and to set the stage for the treatment of disorder, we explicitly take into account the lattice densities of states and Hartree corrections.

The quartic interaction is decoupled in terms of a uniform pairing potential $\Delta=U\mean{\cccc_{i\dn} \cccc_{i\up}}$ and a Hartree chemical potential $\mu^H = U \mean{n}$. 
Up to a constant,
\footnote{In textbooks, BCS mean-field decoupling is often performed by writing
$\cccc_{i\dn} \cccc_{i\up} = \frac{\Delta}{|U|} + (\cccc_{i\dn} \cccc_{i\up} - \frac{\Delta}{|U|} )$,
expanding $\cdag_{i\up} \cdag_{i\dn} \cccc_{i\dn} \cccc_{i\up}$ using the binomial theorem,
and expanding up to first order in the term in parentheses.
However, for decoupling in two or more channels, this method over-counts the Hubbard interaction, and it is not clear what the constant term in the Hamiltonian should be.  The rigorous variational formalism in  \sref{variationalFormalism} resolves these problems.}
	\begin{align}
	H_\text{MF}
	&=
	-	\sum_{ij\sigma} t_{ij} \cdag_{i\sigma} \cccc_{j\sigma}
	- (\mu + \mu^H) \sum_{i\sigma}		n_{i\sigma}
	-	\sum_{i} (\Delta^* \cccc_{i\dn} \cccc_{i\up} + \Delta \cdag_{i\up} \cdag_{i\dn})
	.
	\end{align}
Since the Hamiltonian has translational symmetry, it can be diagonalized by Fourier transforming to momentum space:
	\begin{align}
	H_\text{MF}
	&=\sum_{\kkk} \left(
		\sum_\sigma \xi_{\kkk} \cdag_{\kkk\sigma} \cccc_{\kkk\sigma}
	-	\Delta^* \cccc_{-\kkk\dn} \cccc_{\kkk\up} - \Delta \cdag_{\kkk\up} \cdag_{-\kkk\dn}
		\right)
	\end{align}
where $\xi_\kkk = \vare_\kkk - \mu - \mu^H$ and $\vare_\kkk = -2t(\cos k_x + \cos k_y)$ as defined earlier.  This can be written in a $2\times 2$ matrix form.  Up to a constant,
	\begin{align}
	H_\text{MF}
	&=\sum_{\kkk}
		\pmat{\cdag_{\kkk\up} & \cccc_{-\kkk\dn} }
		\pmat{\xi_{\kkk} & -\Delta \\ -\Delta & -\xi_{\kkk} }
		\pmat{\cccc_{\kkk\up} \\ \cdag_{-\kkk\dn} }		
	.
	\end{align}
The matrix can be further diagonalized by a Bogoliubov transformation of the fermion operators to bogolon creation and annihilation operators $\gamma$,
	\begin{align}
		\pmat{\cccc_{\kkk\up} \\ \cdag_{-\kkk\dn} }		
	&=
		\pmat{u_{\kkk} & v_{\kkk} \\ -v_{\kkk} & u_{\kkk} }
		\pmat{\gamma_{\kkk\up} \\ \gamma^\dag_{-\kkk\dn} }		
	\end{align}
where
$u_\kkk = \cos \theta_\kkk$, 
$v_\kkk = \sin \theta_\kkk$,
$E_{\kkk}=\sqrt{\xi_\kkk{}^2 + \Delta^2}$,
and
$\tan 2\theta_\kkk =  \frac{\Delta}{E_\kkk}$.
The Hamiltonian is bilinear in the bogolon operators:
	\begin{align}
	H_\text{MF}
	&=\sum_{\kkk}
		\pmat{\gamma^\dag_{\kkk\up} & \gamma_{-\kkk\dn} }
		\pmat{E_\kkk & 0 \\ 0 & -E_\kkk}
		\pmat{\gamma_{\kkk\up} \\ \gamma^\dag_{-\kkk\dn} }		
	=\sum_{\kkk\sigma}
		E_{\kkk} \gamma^\dag_{\kkk\sigma} \gamma_{\kkk\sigma}
	\end{align}
(up to a constant).  Therefore, expectations of bogolon operators are simply Fermi occupation factors:
$	\mean{\gamma^\dag_{\kkk\sigma} \gamma_{\kkk\sigma}}
	=f_\kkk = f(E_\kkk) = \half - \half \tanh \frac{\beta}{2} E_\kkk
$.

The ``pair density'' (or anomalous Green function) on each site is
	\begin{align}
	F
	&=\mean{\cccc_{i\dn} \cccc_{i\up}}
	=
		\frac{1}{N} \sum_{\kkk} \mean{\cccc_{-\kkk\dn} \cccc_{\kkk\up}}
	\nonumber\\
	&=\int_\kkk
		\mean{
			(-v_\kkk \gamma^\dag_{\kkk\up} + u_\kkk \gamma_{-\kkk\dn})
			(u_\kkk \gamma_{\kkk\up} + v_\kkk \gamma^\dag_{-\kkk\dn})
		}
	\nonumber\\
	&=\int_\kkk
		u_\kkk v_\kkk \mean{
		-	\gamma^\dag_{\kkk\up} \gamma_{\kkk\up}
		+	\gamma_{\kkk\dn} \gamma^\dag_{\kkk\dn} 
		}
	= \int_\kkk
		u_\kkk v_\kkk (1 - 2f_\kkk )
	\nonumber\\
	&=\int_\kkk
		\frac{\Delta}{2E_\kkk}
		\tanh \frac{E_\kkk}{2T}
	\end{align}
where $\int_\kkk \equiv \int_{BZ} \frac{d^2 k}{(2\pi)^2}$.
The pairing amplitude, or order parameter (OP), is self-consistently determined by $\Delta=\left| U \right| F$, leading to the OP equation
\footnote{In the literature, $\Delta$ is often called the ``gap'', and the self-consistent equation for $\Delta$ is called the ``gap equation''.  We will avoid this terminology, because, as we shall see, in dirty superconductors near the SIT, the gap $E_g$ and the order parameter $\Delta$ are completely distinct quantities.}
	\begin{align}
	\frac{1}{\left| U \right|}
	&=\int_\kkk
		\frac{1}{2E_\kkk}
		\tanh \frac{E_\kkk}{2T}
	.
	\end{align}
The zero-temperature order parameter $\Delta_0$ and the mean-field critical temperature $T_c^\text{MF}$ are given by setting $T=0$ and $\Delta=0$ respectively in the OP equation:
	\begin{align}
	\frac{1}{\left| U \right|}
	&=\int_\kkk
		 \frac{1}{2\sqrt{\xi_\kkk{}^2 + \Delta_0 {}^2}} 
	=\int_\kkk
		 \frac{1}{2\xi_\kkk} \tanh \frac{\xi_\kkk}{2T_c^\text{MF}}
	.
	\end{align}
	
The number density on each site is
	\begin{align}
	n
	&= \mean{\cdag_{i\up} \cccc_{i\up} + \cdag_{i\dn} \cccc_{i\dn}}
	=\frac{1}{N} \sum_{\kkk} \mean{
		\cdag_{\kkk\up} \cccc_{\kkk\up}
	+	\cdag_{\kkk\dn} \cccc_{\kkk\dn}
	}
	\nonumber\\
	&=\int_\kkk
		\sum_\sigma \left(
		 u_\kkk {}^2 \mean{\gamma^\dag_{\kkk\sigma} \gamma_{\kkk\sigma}}
	+ v_\kkk {}^2 \mean{\gamma_{\kkk\sigma} \gamma^\dag_{\kkk\sigma}}
	 \right)
	\nonumber\\
	&=
		1+
	 \int_\kkk
		 \frac{\xi_\kkk}{E_\kkk}
		\tanh \frac{E_\kkk}{2T}
	\end{align}
(after some algebra). 
This depends (through $\xi$) on the chemical potential $\mu$ and Hartree potential $\mu^H$.  The OP equation and number equation can be iterated to self-consistency.

\begin{figure*}
	\subfigure[
		Properties as a function of on-site attraction $U$ for $n=0.875$,
			calculated within mean-field theory.
		The blue curve is the zero-temperature order parameter $\Delta_0$.
    The red curve is the mean-field critical temperature $T_c^\text{MF}$.
    The green curve is the zero-temperature superfluid stiffness $\Upsilon_0 \sim \rho_s$, 
			which is proportional to the phase fluctuation temperature $T_\theta$.
		The true $T_c$ (not shown) is bounded above by $T_c^\text{MF}$ and $T_\theta$. 
		\label{cleanscUdependence}
		]{
		\includegraphics[width=0.48\textwidth]{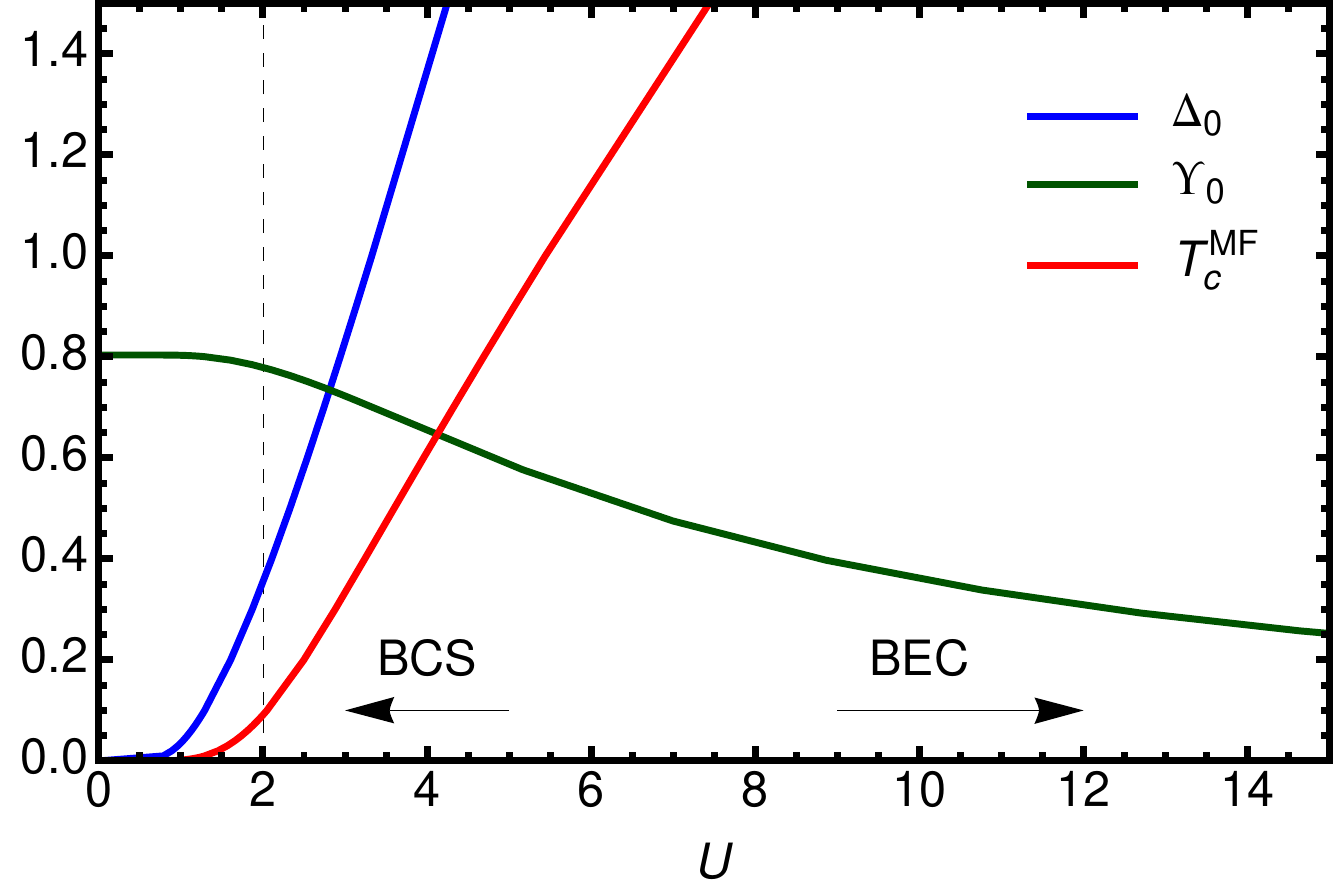}
	}
	\hspace{0.04\textwidth}
	\subfigure[
		Properties as a function of temperature for $n=0.875$ and $\left|U\right|=2t$,
			corresponding to the dashed line in Fig.~\ref{cleanscUdependence}.
		All quantities have an Arrhenius behavior ($e^{-\Delta_0/T}$) at low temperature.
		At $T_c$, $\Delta$ has a square-root singularity, 
			whereas $\Upsilon$ goes linearly to zero.
		The entropy per site $s$ has a kink,
			corresponding to a jump in the specific heat $c$.
		\label{cleanscTdependence}
		]{
		\includegraphics[width=0.48\textwidth]{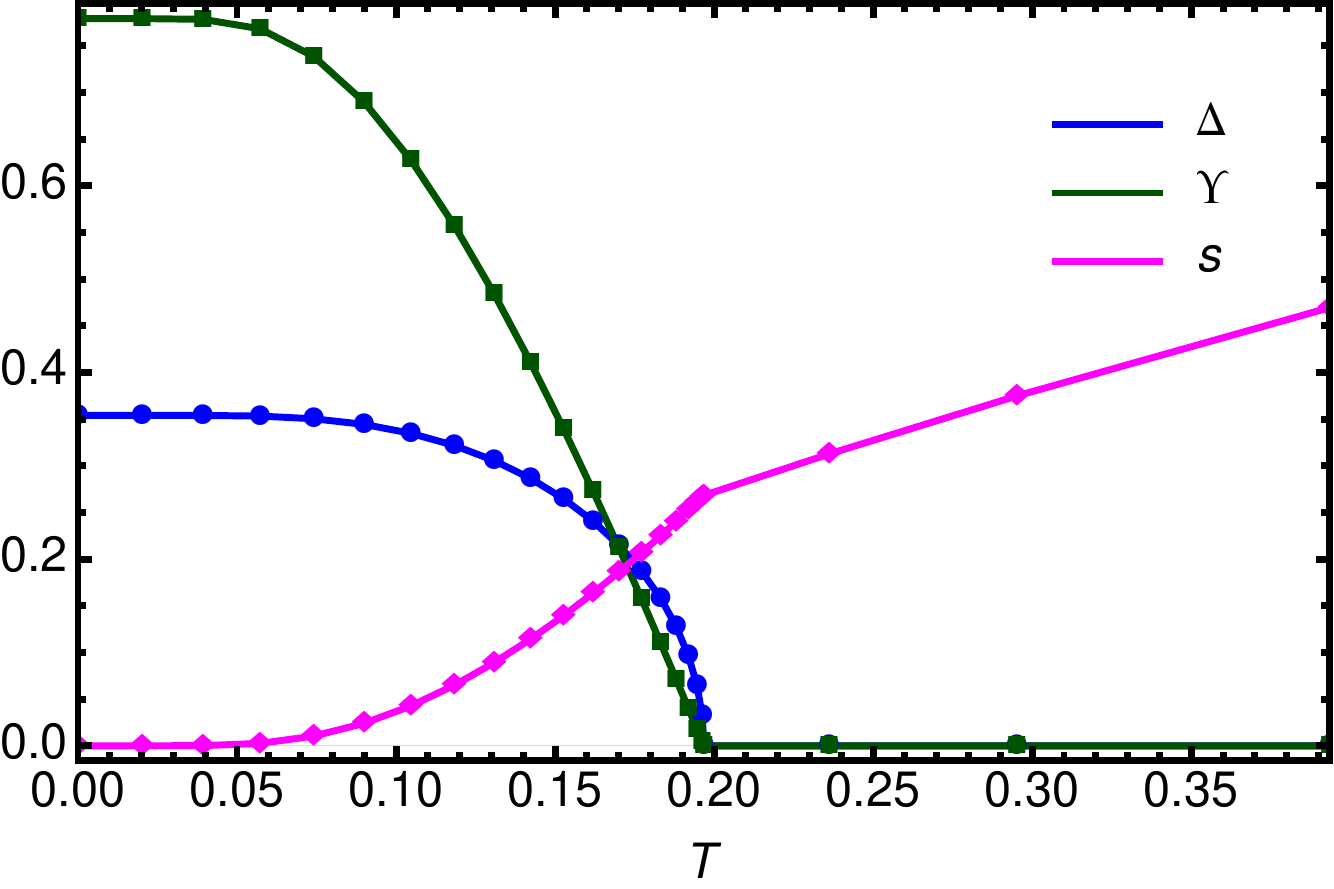}
	}
	\caption{
		Properties of the attractive Hubbard model on a square lattice within mean-field theory.
		All energies are in units of the hopping amplitude $t$.
	}
	\label{cleansc}
	\end{figure*}

Figure~\ref{cleanscUdependence} shows quantities as functions of attraction, $\left|U\right|/t$, at a general filling $n=0.875$.
This illustrates the important dichotomy between amplitude physics (pairing) and phase physics (coherence).
Superconductivity requires both pair formation,
	which occurs below the mean-field critical temperature $T_c^\text{MF}$,
	and phase coherence,
	which is governed by the phase stiffness $\Upsilon$ (or $\rho_s$).
The superconducting critical temperature $T_c$ is determined by the lower of these two energy scales.
In the weak-coupling limit, $T_c^\text{MF} \ll \Upsilon$, and the transition is well described by BCS mean-field theory.
In the strong-coupling limit, $T_c^\text{MF} \gg \Upsilon$, so the transition is dominated by phase fluctuations.  In that case, $T_c^\text{MF}$ is a pseudogap temperature corresponding to pair formation.  This is the BEC limit, in which composite bosons become superfluid at low temperatures.
For the purposes of this chapter we shall work in the BCS limit, which is relevant to the materials used in SIT experiments; however, we will see that phase fluctuations nevertheless become important near the SIT.

Figure~\ref{cleanscTdependence} shows the temperature dependence of the order parameter at $\left|U\right|=2t$ (corresponding to the dashed line in Fig.~\ref{cleanscUdependence}).
The ``strong-coupling ratio'' $2\Delta/T_c$ is about 4.  For comparison, BCS theory in the continuum gives the universal number $2\Delta/T_c = \frac{2\pi}{e^\gamma} \approx 3.53$, where $\gamma$ is the Euler-Mascheroni constant.

\subsection{Single-particle spectrum}
\begin{figure}[!hbt]
	\subfigure[Density of states, showing coherence peaks at $E=\pm\Delta$.
		In this case, the inverse square root singularities are further enhanced
		by the nearby logarithmic van Hove singularity.
	]{
		\includegraphics[width=0.48\textwidth]{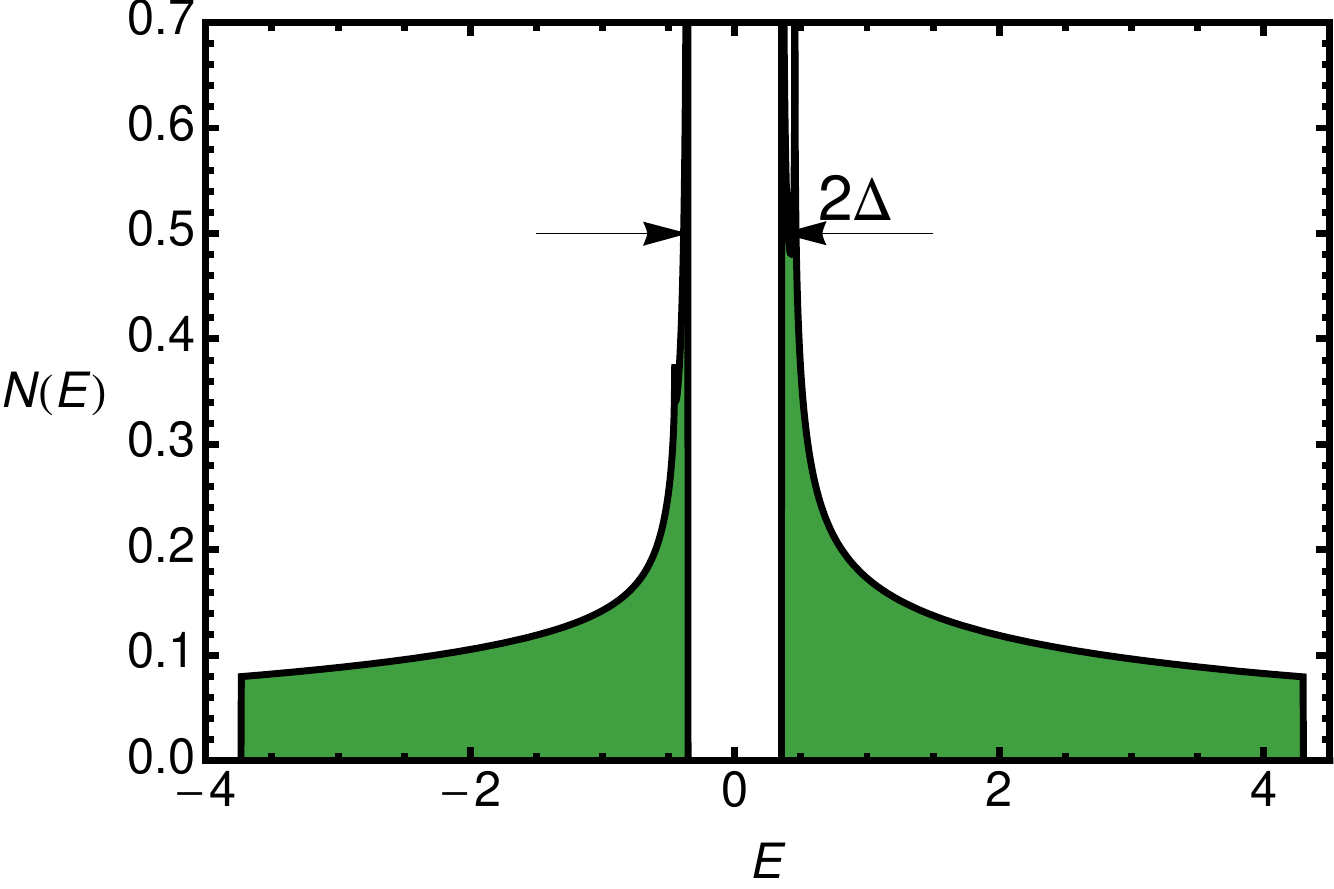}
		\label{cleanscDOS}
	}
	\hspace{0.04\textwidth}
	\subfigure[Spectral function $A(\kkk,\omega)$, 
		showing back-bending due to particle-hole mixing.]{
		\includegraphics[width=0.48\textwidth]{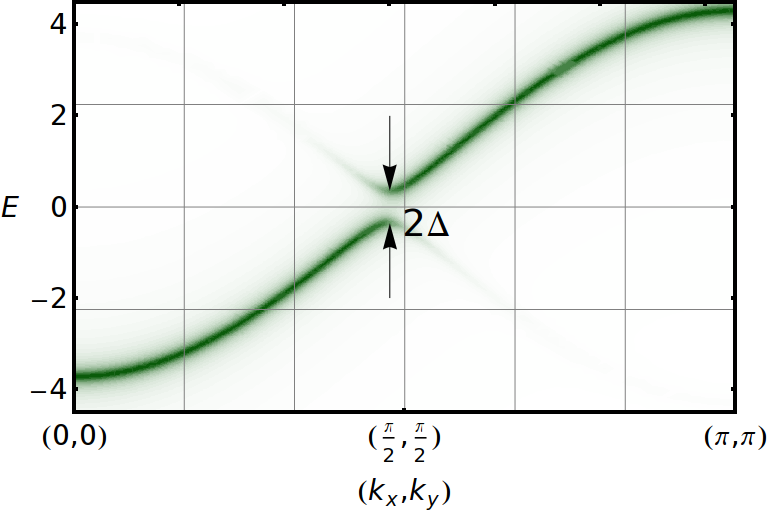}
		\label{BCSSpectralFunction}
	}
	\caption{
		Spectral properties of a clean superconductor within mean-field theory
		for $U=-2$, $n=0.875$, and $T=0$, for which $\Delta=0.354$.
		All energies are in units of the hopping amplitude $t$.
	}
	\end{figure}

We have already derived $\Delta$ and $n$ above.  Now we consider other single-particle properties.
The single-particle Green function is
	\begin{align}
	G_\kkk (\tau)
	&=\mean{  \cccc_{\alpha\tau} \cdag_{\alpha} }
	=u_\kkk {}^2  \mean{ \gamma^\dag_{\kkk\tau} \gamma_\kkk} 
	+ v_\kkk {}^2  \mean{ \gamma_{\kkk\tau} \gamma^\dag_\kkk}
	=u_\kkk {}^2  G_{\alpha\tau}
	- v_\kkk {}^2  G_{\alpha,-\tau}
	\end{align}
where $G_{\alpha\tau}$ is the standard Green function for a fermionic eigenmode as described in \sref{matsubara} and $0<\tau<\beta$.  Fourier transforming, analytically continuing to real frequencies, and taking the imaginary part gives the spectral function,
	\begin{align}
	A_\kkk (E)
	&=u_\kkk {}^2  \delta ( E - E_\alpha )
	+ v_\kkk {}^2  \delta ( E + E_\alpha )
	.
	\end{align}
The spectral function has a pole of strength $u_\kkk {}^2$ at positive frequencies, along the gapped bogolon dispersion relation, and a pole of strength $v_\kkk {}^2$ at negative frequencies.  This structure is illustrated in Fig.~\ref{BCSSpectralFunction}.

The density of states can be obtained by integrating $A_\kkk (E)$ over wavevectors $\kkk$.  Reducing the 2D integral to a 1D integral over the tight-binding DOS and using identities for Dirac delta functions leads to
	\begin{align}
	N(E)
	&=
		\frac{ \sgn E }{2}
		\Bigg[
			\left( \frac{E}{\sqrt{E^2 - \Delta^2}} + 1 \right)
			N_\text{TB} \left( \mu + \sqrt{E^2 - \Delta^2} \right)
	\nonumber\\&~~~~~~~{}
		+
			\left( \frac{E}{\sqrt{E^2 - \Delta^2}} - 1 \right)
			N_\text{TB} \left( \mu - \sqrt{E^2 - \Delta^2} \right)
		\Bigg]
	\end{align}
where $N_\text{TB} (\vare)$ is the tight-binding density of states given in \eref{square-lattice-dos}.  The resulting density of states has BCS coherence peaks (inverse square root divergences) at $E = \pm \Delta$ as well as van Hove singularities at $E=\sqrt{\mu^2+\Delta^2}$ and $E=\sqrt{(\mu \pm 4t)^2+\Delta^2}$, as illustrated in Fig.~\ref{cleanscDOS}.
There is a clearly-defined gap in the spectrum $E_g$, equal to $\Delta$.  Hence, it is common practice to use the terms ``gap'' $E_g$ and ``order parameter $\Delta$'' interchangeably.  However, as we will see, this is very misleading when discussing the SIT.  
\footnote{Another place where it is misleading is in Abrikosov-Gor'kov theory, where a large concentration of magnetic impurities can produce gapless superconductivity with a finite order parameter but zero spectral gap.}

\subsection{Electromagnetic response}
One of the most important properties of a superconductor is its superfluid stiffness or phase stiffness $\Upsilon$.  For a clean SC on a square lattice, the Kubo formula (see Sec.~\ref{AppendixCleanSCKubo}) gives
	\begin{align}
	\Upsilon
	=	\Upsilon_{xx}
	&=\mean{-k_{xx}} - \Lambda_{xx}
	,\\
	\mean{k_{\mu\nu}}
	&=\sum_{\kkk} \frac{\dd^2 \vare_\kkk}{\dd k_\mu \dd k_{\nu}}
			\frac{\xi_\kkk}{E_\kkk}
			(2f_\kkk - 1)
	,\\
	\Lambda_{xx}
	&=
-
		\frac{8}{N} \sum_\ppp
		\sin^2 p_x
	\frac{\dd f(E_\ppp)}{\dd E_\ppp}
	\end{align}
where $\frac{\dd f}{\dd E_\ppp} = -\frac{\beta}{4} \sech^2 \frac{\beta}{2} E_\ppp$.  
The terms $\mean{-k_{xx}}$ and $\Lambda_{xx}$ are often referred to as ``diamagnetic'' and ``paramagnetic'' contributions respectively. 
At $T=0$, $\Lambda_{xx}=0$ vanishes, whereas for $T>T_c$, $\mean{-k_{xx}}=\Lambda_{xx}$ and $\Upsilon=0$ as required.
The behavior of $\Upsilon$ is illustrated in Figure~\ref{cleanscUdependence} and Figure~\ref{cleanscTdependence}.

We refer the reader to SWZ for a discussion of the finite-frequency conductivity, pausing only to mention that a perfectly clean Hubbard model superconductor has pathological properties.

With the exception of $\Lambda_{xx}$, all of the 2D integrals appearing in this section can be reduced to 1D integrals by writing $\int_\text{BZ} \frac{d^2 k}{(2\pi)^2} f(\xi_\kkk) \equiv \int_{-\infty}^{\infty} d\xi~ N_\text{TB} (\xi+\mu) f(\xi)$, where $N_\text{TB} (E)$ is the square lattice tight-binding DOS defined in \eref{square-lattice-dos}.

Rather than dwelling on more sophisticated treatments, we will now go on to the problem of a disordered superconductor.

\section{Dirty Superconductor}
We represent a dirty superconductor by an attractive Hubbard model with a disorder potential,
	\begin{align}
	H
	&=
	-	\sum_{ij\sigma} t_{ij} \cdag_{i\sigma} \cccc_{j\sigma}
	- \sum_{i} \mu_i n_{i\sigma}
	-	U\sum_{i} \cdag_{i\up} \cdag_{i\dn} \cccc_{i\dn} \cccc_{i\up}
	,
	\end{align}
where	$\mu_i = v_i - \mu$, where the disorder potential at each site $v_i$ is picked independently from a uniform distribution on $[-V,+V]$, as before.

\section{Atomic Limit} 
\begin{figure}
		\centering
		\includegraphics[width=0.5\textwidth]{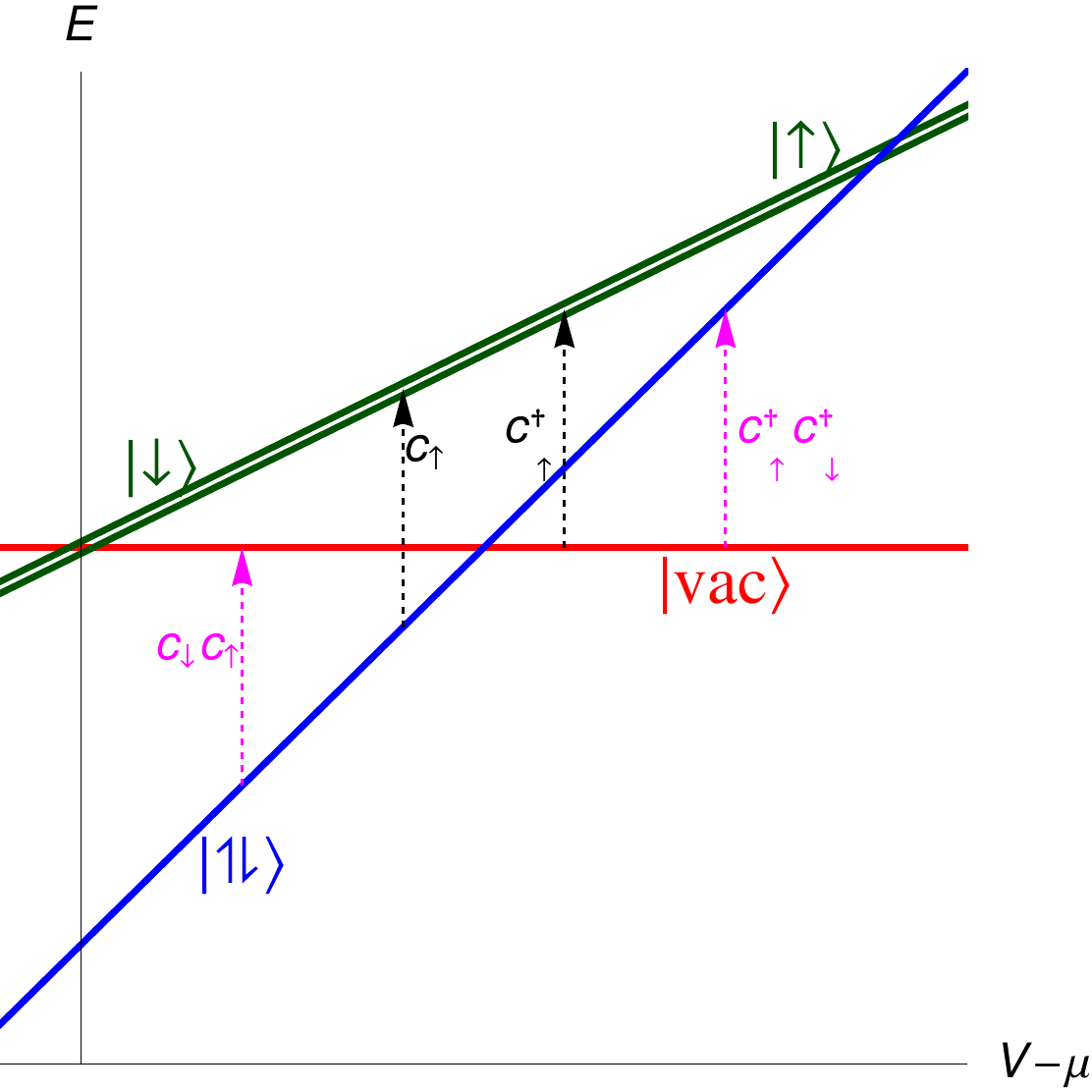}
	\caption{
		\label{AtomicLimit}
		Energy levels $E$ and single-particle and two-particle transition energies
			for a single site with potential $V$, chemical potential $\mu$,
			and attraction $U$ (the atomic limit of the Hubbard model).
	}
	\end{figure}

\slabel{atomicLimit}
In the limit of extreme disorder, the hopping can be neglected, and the system then reduces to an ensemble of single-site Hubbard models, each with the Hamiltonian
	\begin{align}
	H
	&=U n_{\up} n_{\dn}	+ (V - \mu) ( n_{\up} + n_{\dn}	)
	.
	\end{align}
This system has just four Fock states.  The energies of these states are
	$	E_{0}	= 0	$,
	$ E_\up = E_\dn = V - \mu $, 
	and
	$	E_{\updownarrow} = U + 2(V - \mu) $
	.
The four states occur with relative Boltzmann weights $\exp (-\beta E_n)$.
The spectral function (the density of states for single-particle excitations) can be obtained by considering transitions between these four Fock states (amplitudes and energies).
This is illustrated in Fig.~\ref{AtomicLimit}.
Regardless of the on-site potential $V$, single-particle transitions (black arrows) always cost at least $|U|/2$, and therefore the spectrum is always gapped.
\footnote{This is in contrast to the repulsive Hubbard model in the atomic limit, for which the spectrum is only gapped if $U > 2V$.
}
Pair excitations (purple arrows), however, may cost zero energy if $V$ is just right.
This is understood in the literature in terms of the ``parity gap.''
	
We have generalized the above calculation to exact diagonalization of the many-body Hubbard Hamiltonian on small clusters of a few sites, which leads the same conclusions: single-particle excitations are gapped whereas two-particle excitations can be gapless.

We know that a clean s-wave superconductor ($V=0$) has a gap $E_g = \Delta$ given by the BCS gap equation.
We have just found that in the limit of extreme disorder, $V \gg (U,t)$, the gap is finite and large,  $E_g = U/2$.
We willl see that the gap remains finite between these two extremes.

\section{Pairing of Exact Eigenstates (PoEE)}

The above Hamiltonian contains three terms: hopping, disorder, and attraction.  
In typical $s$-wave superconductors, the first two terms have the largest energy scales.
Thus, it makes sense to solve the non-interacting problem first by direct diagonalization, to find the 
disorder eigenvalues and eigenstates $\xi_\alpha$ and $\phi_{i\alpha}$ (as in \sref{anderson}), and then examine the effect of $U$. 
This is very much in the spirit of Anderson's original derivation of Anderson's theorem.
	 
In the basis of exact eigenstates, the Hamiltonian is 
	\begin{align}
	H
	&=
	 \sum_{\alpha}	\xi_\alpha \gamma^\dag_{\alpha\sigma} \gamma_{\alpha\sigma}
	-	U\sum_{\alpha\beta\gamma\delta i}			
		\phi_{i\alpha} \phi_{i\beta} \phi^*_{i\gamma} \phi^*_{i\delta}
		 \cdag_{\alpha\up} \cdag_{\beta\dn} \cccc_{\gamma\dn} \cccc_{\delta\up}
	.
	\end{align}
Following Anderson's suggestion, let us assume that instead of pairing between $\kkk$ and $-\kkk$, we have pairing between time-reversed eigenstates $\alpha$ and $\bar\alpha$ (i.e., complex conjugate eigenfunctions).
Retain only those terms in the Hamiltonian that connect such eigenstates:
	\begin{align}
	H_\text{PoEE}
	&=
	 \sum_{\alpha}	\xi_\alpha \beta^\dag_{\alpha\sigma} \beta_{\alpha\sigma}
	-	U\sum_{\alpha\beta i}			
		\phi_{i\alpha} \phi_{i\bar\alpha} \phi^*_{i\bar\beta} \phi^*_{i\beta}
	 \cdag_{\alpha\up} \cdag_{\bar\alpha\dn} \cccc_{\bar\beta\dn} \cccc_{\beta\up}
	\nonumber\\
	&=
	 \sum_{\alpha}	\xi_\alpha \beta^\dag_{\alpha\sigma} \beta_{\alpha\sigma}
	-	\sum_{\alpha\beta}			
		M_{\alpha\beta}
	 \cdag_{\alpha\up} \cdag_{\bar\alpha\dn} \cccc_{\bar\beta\dn} \cccc_{\beta\up}
	\end{align}
where
$		M_{\alpha\beta}
		=
	U\sum_i \left| \phi_{i\alpha} \right|^2  \left| \phi_{i\beta} \right|^2
$
.
Approximate this by a mean-field Hamiltonian 
	\begin{align}
	H_\text{MF}
	&=
	 \sum_{\alpha}	\xi_\alpha \beta^\dag_{\alpha\sigma} \beta_{\alpha\sigma}
	-	\sum_{\beta} 
	(
		\Delta^*_\beta
    \cccc_{\bar\beta\dn} \cccc_{\beta\up}
  +h.c.
  )
	\end{align}
(up to a constant), where the order parameter is
	\begin{align}
	\Delta^*_\beta
	&=
		U
		\sum_{\alpha}
		M_{\alpha\beta}
	\mean{	 \cdag_{\alpha\up} \cdag_{\bar\alpha\dn}   }
	\end{align}
(assuming that $\xi_\alpha$ have been redefined in this step to include Hartree shifts).

The gap equation works out to be
	\begin{align}
	\Delta_\alpha
	&=
		U\sum_{\beta}
		M_{\alpha\beta}
	\frac{\Delta_\beta}{2E_\beta}
	\tanh \frac{E_\beta}{2T}
	\end{align}
where
$
	E_\beta
	=	\sqrt{\xi_\beta{}^2 + \Delta_\beta{}^2} 
$,
and the chemical potential is determined by the number equation
	\begin{align}
	\mean{n}
	&=	\frac{1}{N}
		\sum_\alpha \left( 1 - \frac{\xi_\alpha}{E_\alpha} \right)
		.
	\end{align}

The PoEE theory can be used in the above form, or one can perform further approximations as follows.  In the low-disorder regime, the disorder eigenstates $\phi_{i\alpha}$ are extended on the scale of the system, so that $M_{\alpha\beta} \approx 1/N$ independent of $\alpha$ and $\beta$.  In this limit Anderson's theorem applies -- the gap equation takes the simple BCS form, and $\Delta$ is spatially uniform.  In the high-disorder regime, on the other hand, the disorder eigenstates are strongly localized with localization lengths $ \xi^\text{loc}_\alpha$, and the $M$ matrix is approximately diagonal,
$M_{\alpha\beta} \approx \delta_{\alpha\beta} \sum_i \left| \phi_{i\alpha} \right|^4
\approx 
\delta_{\alpha\beta} / (\xi^\text{loc}_\alpha)^2
$.

\begin{figure}
	\subfigure[Single-particle gap from PoEE compared with BdG]{
		\includegraphics[width=0.5\textwidth]{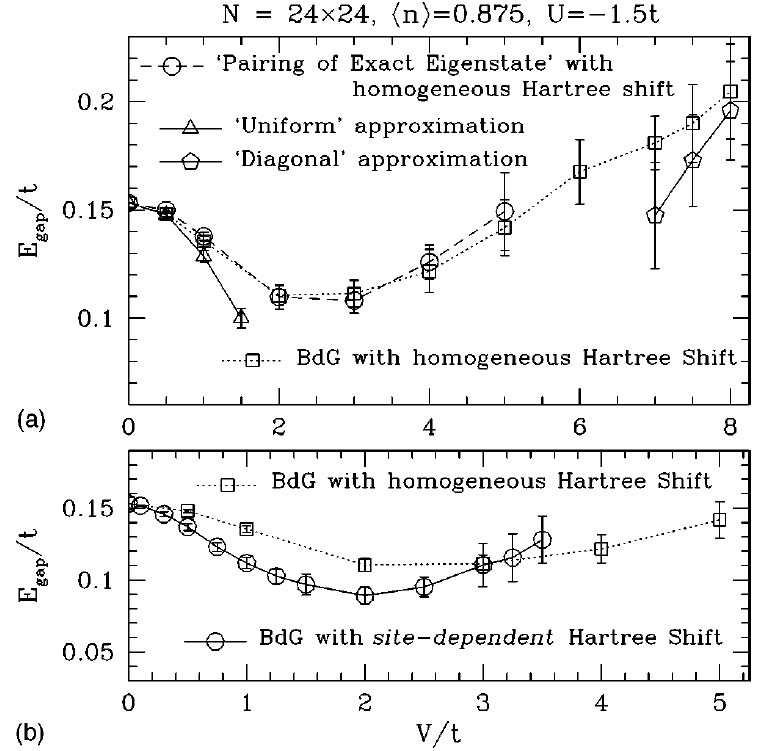}
		\label{ghosalPoEEvsBdG}
	}
	\subfigure[Local density of states from BdG]{
		\includegraphics[width=0.5\textwidth]{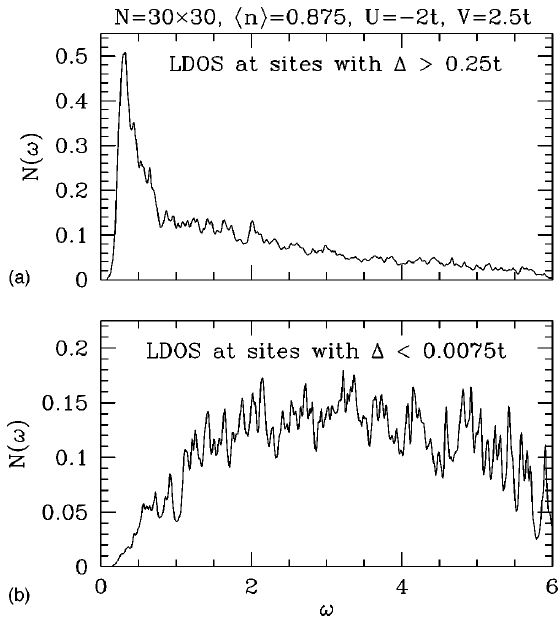}
		\label{ghosalLDOS}
	}
	\caption{
	}
	\end{figure}

The results of the PoEE theory implemented numerically, and in the low-disorder and high-disorder approximations, are compared in Fig.~\ref{ghosalPoEEvsBdG}.  
Surprisingly, \emph{the gap is finite for all values of disorder} $0<V<\infty$.  At large disorder, 
	\begin{align}
	E_g
	&=	\frac{U}{2\xi_\text{loc} {}^2}
	,
	\end{align}
where $\xi_\text{loc}$ is the localization length at the chemical potential.
At extremely high disorder one recovers the atomic limit described in \sref{atomicLimit}.

The PoEE approach is useful for understanding the robustness of the gap,
but it does not give the full story.
It predicts the the BCS coherence peaks in the density of states survive up to infinite disorder, whereas more accurate calculations show that they do not.  
Furthermore, being a mean-field theory, PoEE fails to capture the destruction of phase coherence at SIT due to quantum phase fluctuations.  We now proceed to more sophisticated treatments.

\section{Bogoliubov-de Gennes (BdG)} \slabel{BdG}

In the BdG formalism the Hubbard interaction is decoupled in terms of mean fields that are allowed to take arbitrary values at each site.  The most general decoupling involves six fields at every site (see \sref{variationalFormalism}):
	\begin{align}
	-	U \cdag_{i\up} \cdag_{i\dn} \cccc_{i\dn} \cccc_{i\up}
	\longrightarrow
	&~~{}
		\Delta^*_i \cccc_{i\dn} \cccc_{i\up}
	+	\Delta_i \cdag_{i\up} \cdag_{i\dn}
	\quad\text{(Bogoliubov)}
	\nonumber\\&{}
	+	\mu^H_{i\up} \cdag_{i\up} \cccc_{i\up}
	+	\mu^H_{i\dn} \cdag_{i\dn} \cccc_{i\dn}
	\quad\text{(Hartree)}
	\nonumber\\&{}
	+	h_{i}^+ \cdag_{i\dn} \cccc_{i\up}
	+	h_{i}^- \cdag_{i\up} \cccc_{i\dn}
	\quad\text{(Fock)}
	.
	\end{align}
Although in this section only pairing and density channels are required, 
we nevertheless present a decoupling in three channels (pairing, density, and $z$-magnetization) which we use later in \sref{DirtySCInZeemanField}.
In this formalism the mean-field Hamiltonian is
	\begin{align}
	H_\text{BdG}
	&=
	-	\sum_{ij\sigma} t_{ij} \cdag_{i\sigma} \cccc_{j\sigma}
	- \sum_{i\sigma}	(\mu + \mu^H_{i\sigma}) n_{i\sigma}
	-	\sum_{i} (\Delta^*_i \cccc_{i\dn} \cccc_{i\up} + \Delta_i \cdag_{i\up} \cdag_{i\dn})
	,
	\end{align}
where $\mu^H_{i\up}$, $\mu^H_{i\dn}$, and $\Delta_i$ are $3N$ parameters to be determined self-consistently.  This can be written in a $2N\times 2N$ matrix form:
	\begin{align}
	H_\text{BdG}
	&=\sum_{ij}
		\pmat{\cdag_{i\up} & \cccc_{i\dn} }
		\underbrace{
		\pmat{ - t_{ij} - \mutot_{i\up} \delta_{ij} & -\Delta_i \delta_{ij}
					\\ -\Delta_i \delta_{ij} & t_{ij} + \mutot_{i\dn} \delta_{ij} }
		}_{\mathbf{H}}
		\pmat{\cccc_{j\up} \\ \cdag_{j\dn} }		
	\end{align}
(up to a constant).  This is a real symmetric matrix, so its eigenvalues $E_\alpha$ and eigenvectors $\phi_\alpha$ are real.  It will be convenient to split the eigenvectors into particle parts 
$u_{i\alpha}=\phi_{i1;\alpha}$ and hole parts 
$v_{i\alpha}=\phi_{i2;\alpha}$ (note that the $u$ and $v$ in our 3-channel formalism are different from the u's and v's in traditional BdG).
Then, the fermion operators can be expressed in terms of bogolon operators $\gamma_\alpha$ as
 	\begin{align}
	\cccc_{i\up} &= \sum_\alpha u_{i\alpha} \gamma_\alpha        ,\qquad
	\cccc_{i\dn} = \sum_\alpha v_{i\alpha} \gamma^\dag_\alpha        .
	\end{align}
The Hamiltonian is bilinear in the bogolon operators:
	\begin{align}
	H_\text{BdG}
	&=\sum_{\alpha} E_\alpha \gamma^\dag_\alpha \gamma_\alpha
	\end{align}
(up to a constant), so expectations of bogolon operators are easy to calculate, e.g., 
$\mean{\gamma^\dag_\alpha \gamma_\alpha} = f_\alpha = 1/(\exp \beta E_\alpha + 1)$.
Expectations of fermion operators may be calculated by transforming to the bogolon basis; the derivations are quite similar to those in \sref{anderson}.
The basic recipe for a BdG calculation is as follows:

\begin{enumerate}
\item 
Make initial guesses for the internal fields (the Hartree potentials $\mu^H_{i\sigma}$ and the self-consistent pairing field $\Delta_i$).

\item 
Find the total (effective) fields at site $i$ by combining the external (applied) fields with the internal fields:
$\mutot_i = \mu_i + \mu^H_i$ and 
$\htot_i = h_i + h^H_i$.

\item 
Construct the $2N\times 2N$ Hamiltonian matrix $\mathbf{H}$.

\item 
Find the eigenvalues $E_\alpha$, eigenvectors $(u_{i\alpha}, v_{i\alpha})$, and occupation numbers $f_\alpha$.

\item 
Compute the number densities $n_{i\sigma}$ and the pairing density $F_{i}$ at every site $i$:
	\begin{align}
	n_{i\up} &= \sum_\alpha f_\alpha u_{i\alpha}{}^2
,\\
	n_{i\dn} &= \sum_\alpha (1 - f_\alpha) v_{i\alpha}{}^2
,\\
	F_i &= \sum_\alpha (f_\alpha - 1/2) u_{i\alpha} v_{i\alpha}
.
	\end{align}

\item 
Recompute the internal fields $\Delta_i := U F_i$, $\mu^H_{i\up} := U n_{i\dn}$, and $\mu^H_{i\dn} := U n_{i\up}$.

\item 
Go back to step 2. Repeat till convergence.
\end{enumerate}

There are several possible improvements to the above scheme.  The scheme, as described, uses fixed-point iteration to approach self-consistency.
This may converge slowly, or it may become unstable; convergence can be improved by introducing an empirically determined linear mixing factor $\gamma$:
	\begin{align}
	\Delta_i &:= \Delta_i + \gamma (U F_i - \Delta^i_i )
,\\
	\mu^H_{i\up} &:= \mu^H_{i\up} + \gamma (U n_{i\dn} - \mu^H_{i\up} )
,\\
	\mu^H_{i\dn} &:= \mu^H_{i\dn} + \gamma (U n_{i\up} - \mu^H_{i\dn} )
.
	\end{align}
One can go even further and use the Broyden method for multidimensional root-finding, which converges superlinearly sufficiently close to the solution.

For a given set of internal fields, one can calculate the variational free energy $\Omega$ (see \sref{variationalFormalism} for a derivation),
	\begin{align}
	\Omega &= 
	-T \sum_\alpha \ln (2 \cosh \half\beta E_\alpha) 
	+ \sum_i U(F_i^2 + x_i^2 - m_i^2)
	+ \sum_i 2(\Delta_i F_i + \mu^H_i x_i + h^H_i m_i )
	\end{align}
where $x_i = \half n_i - \half$ and $m_i = \half(n_{i\up} - n_{i\dn})$.
It is good practice to track the value of $\Omega$, to verify that the iteration is converging to a minimum and not to a saddle-point or maximum.

To study systems with fixed average densities $n^\text{target}_\sigma$, one can include Lagrange multipliers $\mu^\text{Lag}_\sigma$ as two additional variables in the self-consistency iteration.  However, in that case, the root-finding problem can no longer be rephrased as the problem of minimizing $\Omega$.

\subsection{Eigenstates}
\begin{figure}[!h]
	\includegraphics[width=\textwidth]{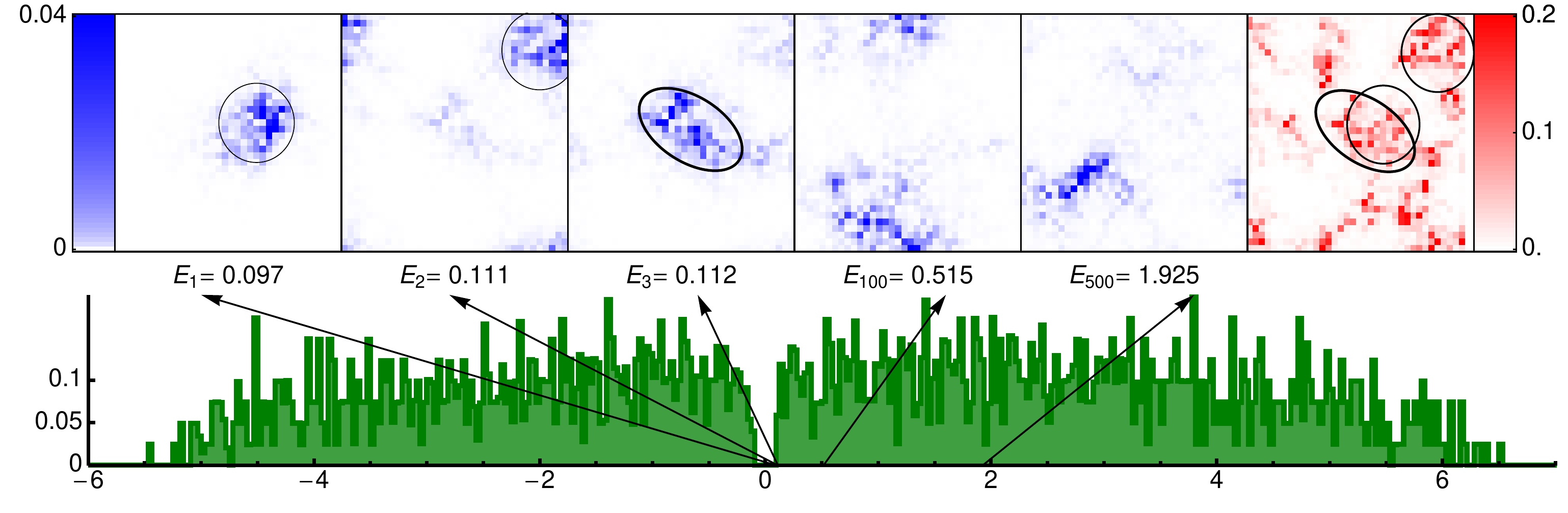}
	\caption{
		The first five panels show the magnitude of five BdG eigenstates (bogolon wavefunctions),
			$\left|u_i\right|^2 + \left|v_i\right|^2$.
		The last panel (red) is a map of the local pairing amplitude $\Delta_i$.
		The low- and high-energy eigenstates are localized, 
		whereas the intermediate-energy eigenstates are quasi-extended.
		In particular, the lowest eigenstates correspond to
		 the locations of the superconducting puddles (where $\Delta_i$ is large).
		The parameters were $U=-1.5t$, $n=0.875$, $N=36\times 36$, $V=3t$.
	}
	\label{bdgEigenstates}
	\end{figure}

Figure~\ref{bdgEigenstates} shows some of the BdG eigenstates (bogolon modes).
The lowest-energy modes are concentrated in the same locations as the superconducting puddles.  
In contrast, the higher-energy excitations correspond to breaking of localized pairs.

\subsection{Single-particle spectrum}
\begin{figure*}[!h]
	\includegraphics[width=\textwidth]{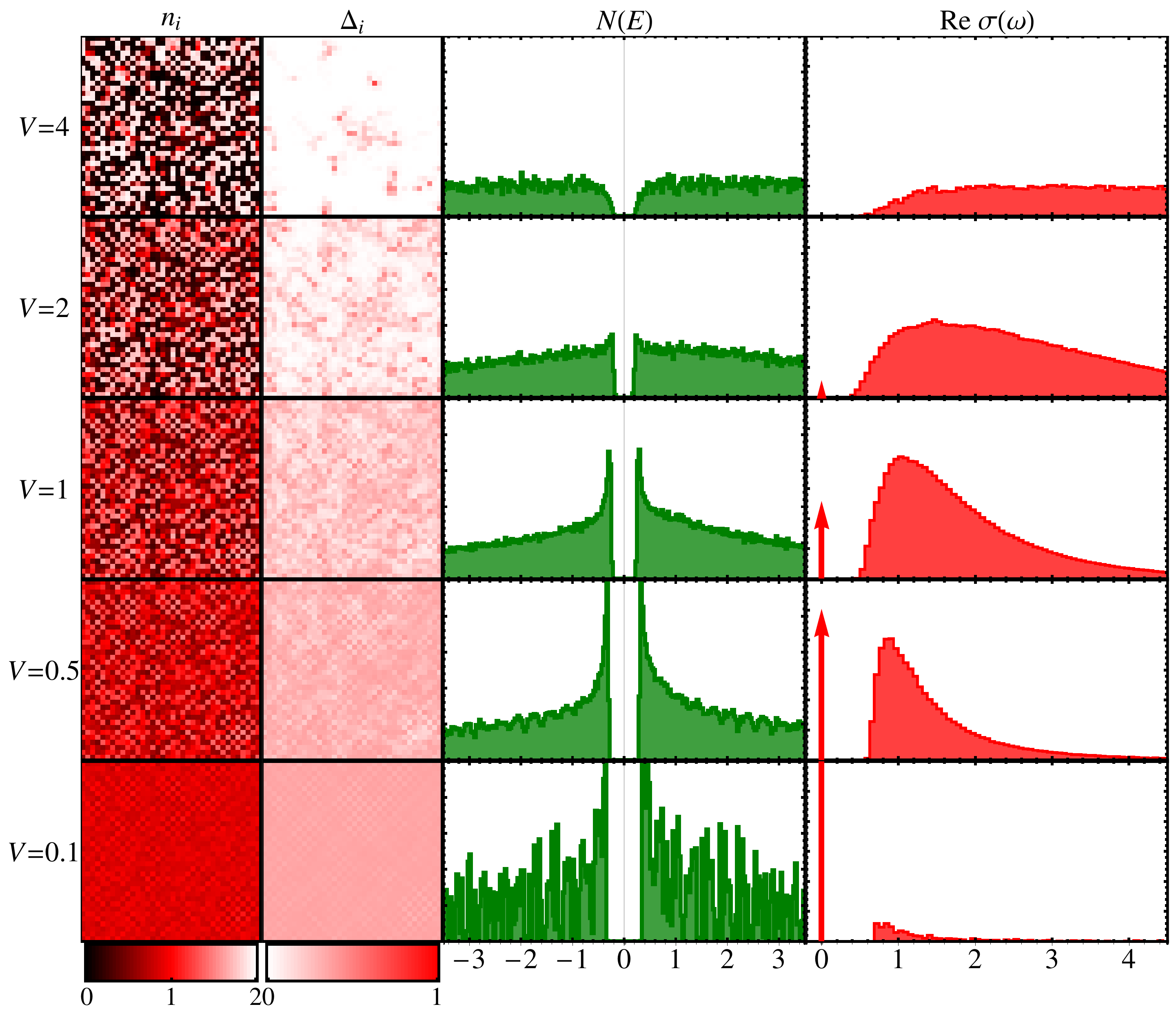}
	\caption{
		The four columns show the number density, pairing density,
			single-particle spectrum, and optical conductivity
			of the attractive Hubbard model with a disorder potential,
			on a $36\times 36$ square lattice 
			with $U=-2t$, $\mean{n}=0.875$, and $T=0$,
			within the BdG approximation.
		$N(E)$ and $\sigma$ are averaged over [10] disorder configurations.
	}
	\label{ghosal-everything}
	\end{figure*}

After the iteration has converged one may calculate further quantities of interest.
In the same manner as in \sref{anderson},
the single-particle Green functions, in the Matsubara frequency domain, are
	\begin{align}
	G_{ij\up} (i\vare_n) 
	&= \mean{\psi_{i\up} (i\vare_n)  \bar\psi_{j\up} (i\vare_n) }
	=	\sum_{\alpha\beta} u_{i\alpha} u_{j\beta}
		\mean{ \gamma_\alpha (i\vare_n) \bar\gamma_\beta (i\vare_n) }
	\nonumber\\
	&=	\sum_\alpha u_{i\alpha} u_{j\alpha} \frac{1}{i\vare_n - E_\alpha}
	,\\*
	G_{ij\dn} (i\vare_n) 
	&=	\sum_\alpha v_{i\alpha} v_{j\alpha} \frac{1}{i\vare_n + E_\alpha}
	,
	\end{align}
so the densities of states for up and down electrons are
	\begin{align}
	N_{i\up} (E) 
	&= \sum_\alpha \delta(E - E_\alpha)
		u_{i\alpha} {}^2
	,\\*
	N_{i\dn} (E) 
	&= \sum_\alpha \delta(E + E_\alpha) 
		v_{i\alpha} {}^2
	.
	\end{align}
Figure~\ref{ghosal-everything} shows the number density, pairing amplitude, and density of states from BdG calculations on $36\times 36$ lattices at various disorder strengths.  
At weak disorder, the number density and pairing amplitude are uniform, and the density of states has coherence peaks.  At strong disorder, most sites are empty ($n_i=0$) or doubly occupied ($n_i=2$) due to the on-site attraction.  
However, there are still locally superconducting puddles with intermediate occupation numbers $n_i$ and finite pairing amplitudes $\Delta_i$.
In fact, the sites with the smallest values of $\Delta_i$ tend to be the ones with the largest values of the local spectral gap $E_g$.

The BdG calculation shows that the spectral gap persists at all disorder strengths, and
the size of the gap is approximately in agreement with the prediction from PoEE (see Fig.~\ref{ghosalPoEEvsBdG}).  
However, unlike in PoEE, the coherence peaks are suppressed by disorder and vanish for $V\geq 2t$ (see Fig.~\ref{ghosal-everything}). 

	\begin{figure}
	\subfigure[$\left| U \right| = 2$]{
		\includegraphics[width=0.5\textwidth]{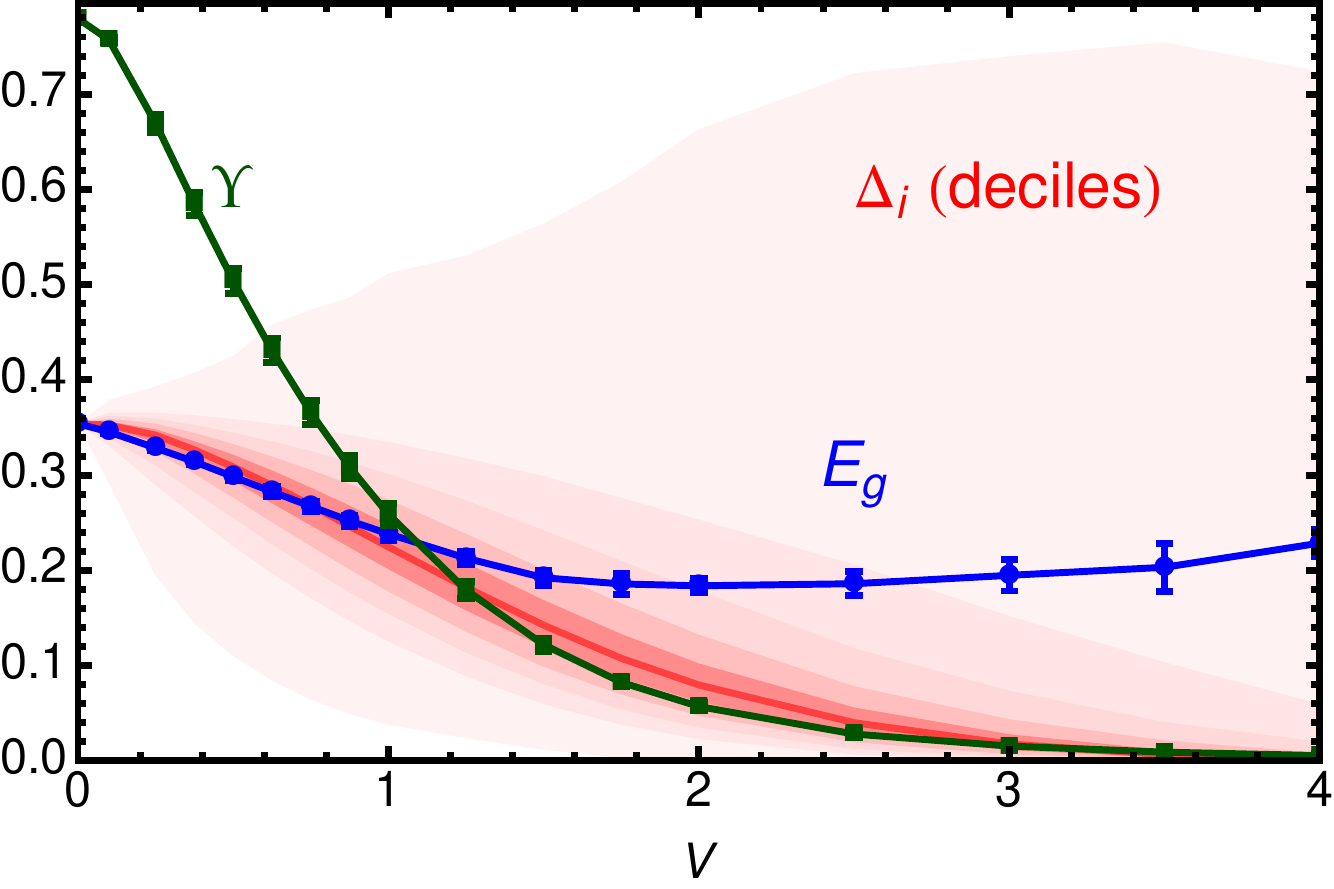}
	}
	\subfigure[$\left| U \right| = 4$]{
		\includegraphics[width=0.5\textwidth]{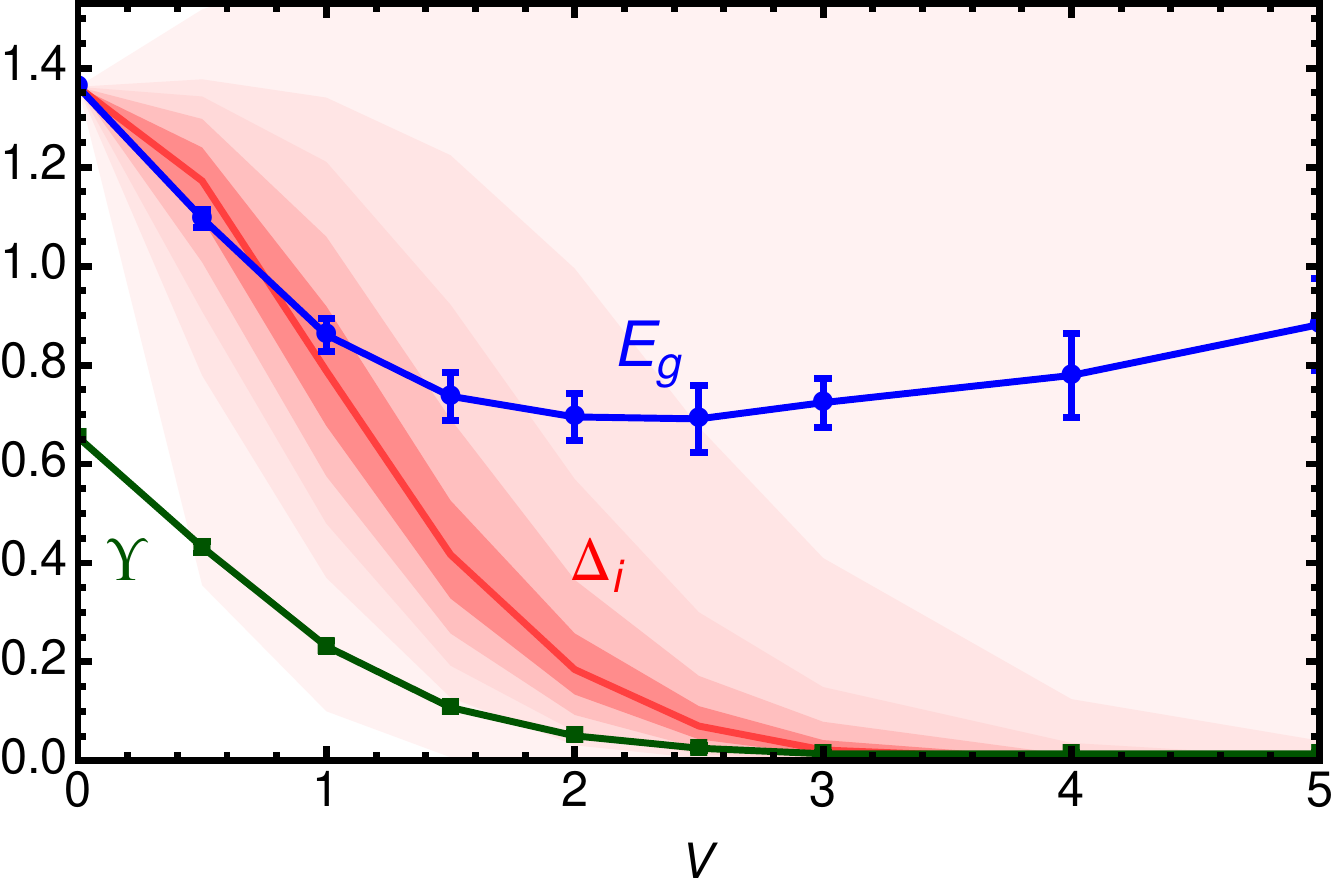}
	}
	\caption{
		BdG results at $T=0$ 
			on a $36\times 36$ lattice.
		The single-particle gap $E_g$ remains finite,
			whereas the stiffness $\Upsilon$ falls to very small values for $V>V_c$.
		The pairing amplitude $\Delta_i$ varies from site to site.
		The distribution, $P(\Delta_i)$, is visualized in terms of deciles; that is,
			successive red-colored bands represent the lowest 10\percent of $\Delta$ values,
			the next 10\percent, and so on up to the highest value of $\Delta$.
		The central red line is the median value of $\Delta$.
		All energy scales are in units of the hopping amplitude $t$.
	}
	\label{GapAndStiffnessVersusDisorder}
	\end{figure}

\subsection{Electromagnetic response}
The electromagnetic response function can be calculated using the Kubo formula (see Sec.~\ref{AppendixDirtySCKubo} for derivation):
	\begin{align}
	\frac{ \Im \Upsilon_{\mu\nu\qqq\omega}}{\pi}
	&= 
		\sum_{\alpha\beta} 
		\Gamma_{\alpha\beta\mu\qqq} \Gamma_{\beta\alpha\nu\bar\qqq}
		(f_\beta - f_\alpha)~
		 \delta(E_\alpha - E_\beta - \omega)
,\\
	\Re \Upsilon_{\mu\nu\qqq\omega}
	&=
		\mean{-k_{\mu\nu}}
+
	\calP \int_{-\infty}^{\infty} \frac{d\omega'}{\pi}~ 
  \frac{ \Im \Upsilon_{\mu\nu\qqq\omega'} }{ \omega - \omega' }
,
	\end{align}
where $\mean{-k}$ and $\Gamma$ now involve both ``spin components'' of the eigenvectors (\eref{BdGEMQuantities}) and the eigenmode indices $\alpha$ and $\beta$ now run from $1$ to $2N$:
	\begin{align}
	\mean{k_{\mu\nu}}
	&=- \sum_{\alpha ij} r_{ij\mu} r_{ij\nu} t_{ij} 
		\left(	
			u_{i\alpha} u_{j\alpha} 
		-	v_{j\alpha} v_{i\alpha}
		\right)
		f_\alpha
,\nonumber\\
	\Gamma_{\alpha\beta\mu \qqq}
	&=\sum_{ij}  
		e^{-i\qqq\cdot\rrr_i} 
		r_{ij\mu}
		it_{ij}
		\left(	
			u_{i\alpha} u_{j\beta} 
		-	v_{j\alpha} v_{i\beta}
		\right)
.
	\elabel{BdGEMQuantities}
	\end{align}

First consider the superfluid stiffness, $\Upsilon \equiv \Upsilon(\omega=0)$, which is plotted in Fig.~\ref{GapAndStiffnessVersusDisorder}.
As disorder increases, $\Upsilon$ decreases, but it never falls to zero.  Even at large disorder, there are rare regions that are relatively disorder-free and have a large order parameter.  These contribute to a finite stiffness within BdG theory.  
This means that \emph{BdG, by itself, does not capture the SIT}.

Now consider the finite-frequency response.  The real part of the dynamical conductivity from BdG calculations is of the form
	\begin{align}
	\Re \sigma(\omega) = \Upsilon_{xx} \pi \delta(\omega) - \frac{1}{\omega} \Im \Lambda_{xx} (\omega)
	.
	\end{align}
This quantity is shown in the rightmost column of Fig~\ref{ghosal-everything}.
The heights of the red arrows indicate the relative weights of the delta functions.
The weight of the delta function is the charge stiffness, which is equal to the superfluid stiffness $\Upsilon$ plotted in Fig.~\ref{GapAndStiffnessVersusDisorder}.
\footnote{In general, one has to be very careful when taking limits of the EM response function 
$\Upsilon_{xx} (\omega, q_x, q_y)$.
Different limits give three quantities -- charge stiffness $D$, superfluid stiffness $D_s$ ($\equiv\Upsilon$), and longitudinal response $\frac{ne^2}{m}$ -- which are, in general, different.
In this case, because the single-particle spectrum is gapped, it has been proven that charge stiffness and superfluid stiffness are equal.\cite{scalapinowhitezhang1993}
}
At weak coupling, the behavior is roughly in accord with Mattis-Bardeen theory.
Within the BdG approximation, $\sigma(\omega)$ has a hard gap that is twice the single-particle gap: $\omega_g = 2E_g$.  The weight above the gap grows as the disorder strength increases.
However, this is not the whole story.  Physically, if the superconductor-insulator transition is a continuous phase transition due to long-range phase coherence between superconducting puddles, one would expect that near the transition there should be low-frequency weight due to charge sloshing around between distant puddles; indeed, near a quantum critical point, one might expect $\sigma(\omega,T)$ to follow a universal scaling form possibly with low-frequency weight.  Thus the BdG results are questionable.  The true behavior of the conductivity is a topic of further research.

The BdG method gives a lot of insight into the nature of the SIT and of the ``Cooper pair insulator'' at large $V$, but it only models the amplitude of the order parameter.  
The fact that the pairing amplitude becomes highly inhomogeneous gives a picture of superconducting islands weakly coupled by Josephson tunneling, and points to the importance of phase fluctuations in driving the SIT, as illustrated in Fig.~\ref{GhosalPuddleCartoon}.
Generally, it appears that phase fluctuations are less important for single-particle properties, but more important for two-particle properties.
To gain a full understanding of the latter, it is necessary to include quantum phase fluctuations.

\section{Self-Consistent Harmonic Approximation}
There are several ways to treat phase fluctuations beyond BdG.  
In Ref.~\cite{ghosal2001}, a 2D quantum XY action in imaginary time was used to describe the dynamics of the phase variables $\theta(\rrr,\tau)$ defined on a coarse-grained square lattice of spacing $\xi$:
\newcommand{\dddelta}{\boldsymbol{\delta}}
	\begin{align}
	S_\theta
	=
	\frac{\kappa \xi^2}{8} \int_0^\beta d\tau \sum_\rrr 
	\left(  \frac{\dd \theta(\rrr,\tau)}{\dd \tau}  \right)^2
	+
	\frac{D_s^0}{4}  \int_0^\beta d\tau \sum_{\rrr\dddelta}
	\big\{    1 - \cos [\theta(\rrr,\tau) - \theta(\rrr+\dddelta,\tau)]
	\big\}
	\end{align}
where $\kappa$ was the mean-field static uniform compressibility
and $D_s^0$ was the mean-field phase stiffness
(referred to in this article as $\Upsilon$, up to a constant factor).
The authors made the approximation of ignoring the spatial variations of $\kappa$ and $D_s^0$.
Then, they performed a self-consistent harmonic approximation (SCHA) by choosing the optimal Gaussian action to minimize the free energy,
	\begin{align}
	S_\theta
	=
	\frac{\kappa \xi^2}{8} \int_0^\beta d\tau \sum_\rrr 
	\left(  \frac{\dd \theta(\rrr,\tau)}{\dd \tau}  \right)^2
	+
	\frac{D_s}{8}  \int_0^\beta d\tau \sum_{\rrr\dddelta}
	\Big[   \theta(\rrr,\tau) - \theta(\rrr+\dddelta,\tau)   \Big]^2
	,
	\end{align}
where the renormalized stiffness was
	\begin{align}
	D_s = D_s^0 \exp (-\mean{ \theta_{ij}^2 }_0 /2).
	\end{align}
Here $\mean{ \theta_{ij}^2 }_0$ is the mean square fluctuation of the nearest-neighbor phase difference
	\begin{align}
	\mean{ \theta_{ij}^2 }_0 
	=
	\frac{2}{\xi} \int_\QQQ  \left(  \frac{\vare_\QQQ}{D_s \kappa}  \right)^{1/2}
	\end{align}
where $\vare_\QQQ = 2(2 - \cos Q_x - \cos Q_y)$ and $\int_\QQQ$ is the average over the Brillouin zone.  Defining the renormalization factor $X = D_s / D_s^0$ and 
	\begin{align}
	\sqrt{\alpha}
	=
	\frac{1}{\xi \sqrt{D_s^0 \kappa} } \int_\QQQ {\vare_\QQQ}^{1/2},
	\end{align}
the SCHA equation becomes 
	\begin{align}
	X = \exp (-\sqrt{\alpha/X})
	.
	\label{GhosalEq23}
	\end{align}

\begin{figure}
	\centering
	\subfigure[
		\label{GhosalPuddleCartoon}
	]{
		\includegraphics[width=0.58\textwidth]{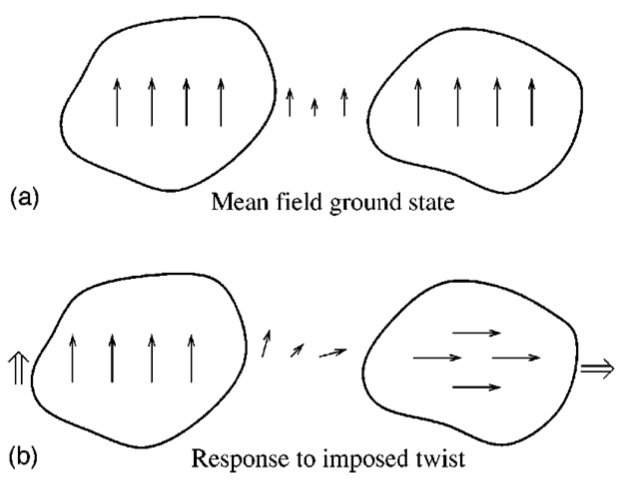}
	}
	\subfigure[
		\label{GhosalSCHAStiffness}
	]{
	\includegraphics[width=0.23\textwidth]{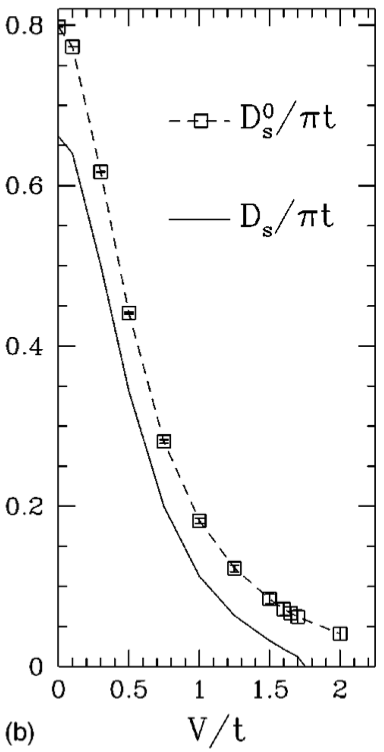}
	}
	\caption{
		(Left) 
		Schematic of a disordered SC in which the nonuniform amplitude
		results in the formation of SC islands.
		The length and direction of each arrow represents the amplitude and phase
			of $\Delta(\rrr)$.
		The phases on different islands are weakly coupled, so that
			the global phase stiffness $D_s^0$ is greatly reduced from
			that in a clean SC.
		(Right)
		Renormalization of the BdG phase stiffness $D_s^0$
		to the SCHA value $D_s$,
		which vanishes for $V > V_c \approx 1.75t$.
		[Figures from Ghosal et al. (2001);		
		parameters were $\left|U \right|=1.5t$ and $\mean{n}=0.875$,
		and the BdG was performed on a $24\times 24$ lattice.]
	}
	\end{figure}

Solving Eq.~\eqref{GhosalEq23} gives the renormalized stiffness $D_s$ as shown in Fig.~\ref{GhosalSCHAStiffness}.
We see that the BdG+SCHA approach \emph{successfully predicts a SIT} -- the renormalized stiffness at $V_c=1.75t$ for the parameters being considered.
The critical disorder obtained from such a calculation is in reasonable agreement with quantum Monte Carlo results for parameter values ($\left| U \right|=4t$) for which such a comparison can be made.

Despite the above success, the SCHA is ultimately a theory of Gaussian fluctuations.  It  predicts a transition at $\alpha_\text{crit}=4e^{-2}$ with a jump discontinuity of $e^{-2}$ in the value of $X$, which is probably an artifact.  The true transition is expected to be a continuous quantum phase transition in the universality class of the disordered Bose-Hubbard model.
\footnote{The actual universality class and critical exponents depend on whether particle-hole symmetry is satisfied exactly, on average, or not at all.  See the chapter by A. M. Goldman for a discussion of experimentally observed scaling.}

\section{Quantum Monte Carlo}
Compared to the PoEE and BdG approaches, 
determinant Quantum Monte Carlo (DQMC) is a computationally intensive but even more powerful computational technique.
\cite{hirsch1986}
It includes thermal and quantum fluctuations of the amplitude and phase,
and for the model being considered, it is free of the sign problem.
For a description of the DQMC algorithm as applied to repulsive Hubbard models, please refer to the chapter entitled ``Numerical Studies of Metal-Insulator Transitions in Disordered Hubbard Models'' by Chiesa, Scalettar, Chakraborty, Denteneer, Paiva, and Story.

\subsubsection{Superfluid stiffness}
\begin{figure}
	\centering
	\subfigure[
		\label{TrivediStiffness}
	]{
		\includegraphics[width=0.3\textwidth,height=4cm]{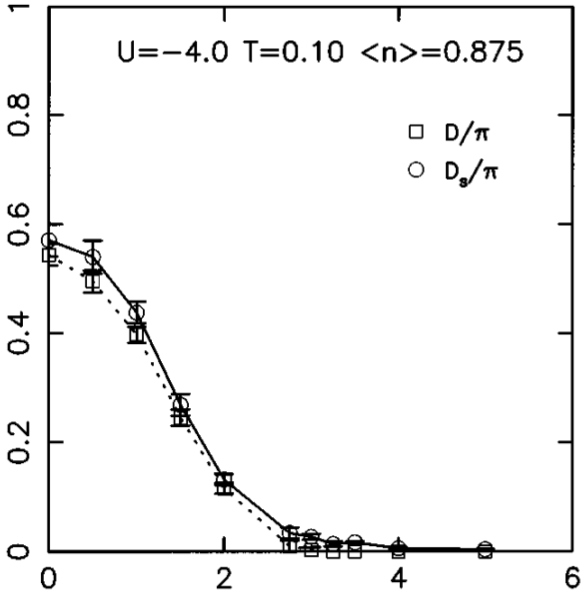}
	}
	\subfigure[
		\label{PhaseDiag}
	]{
		\includegraphics[width=0.3\textwidth,height=4cm]{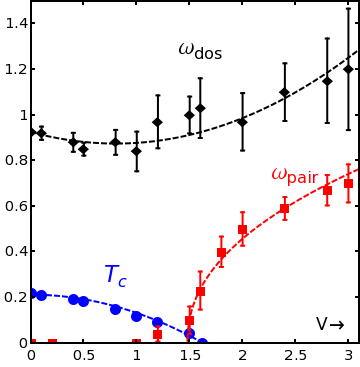}
	}
	\subfigure[
		\label{VscanQMCDOSAllInOne}
	]{
		\includegraphics[width=0.33\textwidth,height=4cm]{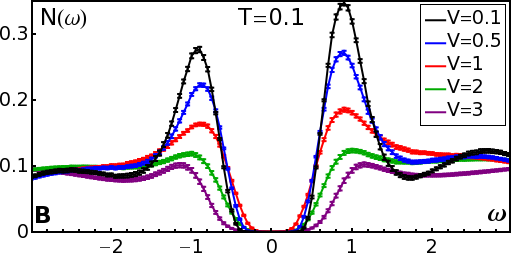}
	}
	\caption{
	(a)
		DQMC results for the low-temperature superfluid stiffness $D_s$ 
			as a function of disorder $2V$
			[Trivedi et al. (1996)].
			$D_s$ falls to zero at $V_c \approx 1.6$.		
	(b)
		DQMC+MEM results for the single-particle gap $E_g$ ($\equiv \omega_\text{dos}$),
			which remains finite for all $V$
			[Bouadim et al. (2011)].
	(c)
		DQMC+MEM results for the density of states $N(\omega)$.  
		With increasing $V$ the gap remains robust, 
			but the coherence peaks disappear beyond $V_c \approx 1.6$.
		All energies are in units of hopping $t$.
	}
	\end{figure}

Using DQMC, 
it was found that the superfluid stiffness $D_s$ does indeed fall to zero at a critical disorder $V=V_c$, as shown in Fig.~\ref{TrivediStiffness}.  Furthermore, the dc conductivity shows a crossing point between superconducting and insulating behavior, confirming that the model does indeed contain a SIT.  See Refs.~\onlinecite{trivedi1996,scalettar1999} for details.

\subsubsection{Single-particle spectrum}
In a more recent DQMC study \cite{bouadim2011arxiv}, 
dynamical quantities were obtained by analytic continuation of imaginary-time correlation functions using the maximum entropy method (MEM).
It was found that the single-particle gap is robust against disorder, and the coherence peaks are more or less correlated with the existence of long-range order (see Figs.~\ref{PhaseDiag} and \ref{VscanQMCDOSAllInOne}).  This is in agreement with the results of BdG+SCHA.

The implication is that the SIT in this model occurs via a \emph{bosonic} mechanism, involving the delocalization/localization of \emph{bound} pairs, rather than via a fermionic mechanism -- the state at large $V$ is a ``Cooper pair insulator.''
Indeed, recent experiments\cite{sacepe2008} provide evidence for such a bosonic SIT in InOx films.

\subsubsection{Two-particle spectrum}
A quantum phase transition is typically characterized by energy and temperature scales that vanish as the transition is approached from either side.  On the superconducting side, we have seen that $D_s$ and $T_c$ go to zero (Figs.~\ref{TrivediStiffness} and \ref{PhaseDiag}).
What are the relevant energy scales on the insulating side?
QMC results suggest that there is a peak in the spectrum of \emph{two-particle excitations} in the insulator, and the location of this peak moves to zero energy as the SIT is approached \cite{bouadim2011arxiv} (because pairs can be inserted into a sueprconductor at no cost).  However, this two-particle peak is not fully understood and is a topic of further research.

\subsubsection{Dynamical conductivity}
Whereas BdG predicts that the conductivity $\Re \sigma(\omega)$ has a hard gap $\omega_\sigma^\text{BdG} = 2E_g^\text{BdG}$, 
preliminary results from DQMC+MEM suggest that there is weight within this gap arising from phase fluctuations (i.e., vertex corrections beyond BdG).
In particular, the integrated low-energy weight $I = \int_0^{2E_g^\text{QMC}} d\omega~  \Re \sigma(\omega)$ was found to be finite, and it had a peak at $V \sim V_c$, suggesting quantum critical fluctuations near the SIT. 
However, there are complications from finite-temperature effects and from the difficulty of analytic continuation, and the study of $\sigma(\omega)$ is still a work in progress.

\section{Finite Temperature}
\begin{figure}
	\includegraphics[width=\textwidth]{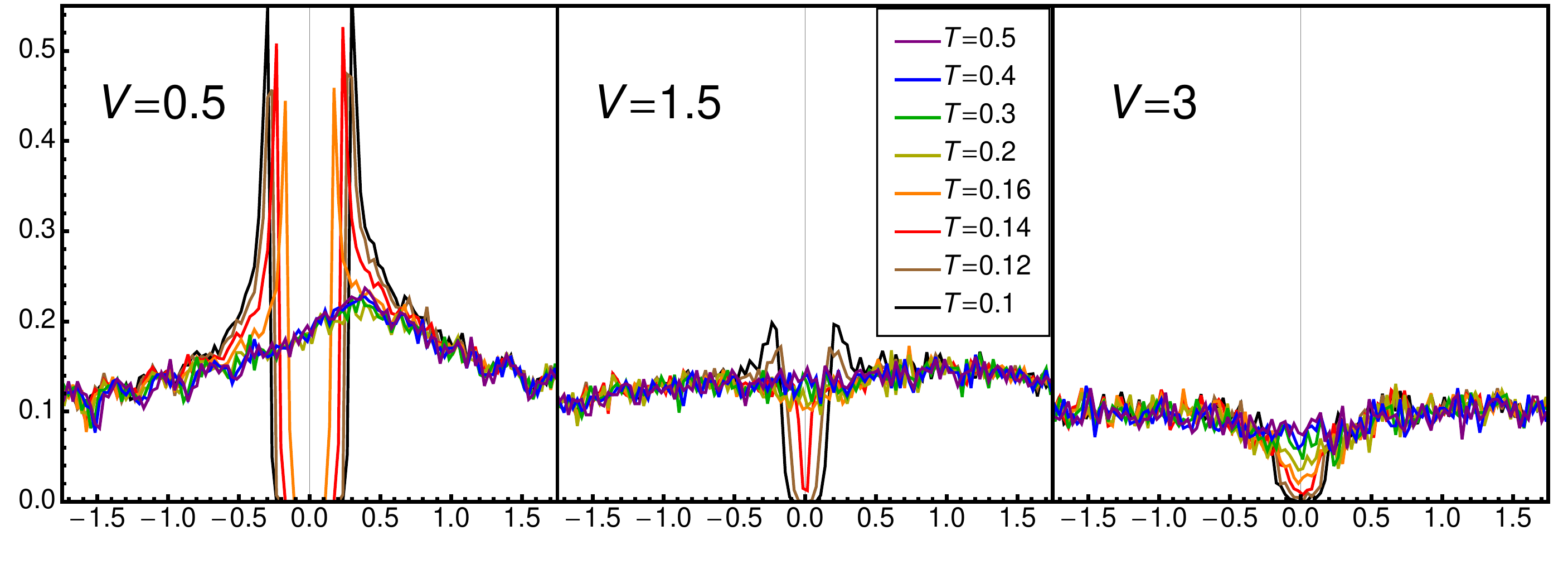}
	\includegraphics[width=0.48\textwidth]{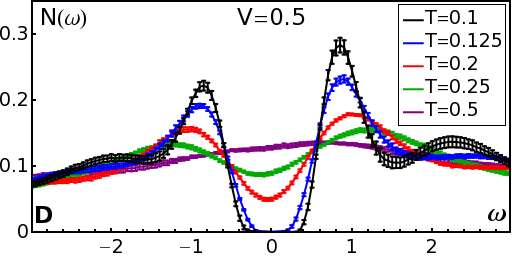}
	\includegraphics[width=0.48\textwidth]{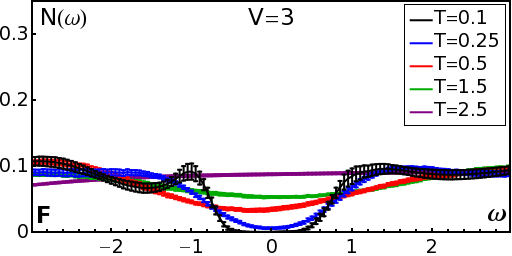}
	\caption{
		\label{BdGDisorderInducedPseudogap}
		\label{QMCDisorderInducedPseudogap}
		(Top)
		BdG results for $\left| U \right|=2$,
			showing the evolution of the density of states $N(\omega)$ 
			with disorder $V$ and temperature $T$.
		For weak disorder ($V=0.5$), the coherence peaks remain while the superconducting gap closes.
		For strong disorder ($V=3$), the insulating gap fills up slowly,
			forming a pseudogap (a suppression of the DOS near the Fermi level).
	(Bottom)
		DQMC results for $\left| U \right|=4$.
		There is an interaction-induced pseudogap at weak disorder,
			but otherwise the physics is essentially the same.
		All energies are in units of the hopping amplitude $t$.
	}
	\end{figure}

The superconductor-insulator transition is a quantum phase transition that occurs at zero temperature, so we have concentrated our effort on the zero-temperature behavior of the model.  
The zero-temperature phase diagram nevertheless has important ramifications for finite-temperature properties.  For example, we have found that the spectral gap $E_g$ at $T=0$ persists even on the insulating side of the SIT.  What is the effect of temperature?  The only sensible scenario is that at finite temperature $0 < T \lesssim E_g$, the gap must fill up gradually to form a pseudogap.  This expectation is borne out by actual BdG calculations at finite temperature, as illustrated in Fig.~\ref{BdGDisorderInducedPseudogap}.

\begin{figure}
	\subfigure[$\left| U \right|=2t$]{
		\includegraphics[width=0.5\textwidth]{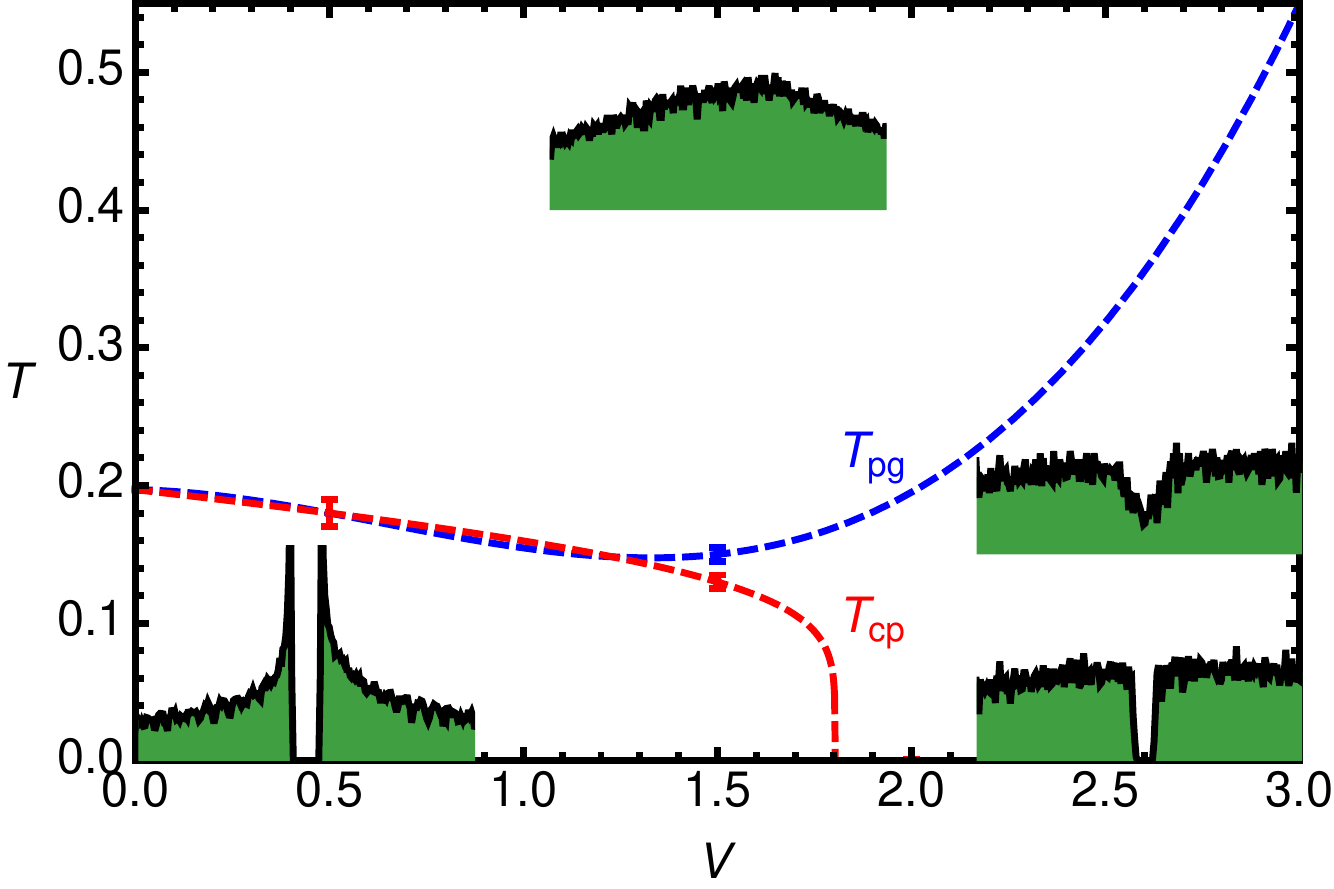}
		\label{BdGTpgAndTcpU2}
	}
	\subfigure[$\left| U \right|=4t$]{
		\includegraphics[width=0.5\textwidth]{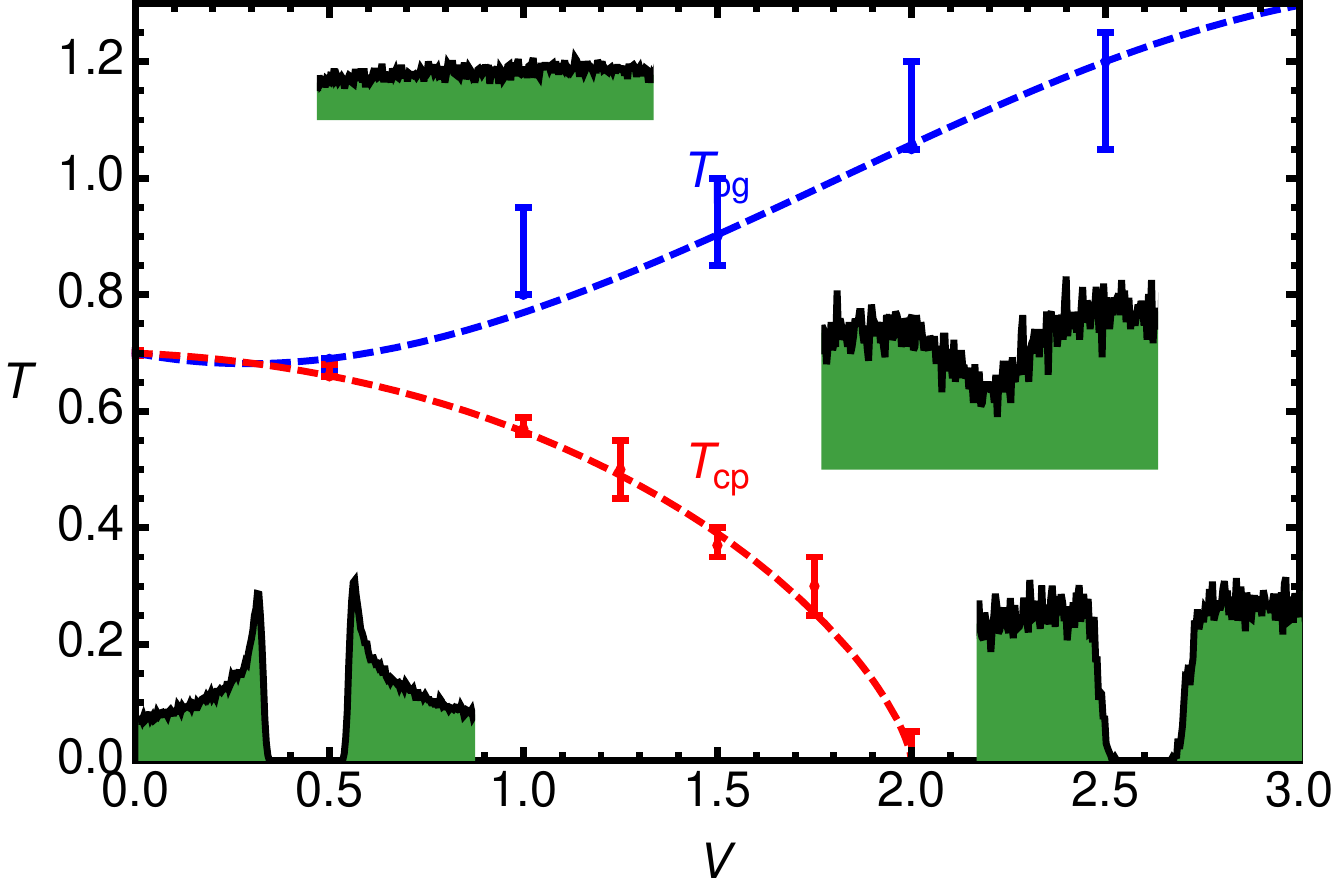}
		\label{BdGTpgAndTcpU4}
	}
	\caption{
		BdG results
			showing the temperature $T_{cp}$ below which the DOS has coherence peaks,
			and the temperature $T_{pg}$ above which the pseudogap disappears.
		These are crossover temperatures, but nevertheless, 
			a clear qualitative trend is visible.
		All energy scales are in units of the hopping amplitude $t$.
	}
	\label{BdGTpgAndTcp}
	\end{figure}

By examining the behavior of the DOS as a function of disorder $V$ and temperature $T$, as in  Fig.~\ref{BdGDisorderInducedPseudogap}, one can identify crossover temperature scales $T_{cp}$ and $T_{pg}$ for features in the DOS (coherence peaks and pseudogaps), as shown in Figs.~\ref{BdGTpgAndTcpU2} and \ref{BdGTpgAndTcpU4}. 
Although the critical temperature $T_c$ for spontaneous order within BdG theory is largely unaffected by disorder (because it is controlled by the existence of rare clean regions), the coherence peaks seem to be an indication of whether the system ultimately has phase coherence or not -- even though BdG neglects phase fluctuations.

Figure~\ref{QMCDisorderInducedPseudogap} shows DQMC+MEM results for the density of states at finite temperature \cite{bouadim2011arxiv}.  Due to the relatively large coupling ($\left| U \right|=4$) there is already an interaction-induced pseudogap at weak disorder.  Nevertheless, it was found that the size of the pseudogap temperature range increased with disorder, in agreement with BdG.
Recent experiments on TiN and NbN films do indeed see a pseudogap up to many times $T_c$ \cite{sacepe2010,mondal2011}. 

It is common practice to characterize superconductors by the so-called strong-coupling ratio $2E_g/T_c$, which is the ratio of two experimentally measurable quantities, the zero-temperature gap $E_{g0}$ (from tunneling) and the critical temperature $T_c$ (from transport).
BCS MFT predicts that $2E_g/T_c = 2\pi/e^\gamma \approx 3.52775$.
In the present situation, 
DQMC+MEM predicts that $E_g$ is robust against disorder whereas $T_c$ is suppressed to zero, so that the ratio $2E_g/T_c$ tends to infinity (see Fig.~\ref{PhaseDiag}). 
This is a strong deviation from the BCS result.
Experiments on InOx do indeed see a divergence of this ratio as the SIT is approached \cite{sacepe2008}.

	\begin{figure}
	\subfigure[
		Temperature dependence in a clean weak-coupling superconductor.
		]{
		\includegraphics[width=0.48\textwidth]{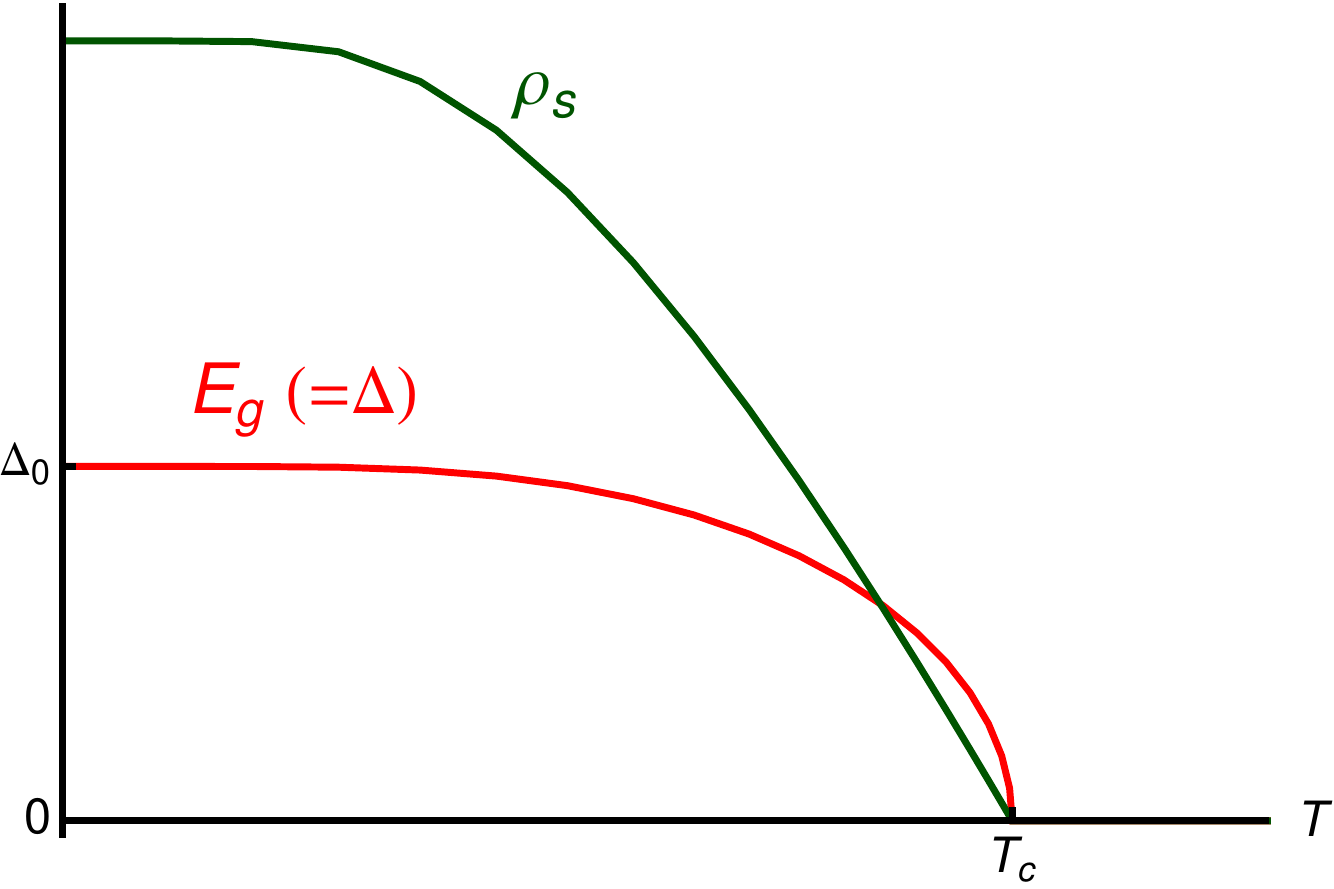} 
		\label{schematicTdependence}
	}
	\hspace{0.04\textwidth}
	\subfigure[
		Superfluid stiffness $\rho_s$ and single-particle gap $E_g$
		as a function of disorder strength $V$.
		]{
		\includegraphics[width=0.48\textwidth]{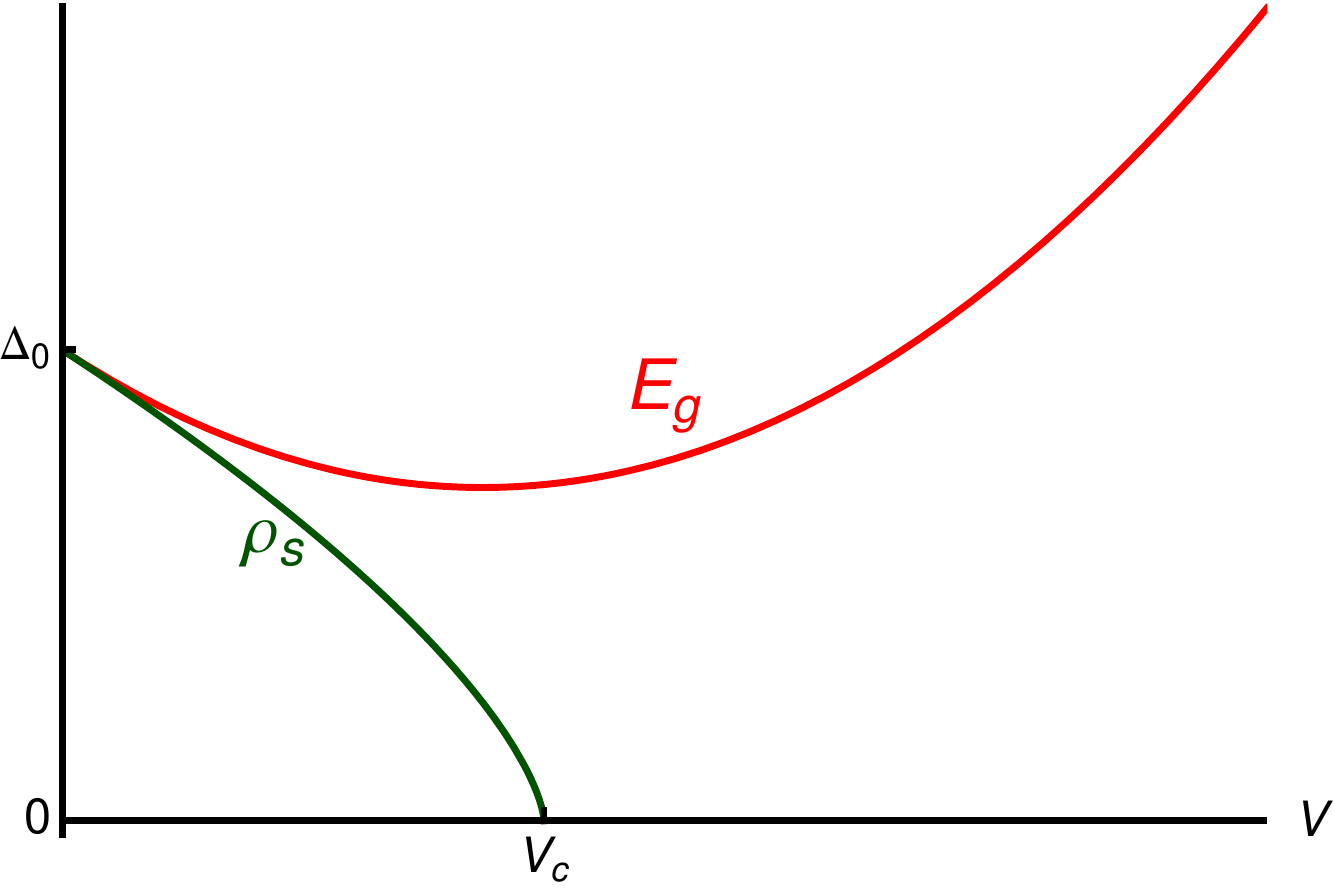} 
		\label{schematicVdependence}
	}
	\subfigure[
		Temperature dependence in a disordered superconductor near the SIT.
		]{
		\includegraphics[width=0.48\textwidth]{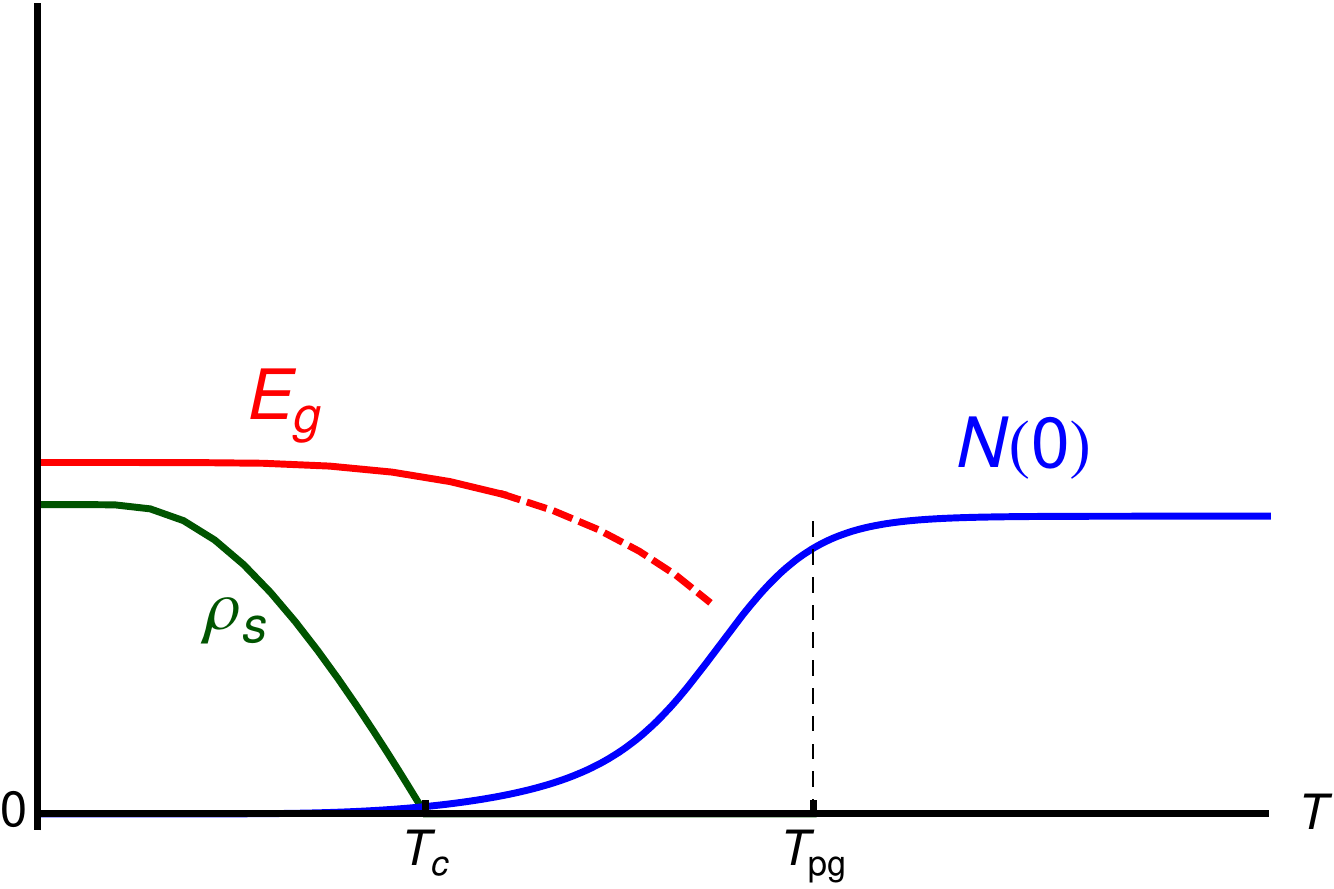} 
		\label{schematicTdependenceDirty}
	}
	\subfigure[
		Phase diagram, showing normal state, superconducting state (SC),
		pseudogap state (PG), and quantum critical fan (QC).  
		Dashed vertical lines correspond 
			to Figs.~\ref{schematicTdependence} and \ref{schematicTdependenceDirty}.
		]{
		\includegraphics[width=0.48\textwidth]{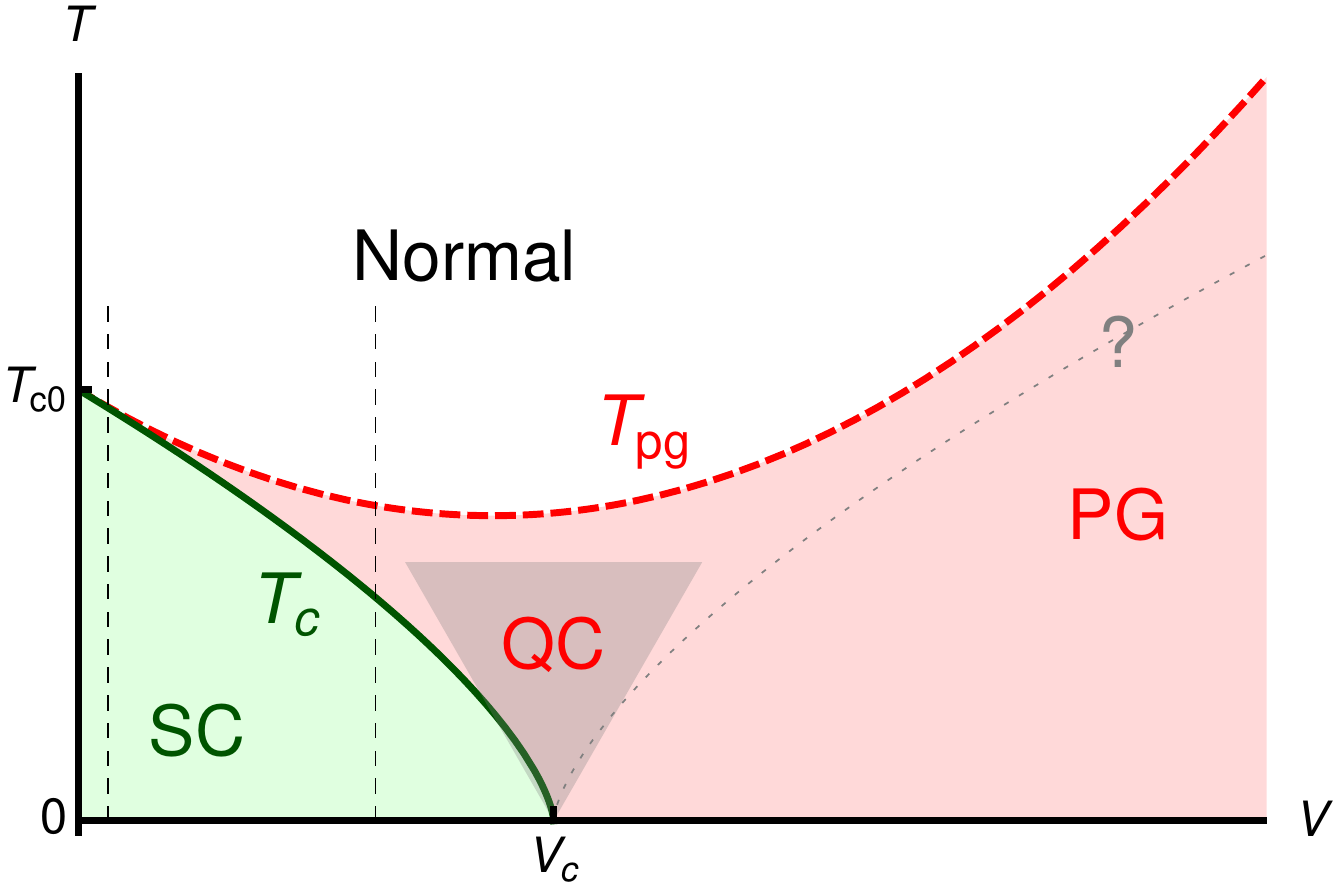} 
		\label{schematicVTphasediagram}
	}
	\subfigure[
		Estimated phase diagram of the attractive Hubbard model at filling $n=0.875$ 
			as a function of attraction $U$ and disorder strength $V$,
			adapted from Ghosal et al. (2001).
		]{
		\includegraphics[width=0.48\textwidth]{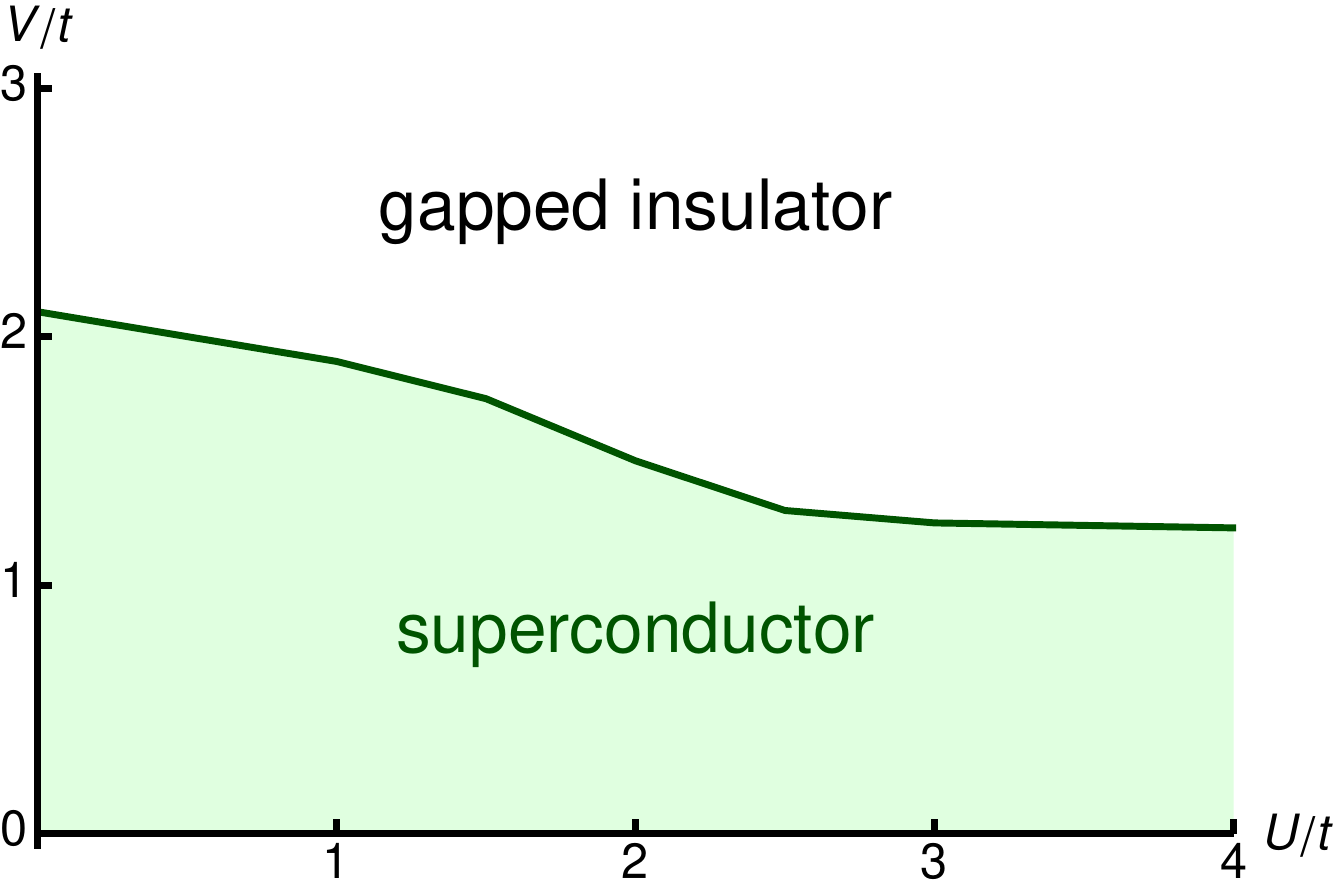} \label{schematicUdependence}
	}
	\hspace{0.04\textwidth}
	\subfigure[
		Three-dimensional visualization of various quantities as functions of $V$ and $T$.
		]{
		\includegraphics[width=0.48\textwidth]{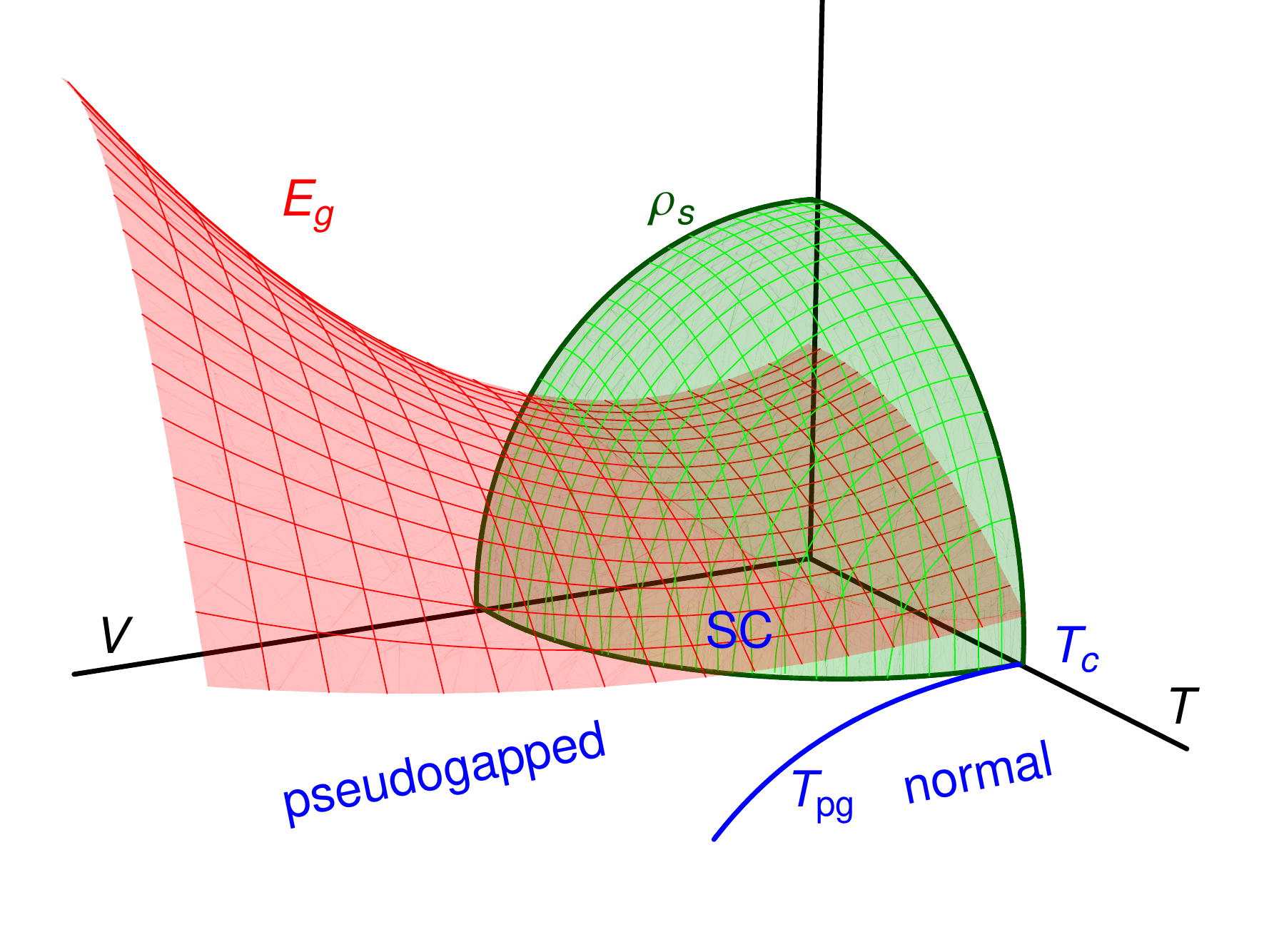} \label{schematicVTdependence3D}
	}
	\caption{
		\label{schematics}
		Schematic diagrams illustrating the physics of the disorder-tuned superconductor-insulator transition
		 in two dimensions.
	}
	\end{figure}
\section{Emerging Picture of the SIT}

A summary of the SIT, within the model of this section, is presented in Fig~\ref{schematics}.
For a weak-coupling superconductor like Al it is known that the superfluid density $\rho_s$ and energy gap $E_g$
decrease with increasing temperature $T$ and vanish at $T_c$ as seen in Fig.~\ref{schematicTdependence}. 
The scale for $\rho_s$ is set by the Fermi energy, whereas the scale for $E_g$ is
exponentially suppressed from the Fermi energy in weak
coupling.
%
The behavior of these two quantities as a function of disorder
at $T=0$ is markedly different.  While $\rho_s$ decreases as expected with
increasing disorder and vanishes at a critical disorder strength
$V_c$,
the gap in the spectrum remains a hard gap for all values of the
disorder (see Fig.~\ref{schematicVdependence}). 
%
The behavior at finite disorder as a function of $T$ is quite distinct
from the behavior at zero disorder.
As seen in Fig.~\ref{schematicTdependenceDirty}, $\rho_s$ vanishes at $T_c(V)$ where $T_c(V)<T_c(0)$.
However, the energy gap, which started as a hard gap at $T=0$, 
starts filling up and finally approaches the normal state value at a
temperature $T^\ast(V)>T_c(V)$. 
Thus, for $T_c < T < T^*$, there is a pseudogap (PG): the density of states at the Fermi level, $N(0)$, is suppressed relative to its normal-state value.
This is a separation between the temperatures for pairing and long-range
phase coherence occurring even in a weak coupling superconductor, this
time produced by the combined effects of interaction and disorder.
Figures~\ref{schematicVTphasediagram} and \ref{schematicVTdependence3D} illustrate 
this picture.
Although the simulations in this section were performed at finite interaction $U$,
it can be argued \cite{ghosal2001} 
that the conclusions remain valid in the limit of infinitesimal interaction (Fig.~\ref{schematicUdependence}).

\section{Summary}

The results of this section are summarized below:

\begin{itemize}
\item 
The pairing-of-exact-eigenstates (PoEE) approximation finds that the single-particle gap remains finite for all values of disorder, but fails to describe the vanishing of the coherence peaks.
\item 
The Bogoliubov-de Gennes (BdG) approach 
finds that the pairing amplitude becomes extremely inhomogeneous with increasing disorder.
It predicts that coherence peaks disappear and that the phase stiffness drops precipitously, but it does not by itself explain the SIT.
At finite temperature, BdG predicts the existence of a disorder-induced pseudogap.
\item
The self-consistent harmonic approximation (SCHA) predicts that quantum phase fluctuations suppress the phase stiffness (and hence the critical temperature) to zero
beyond a certain critical disorder, thus capturing the SIT.
\item
Determinant Quantum Monte Carlo (DQMC), which includes all amplitude and phase fluctuations, confirms the above predictions for the gap, coherence peaks, pseudogap, and stiffness.
\item
The attractive Hubbard model undergoes a bosonic SIT as a function of disorder strength, which is due to localization of Cooper pairs by disorder.
This scenario is in agreement with experiments on InOx, TiN, and NbN.
\end{itemize}

The present model includes attraction and disorder, but ignores the Coulomb repulsion between electrons.  
Is it possible to include the Finkel'stein mechanism (suppression of uniform pairing amplitude by Coulomb and disorder) together with the physics of amplitude inhomogeneity and phase fluctuations, and thereby obtain a quantitative explanation of experiments across all materials and parameter ranges?
This and many other questions (such as the behavior of two-particle spectra and dynamical conductivity) are still unanswered and hopefully will be addressed in future research.

\chapter{Parallel Field-Tuned Superconductor-Insulator Transition} \slabel{ParallelFieldSIT}

\section{Introduction}

The superconducting state is characterized by a large diamagnetic susceptibility due to the Meissner effect and a vanishing paramagnetic susceptibility due to the binding of spins into Cooper pairs.
Conversely, applying a magnetic field to a superconductor raises its free energy by inducing diamagnetic currents and by tending to align the electron spins.
When an external magnetic field is applied to a superconductor, it suppresses superconductivity via both the orbital effect and the Zeeman effect.  We shall consider only parallel fields on thin films, so that the Zeeman effect dominates.
\footnote{
In real materials there are complications arising from perpendicular field components, 
spin-orbit interactions, magnetic impurities, and disorder;
there has, however, been some progress toward realizations of the pure Zeeman physics.
}

\section{Clean Superconductor} \slabel{CleanSCInZeemanField}

\begin{figure}[!h]
	\subfigure[
		]{
			\includegraphics[width=0.3\textwidth]{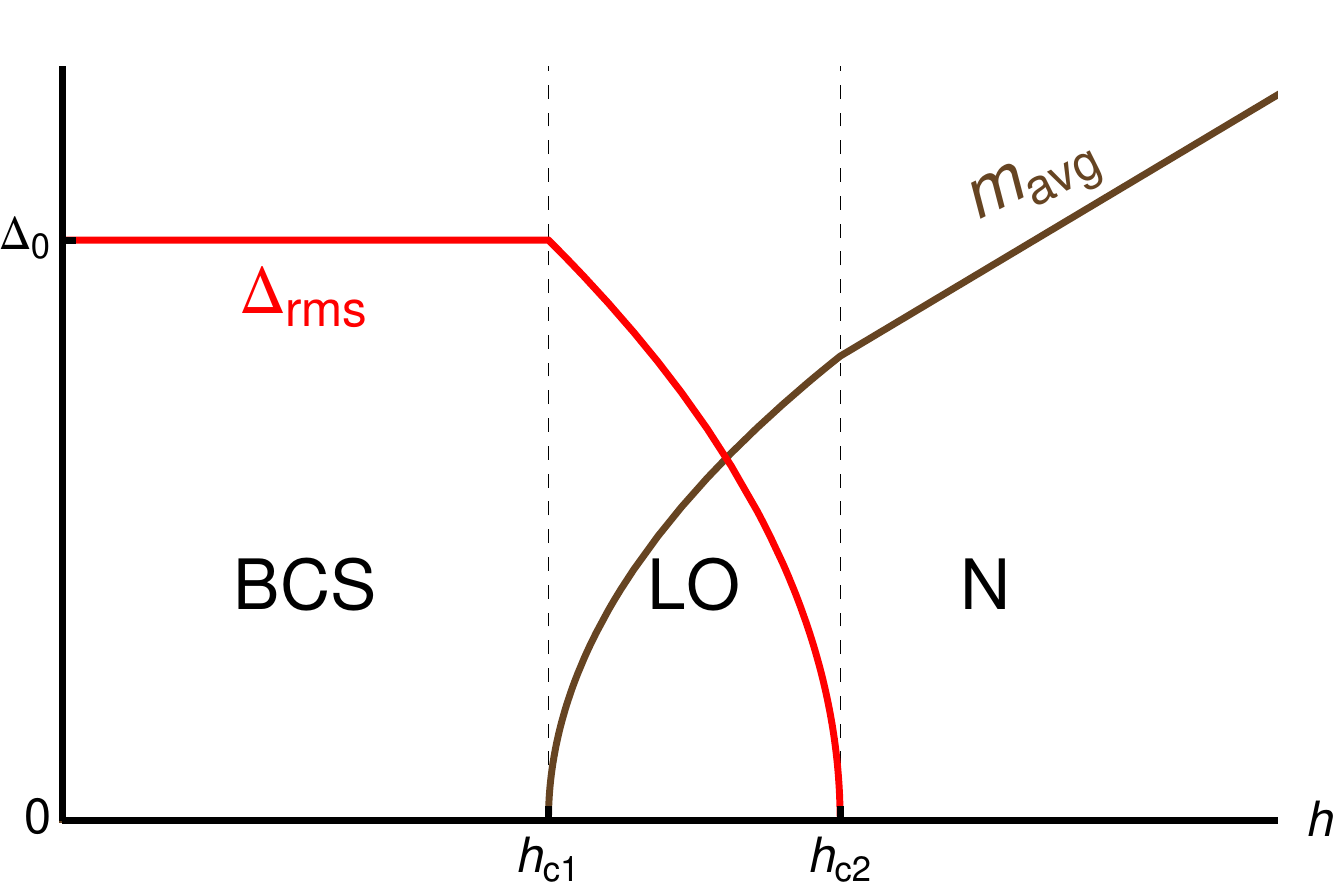}	\label{SchematicCleanLOFieldDependence}
	}
	\subfigure[
		]{
			\includegraphics[width=0.3\textwidth]{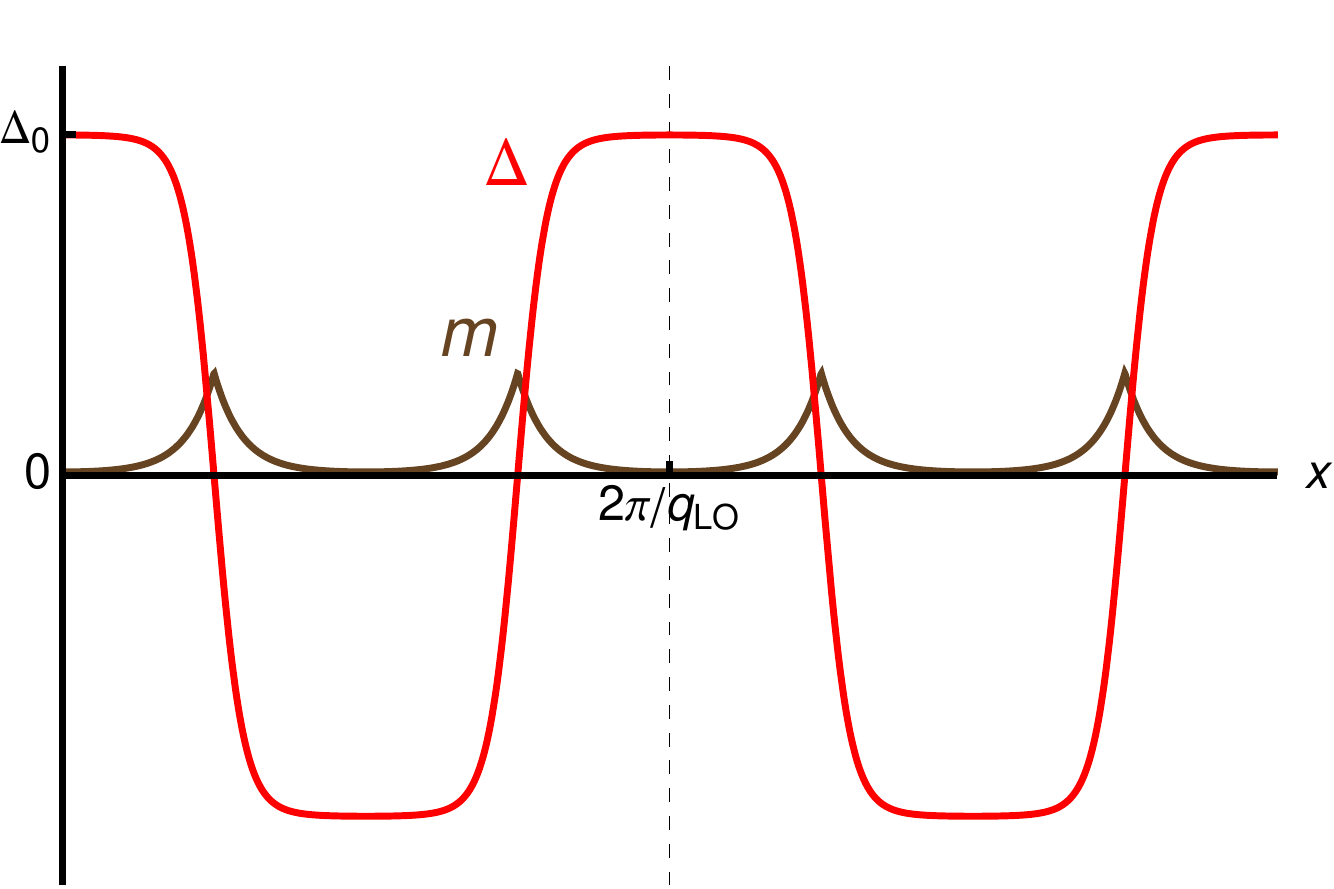}	\label{SchematicStrongLOProfiles}
	}
	\subfigure[
		]{
			\includegraphics[width=0.3\textwidth]{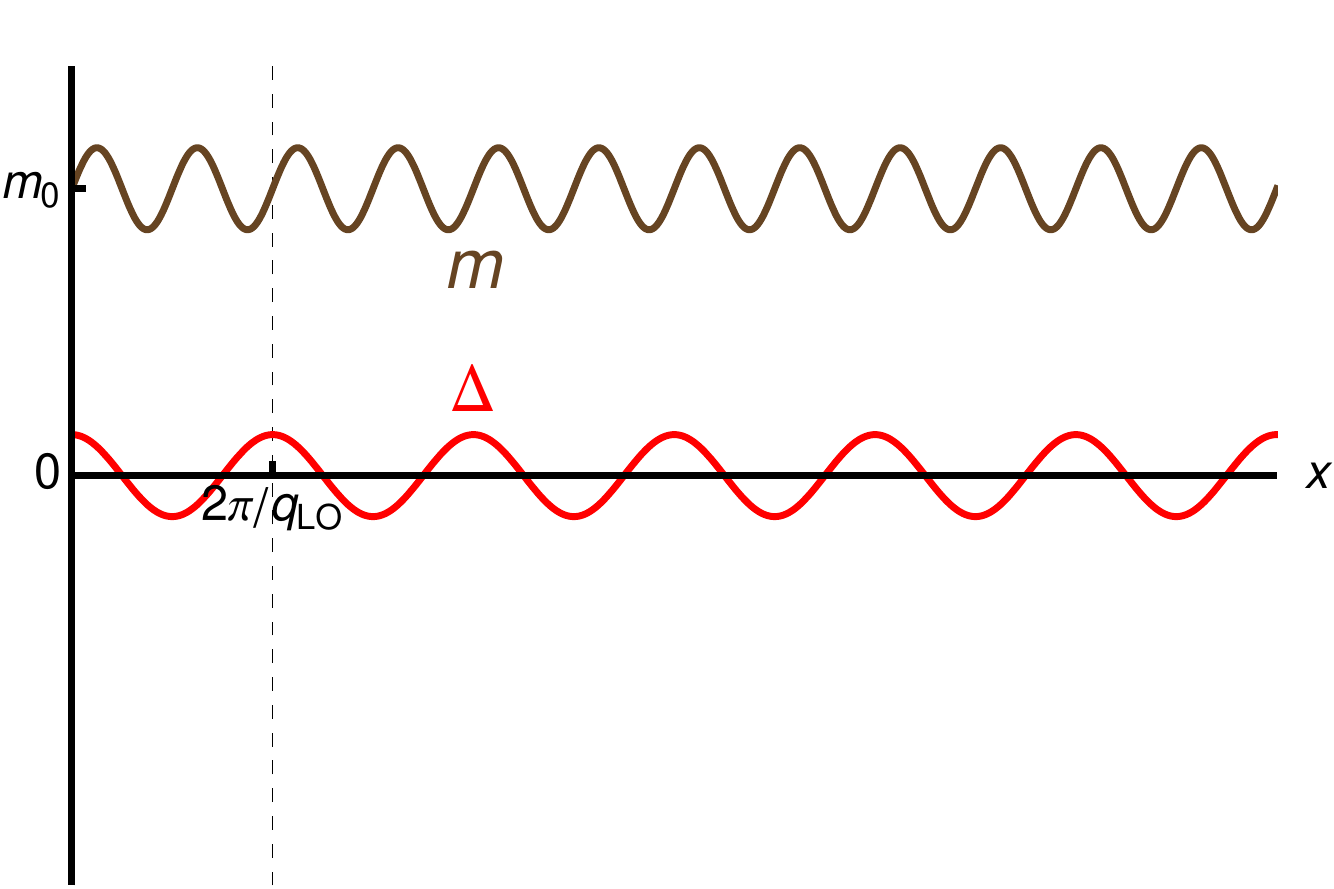}	\label{SchematicWeakLOProfiles}
	}
	\caption{
		\label{SchematicLO}
		Schematic illustration of Larkin-Ovchinnikov physics.
		Fig.~\ref{SchematicCleanLOFieldDependence} shows 
			the root-mean-square pairing amplitude $\Delta_\text{rms}$ and average magnetization $m_\text{avg}$
			in the ground state of a superconductor as a function of Zeeman field $h$, within a mean-field picture.
		At low field the ground state is a uniform BCS superconductor.
		As the field is increased beyond a lower critical field $h_{c1}$, 
			it eventually becomes energetically favorable for magnetization
			to penetrate the system in the form of domain walls at which the order parameter changes sign,
			as illustrated in	Fig.~\ref{SchematicCleanLOFieldDependence} 
		The magnetization (i.e., the excess majority spins) occupies Andreev bound states 
			whose wavefunctions are localized in the domain walls.
		The spatial periodicity $\lambda_\text{LO} = \frac{2\pi}{q_\text{LO}}$ is related to the LO wavevector,	$\qqq_\text{LO}$,
			which is the best nesting vector for the Fermi surfaces of the given numbers of up and down spins
			\emph{in the absence of pairing}.
		As the field increases further, $m$ increases and $\lambda_\text{LO}$ decreases, 
			so that the domain walls begin to overlap and the modulation becomes small and sinusoidal.
		Finally, beyond an upper critical field $h_{c2}$,
			pairing is completely destroyed and a uniform magnetization prevails everywhere.
	}
	\end{figure}

The problem of superconductivity in a Zeeman field has been studied since the 1960's.
The simplest theories assume a uniform order parameter.  With this restriction, there is a first-order transition from a superconductor to a high-field normal metal at the Chandrasekhar-Clogston critical field $h_{CC} = \Delta_0/\sqrt{2} \approx 0.71 \Delta_0$, where $\Delta_0$ is the zero-temperature zero-field gap.
\cite{chandrasekhar1962,clogston1962,sarma1963,gorkovrusinov1964}.
However, allowing the mean-field order parameter to vary in space reveals that in an intermediate range of fields the system can lower its free energy by forming periodic patterns known as Fulde-Ferrell-Larkin-Ovchinnikov (FFLO) states.
Fulde and Ferrell \cite{fulde1964} used the ansatz $\Delta(\rrr) \propto e^{i\QQQ\cdot\rrr}$, which breaks time-reversal symmetry, whereas Larkin and Ovchinnikov studied additional modulation patterns of the form $\Delta(\rrr) \propto \sum_\QQQ \cos \QQQ\cdot\rrr$ \cite{larkin1964}, which break translational symmetry.
It is generally found that LO states are favored over FF states, so we will henceforth simply refer to LO states.
A general LO state consists of regions of positive and negative pairing amplitude $\Delta(\rrr)$ separated by a regular array of domain walls where the majority fermions are concentrated; it exemplifies the emergent phenomenon of ``microscale phase separation''\cite{radzihovsky2009}.
A proper treatment of the LO state requires some amount of numerical work. \cite{machida1984,burkhardt1994,yoshida:063601,koponen:120403,loh2010}
See Ref.~\onlinecite{casalbuoni2004} for a review.
Figure~\ref{SchematicLO} illustrates one version of the story.

\begin{figure}[!h]
	\subfigure[
		Phase diagram	assuming a uniform mean-field pairing amplitude $\Delta$.
		At low temperatures $T$ and fields $h$ the superconductor (SC) is stable.
		At high $T$ or strong $h$ the system becomes a polarized normal Fermi liquid.
		The thin solid curve is a second-order boundary.
		The thick solid curve is the first-order boundary
			at which $\Omega_s = \Omega_n$.
		The dashed curves are the limits of metastability of the normal state
			in the superconductor and vice versa.
		]{
			\includegraphics[width=0.48\textwidth]{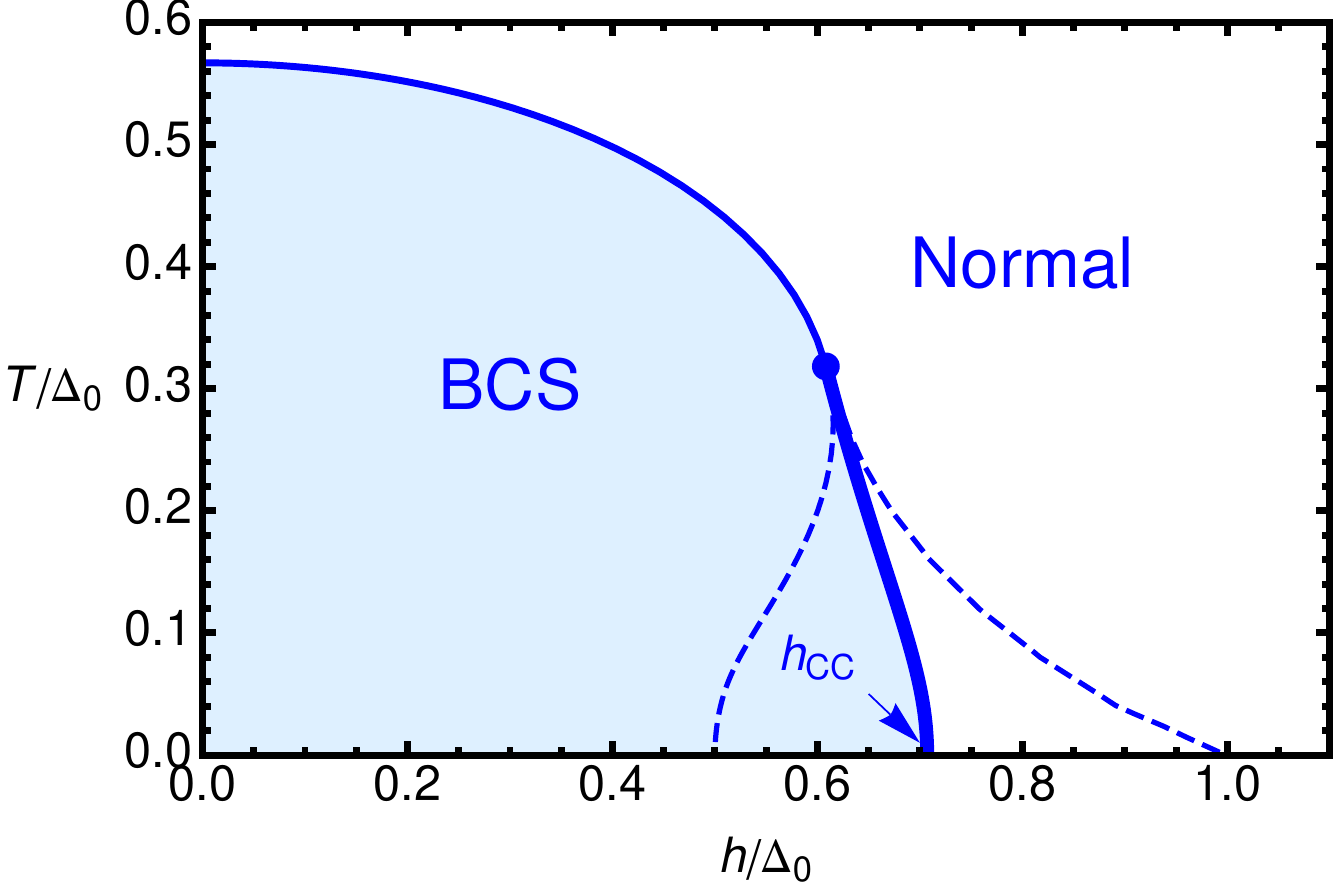}
			\label{BCSZeemanContinuum1}
		}
	\hspace{0.04\textwidth}
	\subfigure[
		Phase diagram allowing inhomogeneous pairing $\Delta(\rrr)$.
		The dotted curve is the hypothetical first-order BCS-normal boundary,
			corresponding to the solid curve in Fig.~\ref{BCSZeemanContinuum1}.
		At the lower critical field $h_{c1}$ 
			(shown schematically),
			magnetization begins to penetrate the superconductor
			in domain walls at which the pairing amplitude changes sign.
		At the upper critical field $h_{c2}=\Delta_0$
			(for the case of a 2D continuum),
			the normal polarized Fermi liquid becomes unstable to FFLO pairing.
		The inset shows the analogous phase diagram in 3D,
			for which $h_{c2}\approx 0.755\Delta_0$.
		]{
			\includegraphics[width=0.48\textwidth]{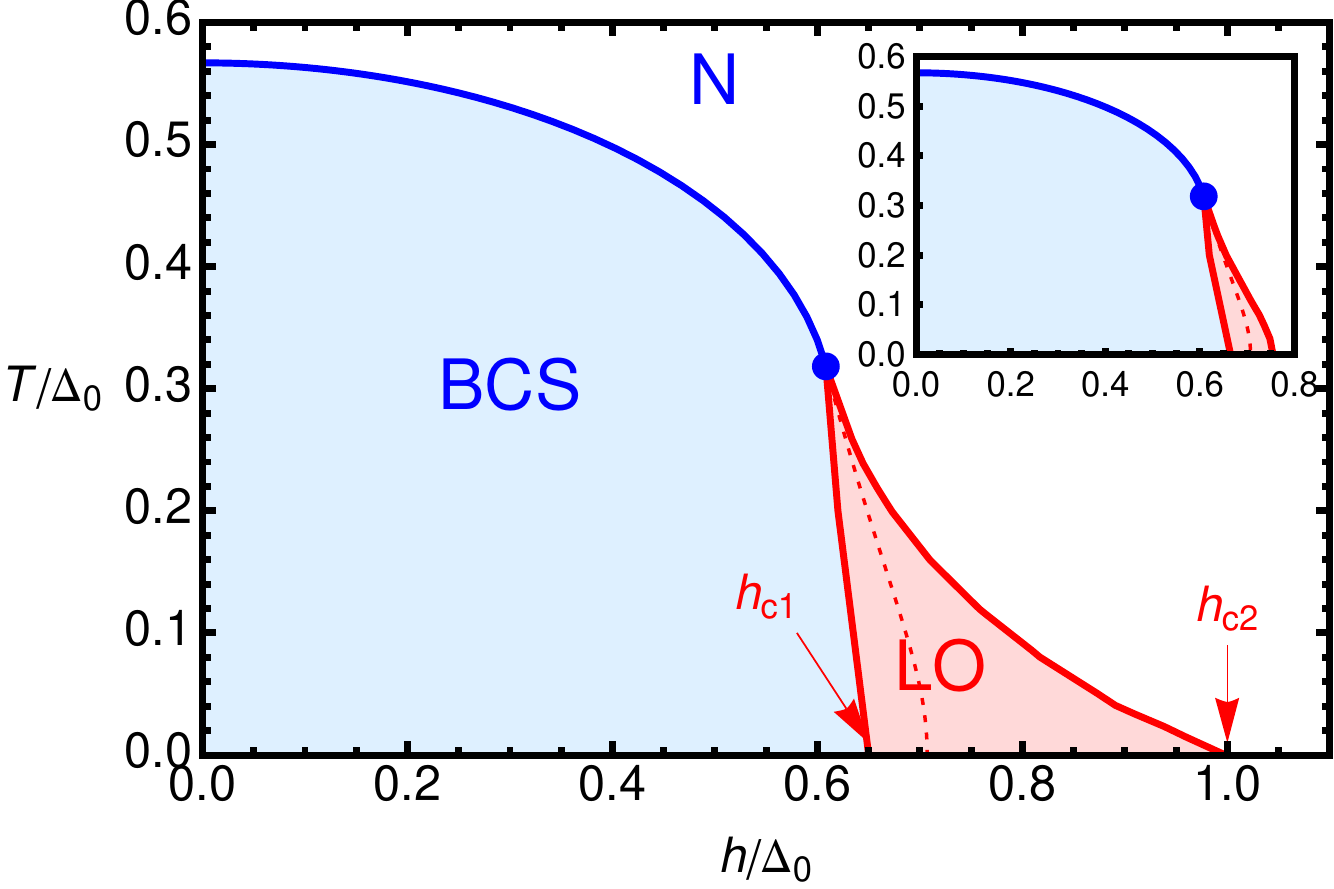}
			\label{BCSZeemanContinuum2DAnd3D}
		}
	\\
	\subfigure[
		Fluctuations destroy the continuum LO state in 2D and in 3D, 
			as illustrated here schematically.
		]{
			\includegraphics[width=0.48\textwidth]{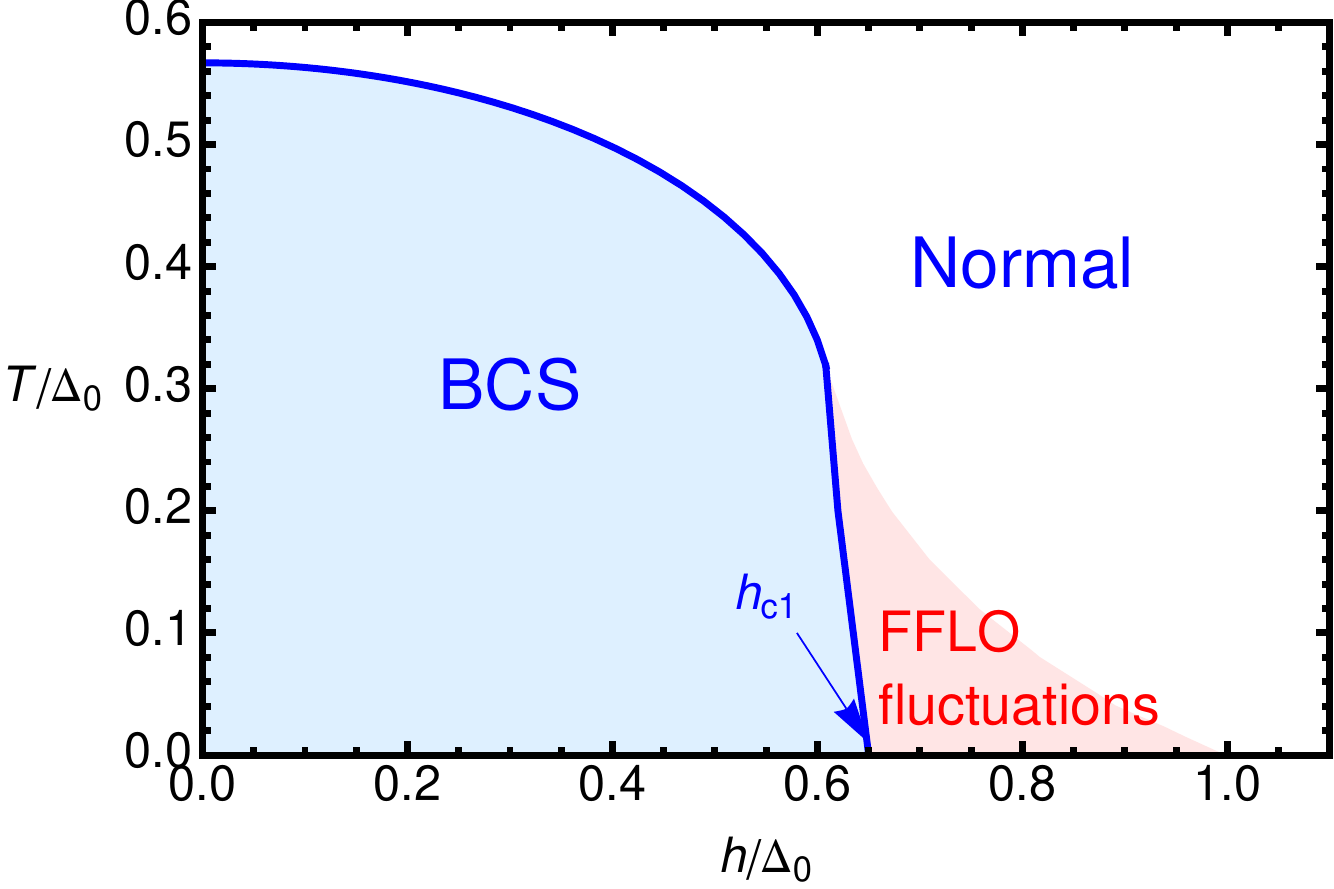}
			\label{BCSZeemanContinuum3}
		}
	\hspace{0.04\textwidth}
	\subfigure[
		BdG phase diagram of the cubic lattice (3D) Hubbard model as a function 
			of field $h$ and chemical potential $\mu$, in units of the hopping amplitude $t$.
		The LO region is considerably larger than for the 3D continuum.
		Furthermore, the lattice suppresses translation and rotation of the LO pattern,
			so quantum fluctuations should be less severe.
	]{
		\includegraphics[width=0.48\textwidth,height=0.32\textwidth]{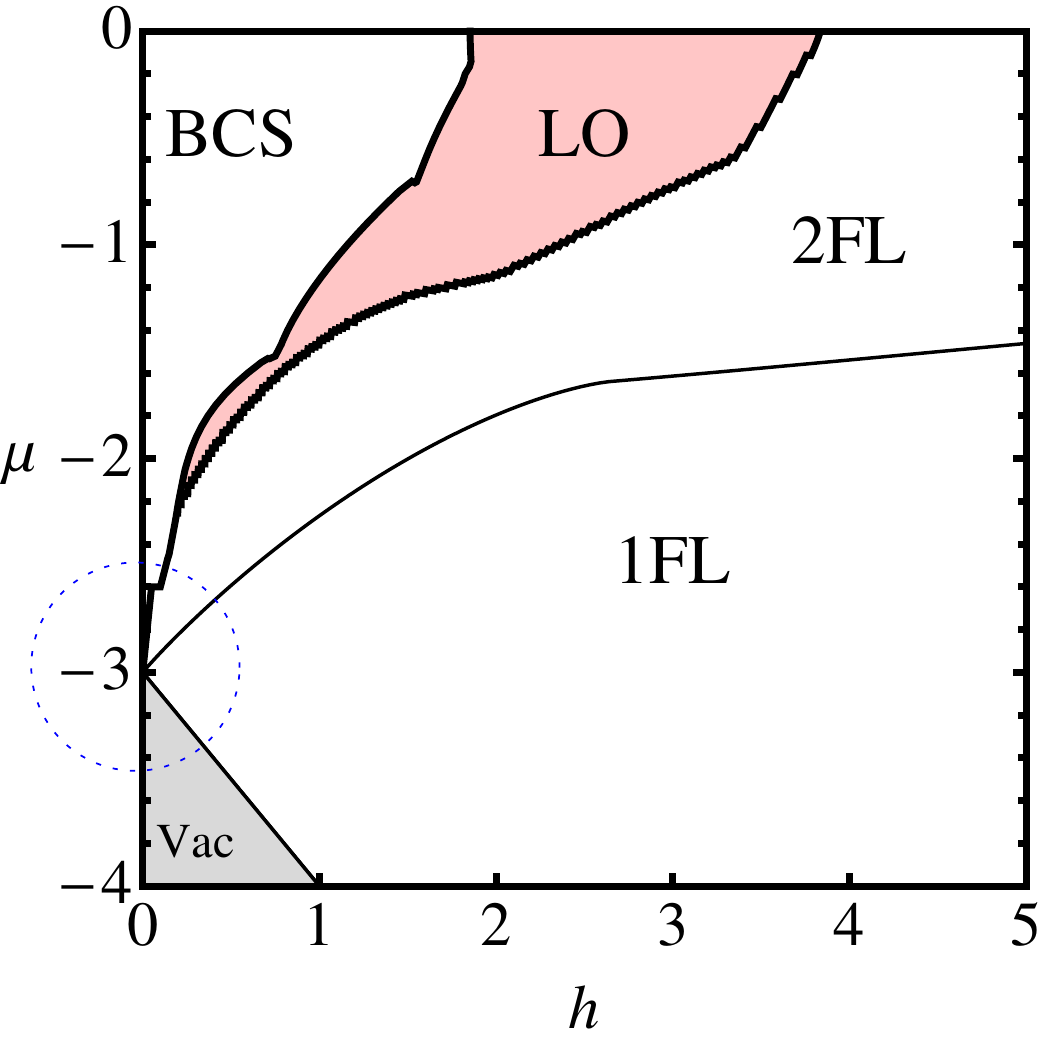} \label{LOPhaseDiagram_h_mu_U6_T0}
	}
	\caption{
		Approximate phase diagrams of a clean $s$-wave superconductor,
			at weak coupling and in the continuum,
			as a function of Zeeman field $h$ and temperature $T$
			in units of the zero-temperature zero-field pairing amplitude $\Delta$.
		\label{BCSZeemanContinuum}
	}
	\end{figure}

In the 3D continuum, FFLO only occupies a tiny sliver of the mean-field phase diagram, between $h_{c1}=0.665$ and $h_{c2}=1$.\cite{sheehy2006}  \todo{[Cite Matuso for hc1.]}
Furthermore, quantum and thermal fluctuations destroy even this sliver \cite{radzihovsky2009}.  
In 2D, the mean-field FFLO region is larger ($h_{c2}=1$), as shown in Fig.~\ref{BCSZeemanContinuum2DAnd3D}, but fluctuations are even more severe.
\todo{[What is hc1 in 2D? Check Burkhardt.]}
Thus, it is not surprising that FFLO order has not been observed except in some reports on layered organic and heavy-fermion superconductors\cite{radovan2003}.
We have attempted to summarize the above discussion in Fig.~\ref{BCSZeemanContinuum}, although it should be acknowledged that the phase diagram is exquisitely sensitive to strong-coupling corrections, material properties, and calculational methods, and that many aspects of FFLO physics continue to be debated;\cite{shimahara1998,matsuo1998,mora2005,bulgac2008}) for example, some authors find first-order transitions to a crystalline LO state with minority spins localized in a superconducting background.

In contrast, cold Fermi gases in optical lattices are a promising arena in which to search for FFLO physics \cite{parish:250403,zhao:063605,koponen:120403,loh2010}.  
For example, the cubic lattice phase diagram [Fig.~\ref{LOPhaseDiagram_h_mu_U6_T0}] has a much larger LO region than the 3D continuum phase diagram [inset of Fig.~\ref{BCSZeemanContinuum2DAnd3D}].
The most favorable systems appear to be coupled tubes or anisotropic lattices at weak-to-intermediate coupling.  Since this chapter is about the field-tuned SIT, we shall not dwell on this topic; we will just present some BdG pictures in order to make conection with the next section, which includes disorder.

The full BdG calculation follows the formalism described in the previous chapter.
\footnote{Due to the breaking of spin symmetry, all $2N$ eigenvalues and eigenvectors are now independent, and certain optimizations are no longer possible.}
BdG calculations on lattices have a strong tendency to give LO states in suitable parameter regimes:
if the system is initialized in a BCS-like state, it may go through many iterations, exploring the free energy landscape, before settling into a reasonable LO state.
	\begin{figure}[!h]
		\includegraphics[width=\textwidth]{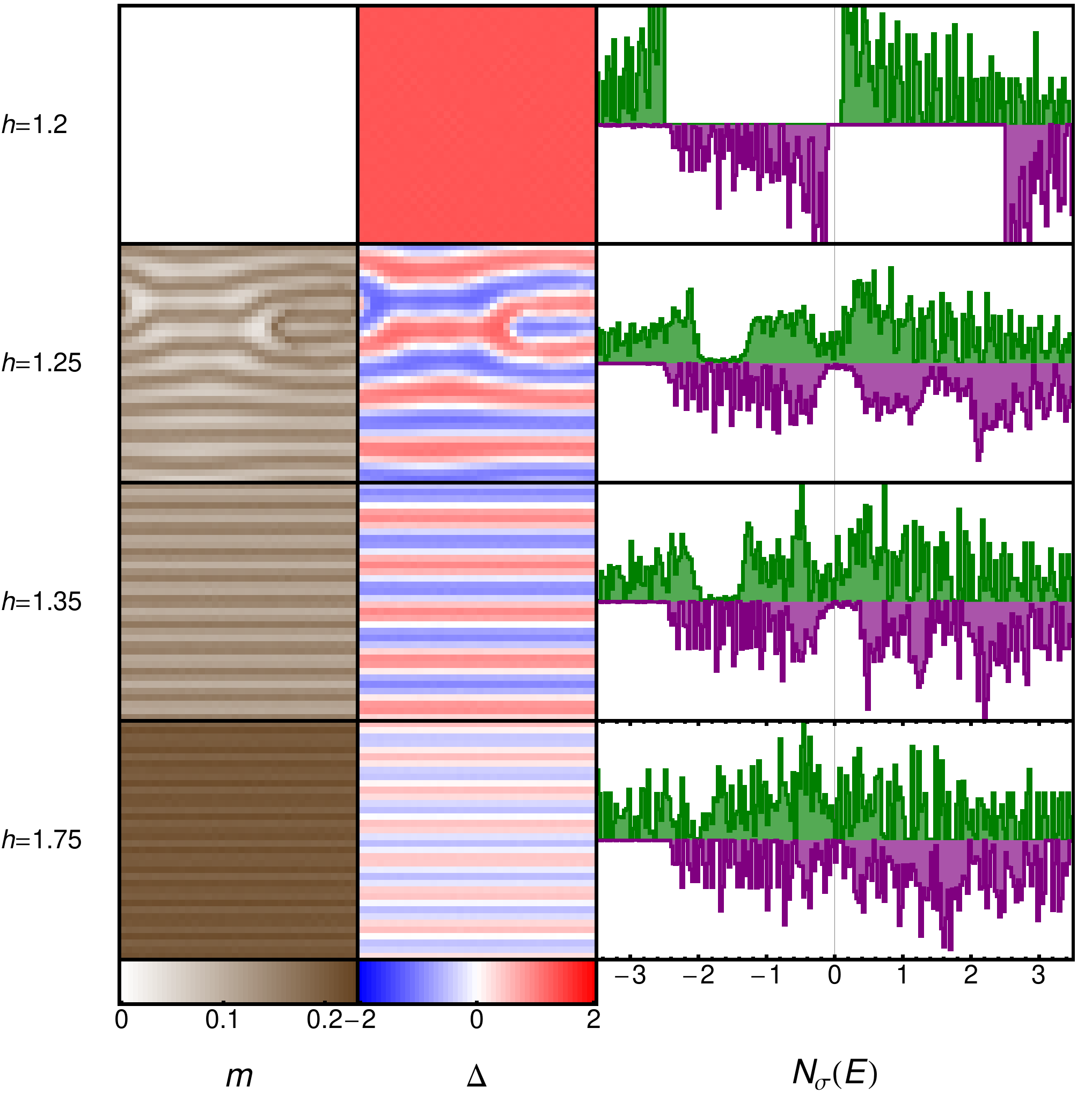} \label{CleanLOEvolution}
	\caption{
		Evolution of a clean superconductor with increasing Zeeman field within BdG
		(the disorder strength was set to a negligible value, $V=0.02$).
		The panels show the results of independent BdG calculations 
			starting from random configurations.
		Because of symmetry, some runs converged to vertical patterns instead of horizontal patterns;
			these are not shown.
		For $h=1.25$, the smectic LO pattern is disrupted by a pair of dislocations.
		In this case, the BdG procedure had become trapped in a local minimum of the free energy,
			and was unable to find the global minimum.
		This does, however, reflect actual physics that may happen 
			in experiments on condensed matter or cold atoms.
		All energy scales are in units of the hopping amplitude $t$.
		\label{CleanLOVariousStuff}
	}
	\end{figure}

\section{Dirty Superconductor} \slabel{DirtySCInZeemanField}
In the presence of disorder potential as well as a parallel field, diagrammatic calculations by Zhou and Spivak\cite{zhou1998} suggested that the superconductor becomes like an ``XY glass'' with random positive and negative Josephson couplings, so that there are many metastable states, in which the superconducting order parameter is positive and negative in different places.

\begin{figure}[!h]
	\centering
		\includegraphics[width=0.48\columnwidth]{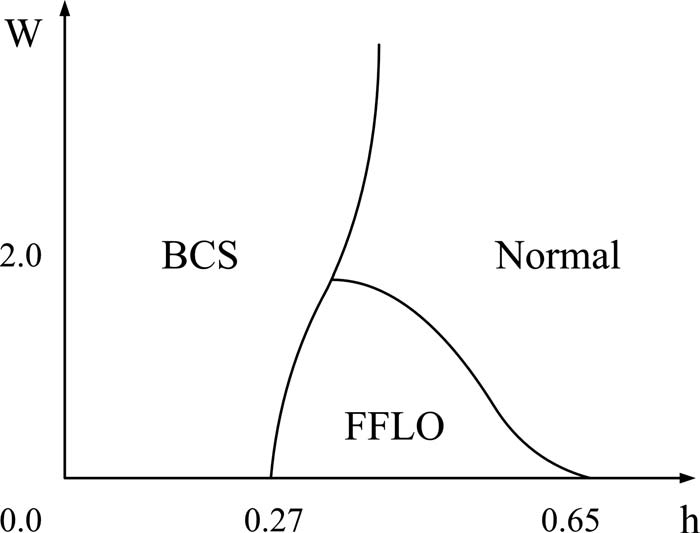}
	\caption{
		Approximate phase diagram of a superconductor in a Zeeman field $h$
		and a random potential of strength $W$ [Cui et al. (2008)].
		A disordered (FF)LO ground state exists over a significant field range $h_{c1} < h < h_{c2}$,
			provided that disorder is not too large.
		\label{CuiYangPhaseDiagram}
	}
	\end{figure}

Cui and Yang\cite{cui:054501} did BdG calculations of the kind described in this article.  They found that an LO-like ground state still survives at weak disorder (see Fig.~\ref{CuiYangPhaseDiagram}).  Although the orientation of the LO stripes is disrupted, there is still some sign of periodicity.  

Dubi, Meir, and Avishai \cite{dubi2007,dubi2008} did BdG calculations, as well as Monte Carlo calculations in which the BdG pairing amplitude was treated as a fluctuating auxiliary field.  They found that the superconducting islands are destroyed with the application of a field.  However, they did not report sign changes of the order parameter and the connection to FFLO physics.

\begin{figure}[!h]
	\centering
		\includegraphics[width=0.48\columnwidth]{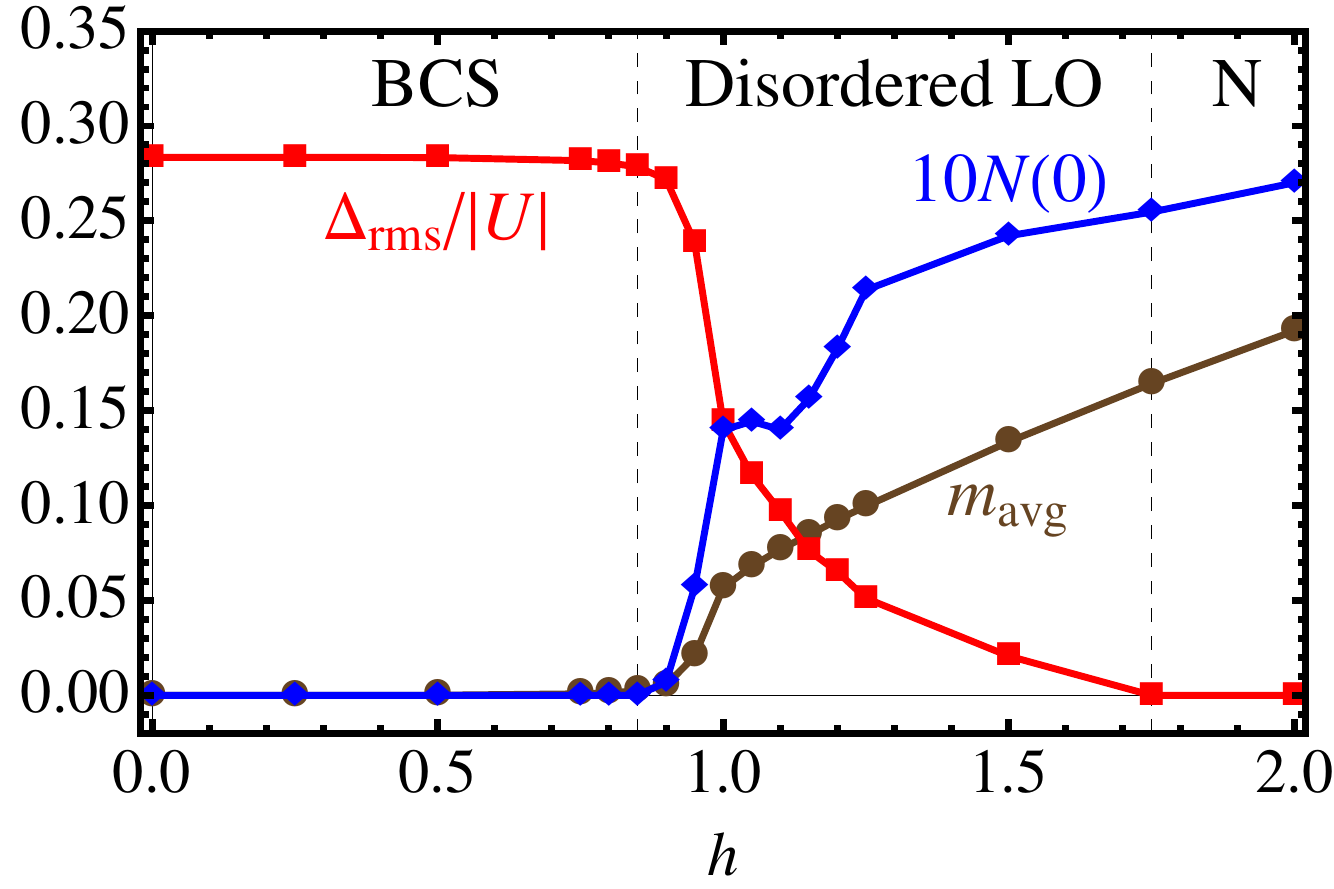}
	\caption{
		Disordered SC in Zeeman field:
		Root-mean-square pairing $\Delta_\text{rms}$ (red squares),
		average magnetization $m_\text{avg}$ (brown circles),
		and Fermi-level density of states $N(0)$ (blue diamonds),
		as functions of Zeeman field $h$.
		Between the two dashed lines there is a disordered LO state 
		with coexistent pairing and magnetization, 
		in which the gap is partially filled in.
		These are results of BdG simulations 
		on a $36\times 36$ square lattice Hubbard model 
		with 
		attraction $\left|U\right| / t = 4$, 
		disorder strength $W/t=1$, 
		and temperature $T/t=0.1$, 
		for different various of the Zeeman field $h$.
		\label{dirtyLOFrmsAndMavg}
	}
	\end{figure}

Figure~\ref{dirtyLOFrmsAndMavg} shows the coexistence of pairing and magnetization as a function of Zeeman field $h$.  As the Zeeman field is increased beyond a critical field, a finite density of states develops within the gap, due to the formation of a disordered Larkin-Ovchinnikov (dLO) state with bound states in domain walls \cite{loh2011arxiv}.
This is illustrated in Fig.~\ref{dirtyLOFrmsAndMavg}.

\begin{figure*}[!h]
	\includegraphics[width=0.98\textwidth]{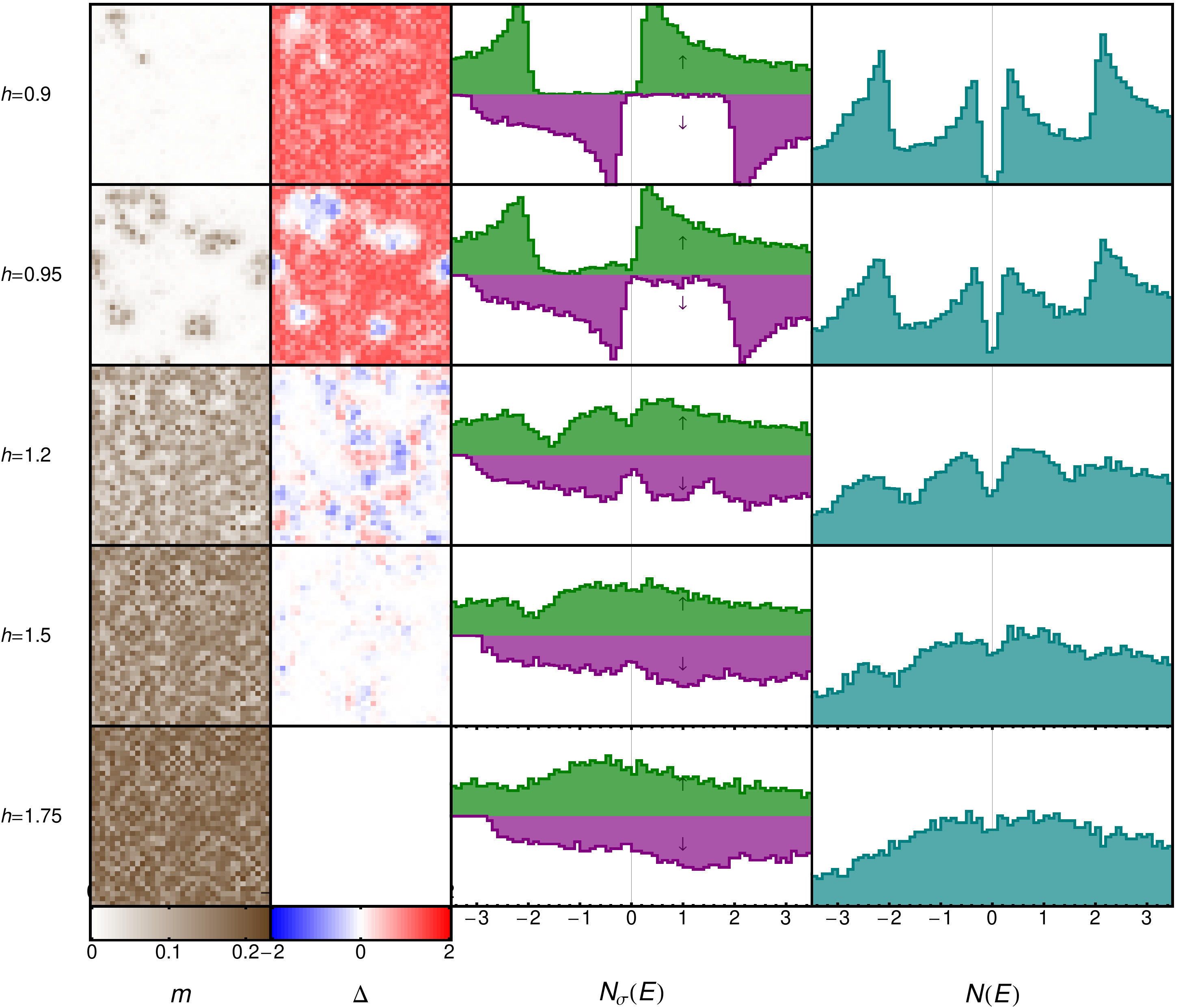}
	\caption{
		The first two columns show spatial maps of 
		the order parameter $\Delta$ 
		and the magnetization $m$.
		The last two columns show 
		the densities of states (DOS's) of up and down electrons $N_\sigma(E)$
		as well as the total density of states $N(E)$.
		For intermediate fields the system exhibits disordered Larkin-Ovchinnikov states
		with domain walls at which $m$ is finite 
		and the sign of $\Delta(\rrr)$ varies in space (as indicated by the red and blue patches).
		The appearance of such states is associated with the filling-in of the gap in the DOS.
		Parameters are as in Fig.~\ref{dirtyLOFrmsAndMavg}.
		\label{EvolutionOfDOSWithZeemanField}
	}
	\end{figure*}

Figure~\ref{EvolutionOfDOSWithZeemanField} illustrates the order parameter $\Delta(\rrr)$, the magnetization density $m(\rrr) = \half \left[ n_\up(\rrr) - n_\dn(\rrr) \right]$, and the spatially averaged densities of states of up and down spins $A_\sigma(E)$, for various values of field.
The disorder strength is $W=1$, which corresponds to a normal-state sheet resistance 
$R_\square$ of the order of $0.3 R_Q$ (the sheet resistance in zero field at temperatures somewhat above $T_c$), where $R_Q = \frac{h}{4e^2} \approx  6.4\ \rm{k\Omega}$ is the quantum resistance appropriate to Cooper-paired systems.

At low fields the system is a BCS superconductor with a nearly uniform order parameter $\Delta(\rrr) \approx \Delta_0$, whose density of states contains coherence peaks at $\pm \Delta \pm h$ that are slightly broadened by inhomogeneous Hartree shifts\cite{ghosal1998,ghosal2001}.
At high fields the system is a normal metal with nearly uniform magnetization.
\footnote{In principle, weak localization corrections in 2D would cause the system to be an insulator; such effects are not visible on the scale of the simulated systems.  Also, in real films, the interplay of disorder and Coulomb repulsion (which is absent from our model) produces Altshuler-Aronov corrections that suppress the Fermi-level density of states.}
However, at intermediate fields, the simulations find inhomogeneous states where the order parameter $\Delta(\rrr)$ has sign changes and there are patches of finite magnetization.  
These can be viewed as disordered Fulde-Ferrell-Larkin-Ovchinnikov (FFLO) states\cite{cui:054501},
in which the order parameter oscillations at wavevector $q_\text{FFLO} \approx 2k_F$ are partially disrupted by the disorder potential.
Since the lowest-energy solutions always have a real order parameter $\Delta(\rrr)$, we will refer to these as disordered ``Larkin-Ovchinnikov'' states, to emphasize the contrast with higher-energy ``Fulde-Ferrell states'' with a more restrictive complex-valued ansatz for $\Delta(\rrr)$.
This is in contrast with the traditional view of a first-order transition directly from a BCS state, with a hard gap, to a gapless Fermi liquid.

To gain some insight into the sign changes in the order parameter, it is interesting to consider a different approach studied by Zhou and Spivak: the effective functional describing the order parameter, $\Delta(\rrr)$, takes the form of an XY spin glass Hamiltonian.\cite{zhou1998}  The ground state of such a spin glass has real $\Delta(\rrr)$ with positive and negative signs.  Zhou and Spivak did not, however, make the connection between XY spin glasses and Larkin-Ovchinnikov physics.

It should be noted that the DOS in the high-field state exhibits a very interesting many-body effect called the pairing resonance, predicted by Aleiner, Altshuler, and Kee, and discovered by Adams et al.  Unfortunately, a full discussion of this phenomenon is beyond the scope of this chapter.

Even though FFLO is susceptible to quantum fluctuations, it still means that fluctuating FFLO physics may be important near the transition.  In particular, the DOS is a probe of local physics at all energies, not of global order, so it is likely that some basic features of the LO state (such as the ``soft gap'') may remain even when the system does not possess long-range order.

\section{Summary}

At intermediate fields, the pairing and magnetization coexist in a disordered pattern with some remnant periodicity (as found by Cui and Yang), with spontaneously occurring ``$\pi$-junctions'' where the magnetization is concentrated.
Furthermore, such dirty LO states are associated with peculiar ``soft gaps'' at the Fermi energy in the minority spin DOS.\cite{loh2011arxiv}

\chapter{Conclusions}

In this article we have examined two types of superconductor-insulator transitions using various analytical and numerical methods (PoEE, BdG, SCHA, and DQMC).
In the disorder-tuned SIT, the Cooper pairs become localized by the disorder potential, leading to an insulating ``Cooper pair glass'' (expected to be adiabatically connected to the Bose glass).  The single-particle gap \emph{remains finite} across the SIT.  The theory predicts a pseudogap at finite temperature, which is indeed observed in experiments.  Many other predictions remain to be tested.

The parallel-field-tuned SIT, from the viewpoint of BdG simulations and with some experimental evidence, proceeds via a disordered Larkin-Ovchinnikov state.  The gap gradually \emph{fills up} due to Andreev bound states.  The pairbreaking effect of the field ultimately leads to a fermionic Anderson insulator.

Clearly, the ``I'' in ``SIT'' can stand for many different types of insulator.
Our studies lead to a classification of electronic ground states based on transport as well as spectral properties,
as shown in Table~\ref{comparison-table}.

\begin{table*}[!h]
\centering
\begin{tabular}{lccc}
\hline
State & Description & $N(\omega)$ & $\sigma(\omega)$ \\ \hline
Fermi liquid        & Fermi sea filled up to $k_F$              & Gapless   & Gapless \\[+1mm]
Band insulator      & Fermions localized by Pauli exc.   & Hard gap  & Hard gap \\[+1mm]
Anderson insulator  & Fermions localized by disorder      & Gapless & Soft gap \\[-2pt]
                    &                                     &          & (WL/MVRH) \\[+1mm]
Fermi MI            & Fermions localized by               & Soft gap & Soft gap \\[-2pt]
                    & Hubbard repulsion                   &          &          \\[+1mm]
Electron glass      & Fermions localized by disorder & Soft gap & Soft gap \\[-2pt]
                    & and Coulomb repulsion          & (ES)     & (ES VRH) \\[+1mm]
Superconductor      & Mobile Cooper pairs                 & Hard gap & $\delta(\omega)$ \\[-2pt]
                    &                                     & + coh peaks & + hard gap   \\[+1mm]
Cooper pair MI      & CPs localized by                  & Hard gap & Hard gap \\[-2pt]
                    & Coulomb blockade                    &          &          \\[+1mm]
Cooper pair glass   & CPs localized by disorder         & Hard gap & Unknown \\[+1mm]
\hline
\end{tabular}
\caption{
	\label{comparison-table}
	Comparison of various fermionic ground states,
	characterized by transport (metallic/superconducting/insulating),
	single-particle density of states $N(\omega)$,
	and dynamical conductivity $\sigma(\omega)$.
	Abbreviations: 
	MI=Mott insulator, CP=Cooper pair, 
	WL=weak localization,
	(M)VRH=(Mott) variable-range hopping,
	ES=Efros-Sh'klovskii.
}
\end{table*}

In this article we have ignored the effects of Coulomb repulsion.  Sufficiently near the SIT, it is likely that the Coulomb effects renormalize away and the phonon-mediated attraction is the dominant effects (as in the theory of ordinary superconductors); however, the Coulomb amplitude suppression mechanism may play an important role in determining the overall shape of the phase diagram.  (See the SIT overview by N. Trivedi.)

\appendix

\chapter{Conventions and Standard Formulas for Imaginary-Time Green Functions}
\begin{figure}[!b]
	\centering
	\subfigure{
		\includegraphics[width=0.35\columnwidth]{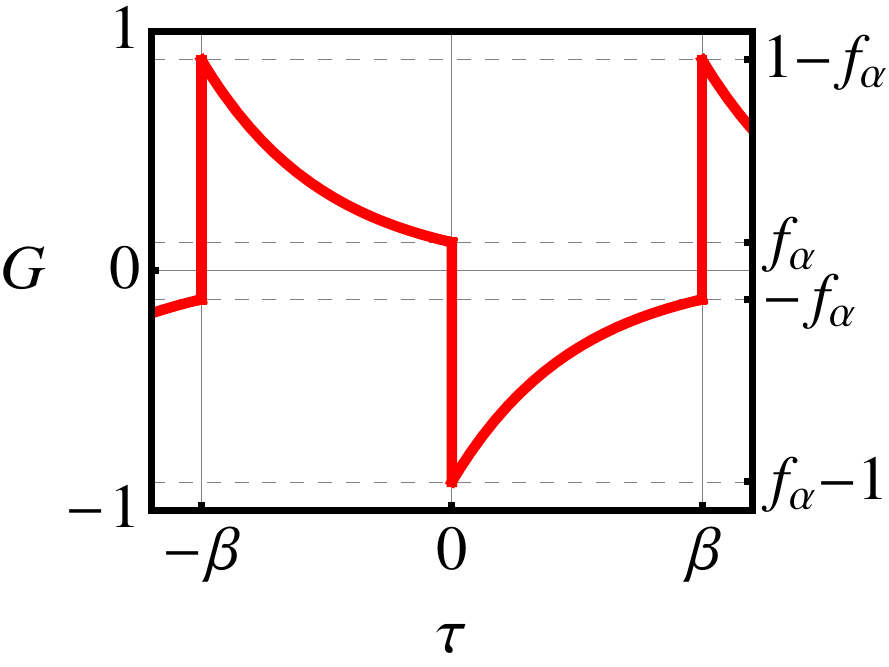}
		\label{schematic-Gtau}
	}
	\subfigure{
		\includegraphics[width=0.35\columnwidth]{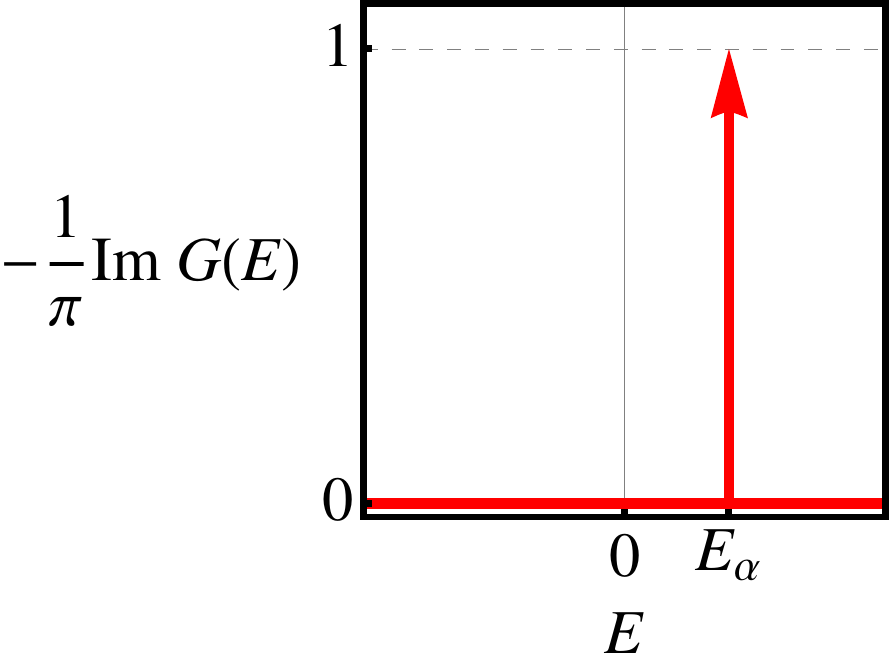}
	}
	\\
	\subfigure{
		\includegraphics[width=0.35\columnwidth]{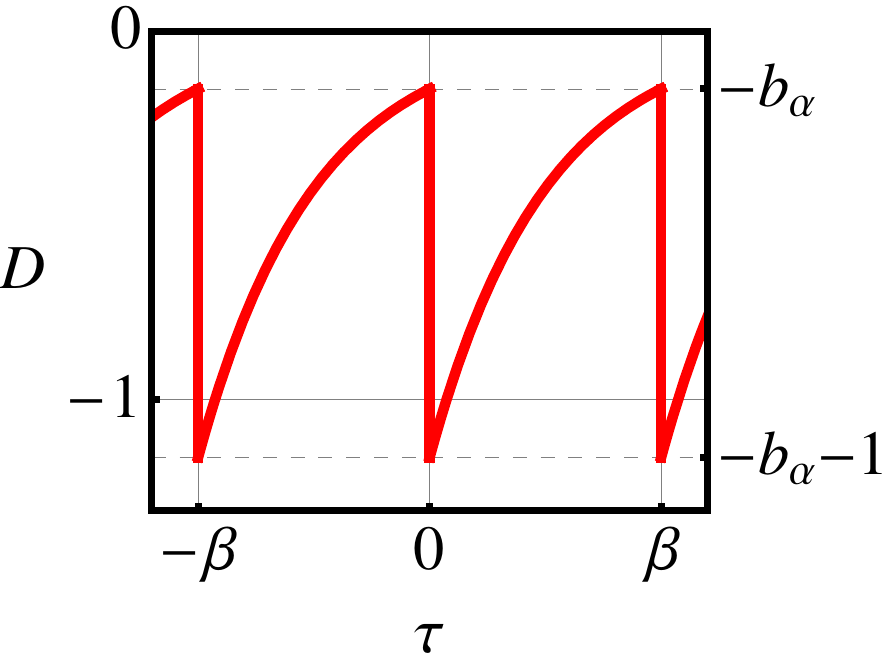}
		\label{schematic-Dtau}
	}
	\subfigure{
		\includegraphics[width=0.35\columnwidth]{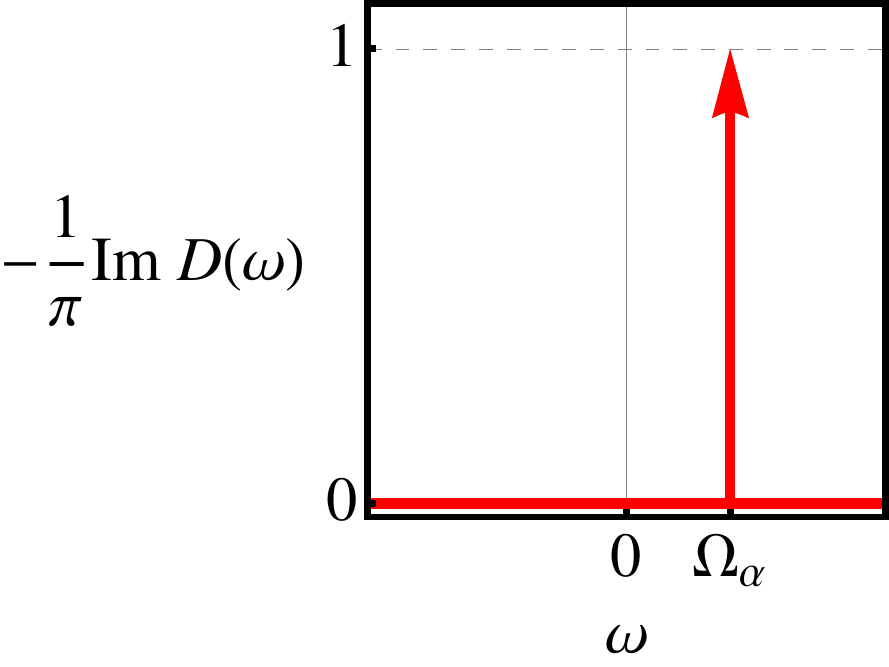}
	}
	\caption{
		Imaginary-time fermion Green function $G_{\alpha\tau}$ and
		boson Green function $D_{\alpha\tau}$
		corresponding to delta-function spectra (single eigenmodes).
	}
	\end{figure}

 \slabel{matsubara}
Consider the Hamiltonian for a single eigenmode, $H = E_\alpha \cdag_\alpha \cccc_\alpha$.
We will define the fermionic imaginary-time Green function as the amplitude for inserting a fermion at zero time and removing it at time $\tau$,
$G_{\alpha\tau} = \mean{ \calT \cccc_{\alpha\tau} \cdag_{\alpha 0}  } 
= \mean{ \psi_{\alpha\tau} \bar\psi_{\alpha 0}  }$,
where $\psi$ and $\bar\psi$ are the Grassmann fields in the coherent state path integral formalism.  This Green function has an exponential behavior and is antiperiodic with period $\beta$ such that
	\begin{align}
	G_{\alpha\tau}
	&=\begin{cases}
		-\beta<\tau<0 & f_\alpha e^{-\tau E_\alpha}  \\
		0<\tau<\beta  & (f_\alpha - 1) e^{-\tau E_\alpha}  \\
		\end{cases}
	\end{align}
where
$f_\alpha = f(\beta E_\alpha) = \frac{1}{e^{\beta E_\alpha} - 1} = \half - \half\tanh\tfrac{\beta}{2}  E_\alpha$, as illustrated in Fig.~\ref{schematic-Gtau}.  In the Matsubara frequency domain this Green function becomes
	\begin{align}
	G_{\alpha} (i\vare_n)
	&=\frac{1}{i\vare_n - E_\alpha}
	.
	\end{align}
Analytically continuing to real frequencies and taking the imaginary part gives
$-\frac{1}{\pi} \Im G(E + i0^+) = \delta_{E - E_\alpha}$.

Similarly, the standard bosonic Green function for a simple harmonic oscillator, $H = \Omega \adag \aaaa$, is
	\begin{align}
	D_{\alpha\tau}
	&=\begin{cases}
		-\beta<\tau<0 & -b_\alpha       e^{-\tau \Omega_\alpha} \\
		0<\tau<\beta  & -(b_\alpha + 1) e^{-\tau \Omega_\alpha} \\
		\end{cases}
	\end{align}
where 
$b_\alpha 
= b(\beta \Omega_\alpha) 
= \half \coth \frac{\beta}{2}\Omega_\alpha - \half
= \frac{1}{e^{\beta\Omega_\alpha} - 1}$.
This corresponds to
	\begin{align}
	D_{\alpha} (i\omega_l)
	&=\frac{1}{i\omega_l - \Omega_\alpha}.
	\end{align}
Analytically continuing to real frequencies and taking the imaginary part gives
$-\frac{1}{\pi} \Im D(E + i0^+) = \delta_{\omega - \Omega_\alpha}$.

	\begin{figure}[!hbtp]
	\centering
	\subfigure[Pairing bubble]{
		\hspace{4mm}
		\includegraphics[width=0.2\columnwidth]{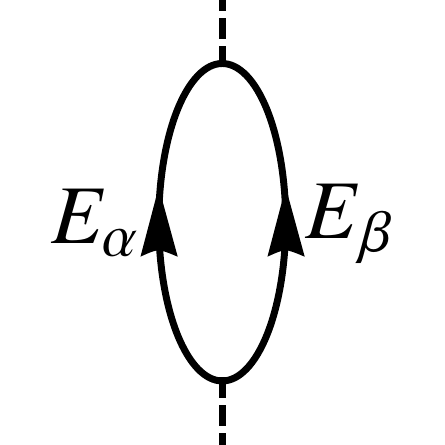}
		\hspace{4mm}
		\label{pairingBubble}
	}
	\subfigure[Polarization bubble]{
		\hspace{4mm}
		\includegraphics[width=0.2\columnwidth]{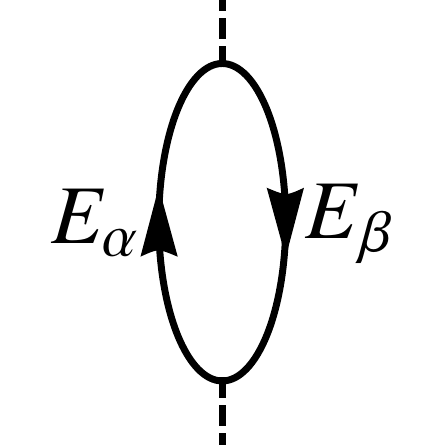}
		\hspace{4mm}
		\label{polarizationBubble}
	}
	\subfigure[Self-energy insertion]{
		\hspace{4mm}
		\includegraphics[width=0.2\columnwidth]{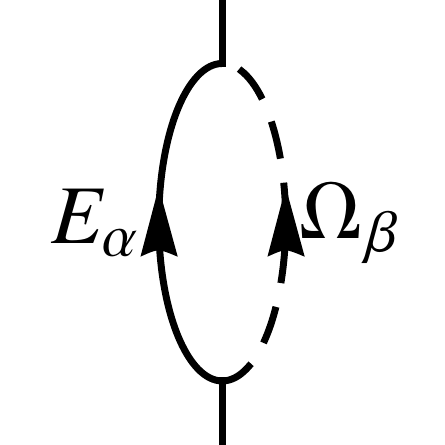}
		\hspace{4mm}
		\label{selfEnergyBubble}
	}
	\caption{
		Two-particle diagrams involving fermion lines (solid) and boson lines (dashed).
	}
	\end{figure}

Now consider the pairing bubble diagram corresponding to the creation of two fermions in eigenmodes $\alpha$ and $\beta$, as shown in Fig.~\ref{pairingBubble}:
 	\begin{align}
	P_\tau 
	&=G_{\alpha\tau} G_{\beta\tau}
	\nonumber\\
	&=(1 - f_\alpha)(1 - f_\beta) e^{-\tau (E_\alpha + E_\beta)}
	\quad\text{(for $0<\tau<\beta$)}
	.
	\nonumber
	\end{align}
Fourier transforming to Matsubara frequencies gives
 	\begin{align}
	P(i\omega_l)
	&=\frac{f_\alpha + f_\beta - 1}{i\omega_l - E_\alpha - E_\beta}
	\end{align}
after some algebra.  

This is a very useful formula, as it tells us the spectrum of a bosonic response in terms of the spectra (eigenenergies) of the fermionic states.
The physical meaning is that the amplitude for creating a fermion pair in modes $\alpha$ and $\beta$ depends on the occupation of those two modes.
The formula automatically satisfies the requirement that 
$-\frac{1}{\pi} \Im P(\omega)$ has the same sign as $\omega$.

Note that $P(\omega=0)$, like $D(\omega=0)$, is negative with these conventions.  Response functions like compressibility are traditionally defined to be positive at zero frequency; therefore, they correspond to the \emph{negatives} of bosonic propagators....

The formula for a polarization (particle-hole) bubble, Fig.~\ref{polarizationBubble}, can be derived simply by changing the signs of $E_{\alpha}$ and $E_{\beta}$:
 	\begin{align}
	\kappa_\tau 
	&=G_{\alpha\tau} G_{\beta,-\tau}
	,
	\end{align}
 	\begin{align}
	\kappa(i\omega_l)
	&=\frac{f_\alpha - f_\beta}{i\omega_l - E_\alpha + E_\beta}
	.
	\end{align}


Analogous formulas can be derived for the self-energy of a fermion due to emission of a boson (Fig.~\ref{selfEnergyBubble}):
 	\begin{align}
	\Sigma_\tau 
	&=G_{\alpha\tau} D_{\beta\tau}
	,
	\end{align}
 	\begin{align}
	\Sigma(i\vare_n)
	&=\frac{-(1 - f_\alpha + b_\beta)}{i\vare_n - E_\alpha - \Omega_\beta}
	.
	\end{align}

\chapter{Derivation of Kubo Formulas for Electromagnetic Response}
\label{kuboDerivation}

\section{General Derivation}
Here we derive the Kubo formula for the electromagnetic response of the Anderson model, a BCS superconductor, and a dirty superconductor.

The response to an electromagnetic field (superfluid response, conductivity, and dielectric polarizability) is an important quantity characterizing the SIT.
We now derive the Kubo formula for the electromagnetic response tensor.
\cite{scalapinowhitezhang1993}
For simplicity we will omit spin indices.
We will work in units where $e=\hbar=1$.  In this system, the unit of magnetic flux is $\hbar/e$, and the unit of conductivity is $e^2/\hbar$.  Thus (for a Cooper-paired system) the flux quantum is $\Phi_Q = h/e=2\pi$ and the conductance quantum is $G_Q = e^2/4h = 2/\pi$.

For non-relativistic charged particles in the continuum, the electromagnetic vector potential $\AAA(\rrr)$ couples to the matter fields in a gauge-invariant way in the kinetic energy term of the Hamiltonian
	\begin{align}
	H_\text{kin}
	&=
		\int d^d r~
		\tfrac{1}{2m} \cdag(\rrr)
		 (-i\nabla - \AAA(\rrr)) \cdot (i\nabla - \AAA(\rrr)) \ccc(\rrr)
	.\nonumber
	\end{align}
For charged particles on a lattice the analogous term is
	\begin{align}
	H_\text{kin}
	&=
	-	\sum_{ij}
			t_{ij} e^{iA_{ij}} \cdag_{i} \cccc_{j}
	\elabel{latticeKE}
	\end{align}
where the lattice vector potential is the line integral of the continuum vector potential,
$	A_{ij}
	=
 	\int_{\rrr_i}^{\rrr_j} \mathbf{dr} \cdot \AAA(\rrr)
$.  
The units of $A$ are $\mathrm{V\ s\ m^{-1}}$, whereas the units of $A_{ij}$ are $\mathrm{V\ s}$, so $\frac{e}{\hbar} A_{ij}$ is dimensionless as required by \eref{latticeKE}.

Following the usual Kubo procedure, we compute the response to a general vector potential that depends on imaginary time and space, $A_{\tau b}$.  It will be convenient to perform this derivation in the coherent state path integral formalism, where fermion operators $c$ are replaced by Grassmann fields $\psi$.  The kinetic energy term in the action is
	\begin{align}
	S_\text{kin} [\psi, \bar\psi, A]
	&=
		\int_\tau	\sum_{ij}
		t_{ij} e^{iA_{ij\tau}} \bar\psi_{i\tau} \psi_{j\tau}
	.
	\elabel{latticeKEaction}
	\end{align}
The full action $S$ may contain other terms that do not depend on $A$.

The current field (corresponding to the current operator $\hat{j}$ in the operator formalism) can be identified as the field that couples linearly to $A$:
	\begin{align}
	j_{ij\tau} [\psi, \bar\psi, A]
	&=\frac{\dd S}{\dd A_{ij}} 	
	= it_{ij} e^{iA_{ij\tau}} \bar\psi_{i\tau} \psi_{j\tau}
	\end{align}
where $i$'s in subscripts are site indices whereas $i$'s in prefactors refer to $\sqrt{-1}$.
(For simplicity in later derivations, we are treating $A_{ij}$ and $A_{ji}$ as independent fields.  So in the above equation, $j_{ij}$ only includes the forward current from site $i$ to site $j$.)

For later convenience, let $k$ be the second derivative of the action,
	\begin{align}
	k_{ij\tau} [\psi, \bar\psi, A]
	&=\frac{\dd^2 S}{\dd A_{ij} ^2} 	
	= -t_{ij} e^{iA_{ij\tau}} \bar\psi_{i\tau} \psi_{j\tau}
	.
	\end{align}

The partition function, or generating functional, is
	\begin{align}
	Z [A]
	&=\int [d\psi\ d\bar\psi] \exp S[\psi, \bar\psi, A]
	.
	\end{align}
The electromagnetic response function is the second derivative of the free energy:
	\begin{align}
	\Upsilon_{ijkl\tau\tau'}
	&=
		-\frac{\dd \mean{j_{ij\tau}}}{\dd A_{kl\tau'}}
	=
		-T \frac{\dd^2 \ln Z}{\dd A_{ij\tau} \dd A_{kl\tau'}}
\nonumber\\
	&=
		-T\frac{\dd^2 }{\dd A_{ij\tau} \dd A_{kl\tau'}}
		\ln 
		\int [d\psi\ d\bar\psi] \exp S[\psi, \bar\psi, A]
	.
	\end{align}
	\newcommand{\bigmean}{\mean}
Differentiating the $\ln$ and $\exp$ functions produces three terms:
	\begin{align}
	\beta\Upsilon_{ijkl\tau\tau'}
	&=
	-	\mean{\frac{\dd^2 S}{\dd A_{ij\tau}~ \dd A_{kl\tau'}} }
	- \mean{\frac{\dd S}{\dd A_{ij\tau}} 
				 ~\frac{\dd S}{\dd A_{kl\tau'}} }
			\nonumber\\&~~~~{}
	+ \mean{\frac{\dd S}{\dd A_{ij\tau}}  }
		\mean{\frac{\dd S}{\dd A_{kl\tau'}} }
	.
	\end{align}
We shall be studying the linear response to an infinitesimal perturbation, so we set $A_{ij}=0$.  Then, the last term in \eref{EMResponseOfLattice} vanishes.  Also, the response function depends only upon $\tau-\tau'$, so we can set $\tau'=0$ without loss of generality.  In terms of $k$ and $j$,
	\begin{align}
	\beta\Upsilon_{ijkl\tau}
	&=
		\delta_{ik} \delta_{jl} \delta(\tau)
		\mean{-k_{ij}} 		
	- \mean{j_{ij\tau} j_{kl} }
	,
	\end{align}
where $k_{ij} = -t_{ij} \bar\psi_{i} \psi_{j}$ is the bond kinetic energy
and $j_{ij\tau}	= it_{ij} \bar\psi_{i\tau} \psi_{j\tau}$ is the current field.
(We are using the convention that omitted $\tau$ indices mean $\tau=0$, i.e., $\psi_{i} \equiv \psi_{i} (\tau=0)$, just as in the operator formalism, where $\cccc_{i} \equiv \cccc_{i} (\tau=0)$.)

Ultimately, we wish to find the electromagnetic (EM) response tensor 
$\Upsilon_{\mu\nu\qqq\omega} 
 = 
 -\dd j_{\nu\qqq\omega} / \dd A_{\mu\qqq\omega}$, which is the current response to an applied vector potential with wavevector $\qqq$, frequency $\omega$, and polarization $\mu$.  Here we consider the transformations from the lattice basis $ij$ to the polarization-wavevector basis $\mu\qqq$ (omitting $\tau$ for the moment):
	\begin{align}
	A_{ij} &= \sum_{\mu\qqq} e^{i \qqq\cdot\rrr_i} r_{ij\mu} A_{\mu\qqq}
	,\\
	j_{\mu\qqq} 
	&= \sum_{ij} e^{-i \qqq\cdot\rrr_i} r_{ij\mu} j_{ij} 
	= \sum_{ij} e^{-i \qqq\cdot\rrr_i} r_{ij\mu} it_{ij} \bar\psi_{i} \psi_{j}
	,\\
	-k_{\mu\nu} 
	&= -\sum_{ij} r_{ij\mu} r_{ij\nu} k_{ij} 
	=  \sum_{ij} r_{ij\mu} r_{ij\nu} t_{ij} \bar\psi_{i} \psi_{j}
	,
	\end{align}
where $\rrr_i$ is the position vector of site $i$
and $\rrr_{ij} = \rrr_j - \rrr_i$ is the displacement vector from site $i$ to site $j$.
In terms of these quantities,
	\begin{align}
	\Upsilon_{\mu\nu\qqq\tau}
	&=
		\delta(\tau)
		\mean{-k_{\mu\nu}} 		
	- \mean{j_{\mu\qqq\tau} j_{\nu\bar\qqq} }
	.
	\elabel{EMResponseOfLattice}
	\end{align}
\eref{EMResponseOfLattice} is a Kubo formula for the electromagnetic linear response function in terms of the current-current correlation.  This is an example of the fluctuation-dissipation theorem.   
The first term is the kinetic energy for each bond weighted by a geometrical factor depending on the hopping distance, and it gives a diamagnetic response (the induced $\jjj$ is opposite to the applied $\AAA$).  The second term is the current-current correlation function, which gives a paramagnetic contribution.

The derivation of \eref{EMResponseOfLattice} is very general.  It can easily be adapted for XY models instead of fermions, for example, by considering an XY model action coupled to an A field, $S[\theta,A]=\sum_{ij} \beta J_{ij} \cos(\theta_i - \theta_j + A_{ij})$, and carrying out the differentiations to obtain $j$, $k$, and $\Upsilon$.

\section{Electromagnetic Response of Anderson Model} \label{AppendixAndersonKubo}
Now let us derive the Kubo formula explicitly for the Anderson model.
Return to the operator formalism (replace $\bar\psi_i \psi_j$ by $\cdag_i \cccc_j$), and transform the fermion operators $c_{i}$ from the site basis to the eigenmode basis, $\gamma_{\alpha}$: 
	\begin{align}
	c_i
	&=\sum_\alpha \phi_{i\alpha} \gamma_{\alpha}
,\nonumber\\
	k_{\mu\nu}
	&=-\sum_{\alpha\beta ij} r_{ij\mu} r_{ij\nu} t_{ij} 
		\phi^*_{i\alpha} \phi_{j\beta} \gammadag_{\alpha} \gamma_{\beta}
,\nonumber\\
	j_{\mu\qqq\tau} 
	&= \sum_{\alpha\beta} \sum_{ij} e^{-i \qqq\cdot\rrr_i} r_{ij\mu}
		it_{ij}
	 \phi^*_{i\alpha} \phi_{j\beta} 
	\gammadag_{\alpha\tau} \gamma_{\beta\tau}
	= 
		\sum_{\alpha\beta} \Gamma_{\alpha\beta\mu\qqq}
	\gammadag_{\alpha\tau} \gamma_{\beta\tau}
	,
	\end{align}
where
	\begin{align}
	\Gamma_{\alpha\beta\mu \qqq}
	&=\sum_{ij}  
		e^{-i\qqq\cdot\rrr_i} 
		r_{ij\mu}
		it_{ij}
		\phi^*_{i\alpha} \phi_{j\beta}
	\end{align}
are matrix elements for the coupling between electromagnetic plane waves and particle-hole excitations in disorder eigenstates.
Since the Hamiltonian is bilinear and diagonal in the eigenbasis, expectations of products of $\gamma$'s can be conveniently reduced to products of Green functions using Wick's theorem.  Using $\mean{\gammadag_{\alpha} \gamma_{\beta} } = f_\alpha \delta_{\alpha\beta}$, $\mean{\gamma_{\alpha\tau} \gammadag_{\beta} } = G_{\alpha\tau} \delta_{\alpha\beta}$, etc., we obtain 
	\begin{align}
	\mean{-k_{\mu\nu}}
	&=\sum_{\alpha ij} r_{ij\mu} r_{ij\nu} t_{ij} 
		\phi^*_{i\alpha} \phi_{j\alpha} f_\alpha
,\nonumber\\
	\mean{j_{\mu\qqq\tau} j_{\nu\bar\qqq} }
	&=
		-\sum_{\alpha\beta}
		\Gamma_{\alpha\beta\mu\qqq} \Gamma_{\beta\alpha\nu\bar\qqq}
		G_{\alpha,-\tau}	G_{\beta\tau}
	,
	\end{align}
where $G_{\alpha\tau}$ is the Green function for inserting a fermion into an eigenmode $\alpha$ with energy $E_\alpha$ (see \sref{matsubara}).
\footnote{In general, one has to be careful with infinitesimal time shifts in the path-integral formalism that arise from anticommutation relations in the operator formalism.  In this particular working, however, the $1$'s arising from commutation relations drop out.}
Transforming to the Matsubara frequency domain and analytically continuing to real frequencies gives
	\begin{align}
	\Upsilon_{\mu\nu\qqq\omega}
	&= 
		\mean{-k_{\mu\nu}}
	+
		\sum_{\alpha\beta} 
		\Gamma_{\alpha\beta\mu\qqq} \Gamma_{\beta\alpha\nu\bar\qqq}
		\frac{f_\beta - f_\alpha}{E_\alpha - E_\beta - \omega}
	.
	\elabel{AndersonModelUpsilon}
	\end{align}
Taking the imaginary part leads to an expression for the spectral (dissipative) part of the EM response, which can be computed efficiently by accumulating delta function weights in bins:
	\begin{align}
	\frac{ \Im \Upsilon_{\mu\nu\qqq\omega}}{\pi}
	&= 
		\sum_{\alpha\beta} 
		\Gamma_{\alpha\beta\mu\qqq} \Gamma_{\beta\alpha\nu\bar\qqq}
		(f_\beta - f_\alpha)~
		 \delta(E_\alpha - E_\beta - \omega)
	.
	\elabel{AndersonModelImUpsilon}
	\end{align}
	\begin{figure}[!h]
		\centering
		\includegraphics[width=0.2\columnwidth]{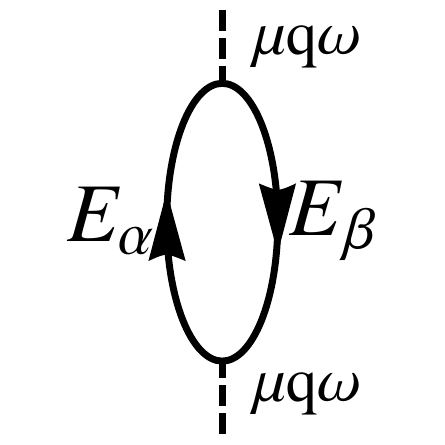}
	\caption{
		Diagrammatic visualization of the Kubo formula for the electromagnetic response,
		\eref{AndersonModelImUpsilon}.
	\label{conductivityBubble}
	}
	\end{figure}
This equation is visualized diagrammatically in Fig.~\ref{conductivityBubble}.
		The vertex factors $\Gamma_{\alpha\beta\mu\qqq}$ 
		are matrix elements connecting plane EM waves with disorder eigenstates,
		the delta function imposes energy conservation,
		and the Fermi occupation factors affect
		the amplitude of creating a particle and hole in disorder eigenstates.

To obtain the reactive response as a function of frequency, it is most efficient to infer it using the Kramers-Kronig relation:
	\begin{align}
	\Re \Upsilon_{\mu\nu\qqq\omega}
	&=
		\mean{-k_{\mu\nu}}
+
	\calP \int_{-\infty}^{\infty} \frac{d\omega'}{\pi}~ 
  \frac{ \Im \Upsilon_{\mu\nu\qqq\omega'} }{ \omega - \omega' }
	.
	\elabel{AndersonModelReUpsilon}
	\end{align}
\eref{AndersonModelImUpsilon} and \eref{AndersonModelReUpsilon} provide an efficient way to calculate the EM response.

To make contact with the notation in the literature,\cite{scalapinowhitezhang1993}
 observe that for a square lattice with lattice spacing $a$ and nearest-neighbor hopping $t$, the equations for $k$ and $\Gamma$ can be written
	\begin{align}
	\mean{k_x}
	&=t a^2
		\sum_{i\alpha}
		f_\alpha (\phi^*_{i\alpha} \phi_{i+\hat{x}, \alpha} + h.c.)
,\\
	\Gamma_{\alpha\beta x\qqq}
	&=i t a \sum_{i}  
		e^{i\qqq\cdot\rrr_i} 
		(\phi^*_{i\alpha} \phi_{i+\hat{x}, \beta} - h.c.)
	\end{align}
where $i+\hat{x}$ refers to the neighbor of site $i$ in the $x$ direction.

\section{Electromagnetic Response of Clean Superconductor} \label{AppendixCleanSCKubo}
Let us now derive the Kubo formula for the EM response of a clean superconductor using the formalism in Sec. 1.3.1.  The diamagnetic term $k$ (which involves bond kinetic energies) and bond current $j$ now include a sum over both spin species,
	\begin{align}
	k_{ij} 
	&=-\sum_{\sigma} t_{ij} \cdag_{i\sigma} \cccc_{j\sigma}
,\\
	j_{ij}
	&=\sum_\sigma  it_{ij} 			\cdag_{i\sigma} \cccc_{j\sigma}
	.
	\end{align}
Transform this into the wavevector-polarization basis:
	\begin{align}
	k_{\mu\nu} 
	&= -\sum_{ij\sigma} r_{ij\mu} r_{ij\nu} t_{ij} \cdag_{i\sigma} \cccc_{j\sigma}
,\\
	j_{\mu\qqq} 
	&= \sum_{ij\sigma} e^{-i \qqq\cdot\rrr_i} r_{ij\mu} it_{ij} 
		\cdag_{i\sigma} \cccc_{j\sigma}
	.
	\end{align}
Using the Fourier relations
	\begin{align}
	\cccc_{j\sigma} &= \sum_\kkk e^{-i\kkk\cdot\rrr_j} \cccc_{\kkk\sigma}
,\\
	t_{ij} &=  \sum_\kkk  e^{i\kkk\cdot(\rrr_j - \rrr_i)} \vare_\kkk 
,\\
	r_{ij\mu} it_{ij} &=  \sum_\kkk  e^{i\kkk\cdot(\rrr_j - \rrr_i)}  
		\frac{\dd \vare_\kkk}{\dd k_\mu}
,
	\end{align}
one obtains
	\begin{align}
	k_{\mu\nu} 
	&=\sum_{\kkk} \frac{\dd^2 \vare_\kkk}{\dd k_\mu \dd k_{\nu}}
		\left(
			\cdag_{\kkk\up} \cccc_{\kkk\up}
		+	\cdag_{\kkk\dn} \cccc_{\kkk\dn}
		\right)
,\\
	j_{\mu\qqq} 
	&=\sum_{\kkk} \frac{\dd \vare_\kkk}{\dd k_\mu}
		\left(
			\cdag_{\kkk\up} \cccc_{\ppp\up}
		+	\cdag_{\kkk\dn} \cccc_{\ppp\dn}
		\right)
	\end{align}
where $\ppp$ is shorthand for $\ppp=\kkk+\qqq$.
\todo{(signs?)}
Now, transform the fermion fields into the bogolon basis using
	\begin{align}
	\cccc_{\kkk\up}
	&=u_\kkk \gamma_{\kkk\up} + v_\kkk \gammadag_{-\kkk\dn}
\\
	\cccc_{\kkk\dn}
	&=u_\kkk \gammadag_{\kkk\dn} - v_\kkk \gamma_{-\kkk\up}
	\end{align}
(assuming $u_\kkk=u_{-\kkk}$).  This gives
	\begin{align}
	k_{\mu\nu} 
	&=\sum_{\kkk} \frac{\dd^2 \vare_\kkk}{\dd k_\mu \dd k_{\nu}}
		\left[
			(u_\kkk \gammadag_{\kkk\up} + v_\kkk \gamma_{-\kkk\dn})
			(u_\kkk \gamma_{\kkk\up} + v_\kkk \gammadag_{-\kkk\dn})
		+
			(u_\kkk \gamma_{\kkk\dn} - v_\kkk \gammadag_{-\kkk\up})
			(u_\kkk \gammadag_{\kkk\dn} - v_\kkk \gamma_{-\kkk\up})
		\right]
,\nonumber\\
	j_{\mu\qqq} 
	&=\sum_{\kkk} \frac{\dd \vare_\kkk}{\dd k_\mu}
		\left[
			(u_\kkk \gammadag_{\kkk\up} + v_\kkk \gamma_{-\kkk\dn})
			(u_\ppp \gamma_{\ppp\up} + v_\ppp \gammadag_{-\ppp\dn})
		+
			(u_\kkk \gamma_{\kkk\dn} - v_\kkk \gammadag_{-\kkk\up})
			(u_\ppp \gammadag_{\ppp\dn} - v_\ppp \gamma_{-\ppp\up})
		\right]
	.\nonumber
	\end{align}
Now, the prefactor  $\frac{\dd \vare_\kkk}{\dd k_\mu k_{\nu}}$ is even in $\kkk$ (unchanged under the transformation $\kkk \rightarrow -\kkk$), whereas $\frac{\dd \vare_\kkk}{\dd k_\mu}$ is odd in $\kkk$.  Carefully collecting terms and simplifying by substituting dummy indices, being mindful of the abovementioned symmetries and the anticommutation relations, leads to 
	\begin{align}
	k_{\mu\nu} 
	&=\sum_{\kkk} \frac{\dd^2 \vare_\kkk}{\dd k_\mu \dd k_{\nu}}
		\left[
			(u_\kkk {}^2 - v_\kkk {}^2)
			(
				\gammadag_{\kkk\up} \gamma_{\kkk\up}
			+	\gamma_{\kkk\dn} \gammadag_{\kkk\dn}
			)
		+ \text{(terms involving $\gamma\gamma$ and $\gammadag\gammadag$)}
		\right]
,\nonumber\\
	j_{\mu\qqq} 
	&=\sum_{\kkk} \frac{\dd \vare_\kkk}{\dd k_\mu}
		\Big[
			(u_\kkk u_\ppp + v_\kkk v_\ppp)^2
			(
				\gammadag_{\kkk\up} \gamma_{\ppp\up}
			+	\gammadag_{\kkk\dn} \gamma_{\ppp\dn}
			)
		+
			(u_\kkk v_\ppp - v_\kkk u_\ppp)^2
			(
				\gammadag_{\kkk\up} \gammadag_{\ppp\dn}
			+	\gamma_{\kkk\up} \gamma_{\ppp\dn}
			)
		\Big]
	.\nonumber
	\end{align}
The expression for $j$ contains combinations of $u$ and $v$ known as ``Case II coherence factors''\cite{tinkham}.
We can now take the expectation to obtain the diamagnetic term in the Kubo formula,
	\begin{align}
	\mean{k_{\mu\nu}}
	&=\sum_{\kkk} \frac{\dd^2 \vare_\kkk}{\dd k_\mu \dd k_{\nu}}
			\frac{\xi_\kkk}{E_\kkk}
			(2f_\kkk - 1)
	.
	\end{align}
The factor $\frac{\xi_\kkk}{E_\kkk}$ is due to the binding of electrons into Cooper pairs, which lowers their potential energy at the expense of a gain in kinetic energy.  (For continuum electrodynamics, the diamagnetic term is simply proportional to the electron density, and this factor is absent.)
For the square lattice, putting in the explicit forms of the dispersion relation gives
	\begin{align}
	\mean{k_{xx}}
	&=\int_\kkk
		(2\cos k_x)
		\frac{\xi_\kkk}{E_\kkk}
			(2f_\kkk - 1)
	.
	\end{align}
For the paramagnetic term, Wick contraction of $\gamma$ and $\gammadag$ operators leads to
	\begin{align}
	\mean{j_{\mu\qqq\tau} j_{\nu\bar\qqq} }
	&=\sum_{\kkk} 
		\frac{\dd \vare_\kkk}{\dd k_\mu}
		\frac{\dd \vare_\kkk}{\dd k_\nu}
		\Big[
			(u_\kkk u_\ppp + v_\kkk v_\ppp)^2
			(
				G_{\kkk\tau} G_{\ppp\bar\tau}
			+	G_{\kkk\tau} G_{\ppp\bar\tau}
			)
\nonumber\\&{}~~~~~~~~~~~
		+
			(u_\kkk v_\ppp - v_\kkk u_\ppp)^2
			(
				G_{\kkk\tau} G_{\ppp\bar\tau}
			+	G_{\kkk\tau} G_{\ppp\bar\tau}
			)
		\Big]
	.
	\end{align}
\todo{check}
Fourier-transforming to Matsubara frequencies and using the explicit forms of $\vare_\kkk$, $u_\kkk$ and $v_\kkk$ gives the result from SWZ~\cite{scalapinowhitezhang1993},
	\begin{align}
	\Lambda_{xx} (q_y, \omega_m)
	&=
		\tfrac{4}{N} \sum_\ppp
		\sin^2 p_x
\Bigg\{
	\left[
		\tfrac{1}{2} 
		\left( 1 - \tfrac{\xi_\ppp\xi_{\ppp+\qqq} + \Delta^2}{E_\ppp E_{\ppp+\qqq}} \right)
	\right]
	\left[
		\tfrac{1 - f_\ppp - f_{\ppp+\qqq}}{E_\ppp + E_{\ppp+\qqq} + i\omega_m}
	+	\tfrac{1 - f_\ppp - f_{\ppp+\qqq}}{E_\ppp + E_{\ppp+\qqq} - i\omega_m}
	\right]
			\nonumber\\&{}~~~~~~~~~~~~~~~~~~~~
+
	\left[
		\tfrac{1}{2} 
		\left( 1 + \tfrac{\xi_\ppp\xi_{\ppp+\qqq} + \Delta^2}{E_\ppp E_{\ppp+\qqq}} \right)
	\right]
	\left[
		\tfrac{f_{\ppp+\qqq} - f_\ppp}{E_\ppp - E_{\ppp+\qqq} + i\omega_m}
	+	\tfrac{f_{\ppp+\qqq} - f_\ppp}{E_\ppp - E_{\ppp+\qqq} - i\omega_m}
	\right]
\Bigg\}
		.\nonumber
	\end{align}
Finally, combining diamagnetic and paramagnetic contributions gives
	\begin{align}
	\Upsilon_{xx\qqq\omega}
	&= \mean{ -k_{xx} } - \Lambda_{xx\qqq\omega}
		.
	\end{align}
The above Kubo formula can be visualized roughly in terms of a two-fluid picture, where the superfluid exhibits a diamagnetic response, whereas the normal fluid gives a paramagnetic contribution arising from quasiparticle excitations.
The static uniform superfluid stiffness is obtained in the limit $\omega_m=0, \qqq\rightarrow 0$:
	\begin{align}
	\Upsilon_{xx}
	&= \mean{ -k_{xx} } - \Lambda_{xx}
				\nonumber\\			\text{where} 	 \quad
	\Lambda_{xx}
	&=
-
		\frac{8}{N} \sum_\ppp
		\sin^2 p_x
	\frac{\dd f(E_\ppp)}{\dd E_\ppp}
	\end{align}
where $\frac{\dd f}{\dd E_\ppp} = -\frac{\beta}{4} \sech^2 \frac{\beta}{2} E_\ppp$.

\section{Electromagnetic Response of Dirty Superconductor} \label{AppendixDirtySCKubo}
Let us follow the derivation of \eref{EMResponseOfLattice}, but including spin indices.  For later convenience, begin by anticommuting the down spin fermion operators:
	\begin{align}
	j_{ij} 
	&=it_{ij} 
		\left(	
			\cdag_{i\up} \cccc_{j\up}  
		-	\cccc_{j\dn} \cdag_{i\dn} 
		\right)
	,\\
	k_{ij} 
	&=-t_{ij} 
		\left(	
			\cdag_{i\up} \cccc_{j\up}  
		-	\cccc_{j\dn} \cdag_{i\dn} 
		\right)
	.
	\end{align}
Transforming the fermion fields from the site basis, $\cccc_{i\tau}$, to the eigenmode basis, $\gamma_{\alpha\tau}$ using
	\begin{align}
	\cccc_{i\up}
	&=\sum_\alpha u_{i\alpha} \gamma_{\alpha}
,\nonumber\\
	\cccc_{i\dn}
	&=\sum_\alpha v_{i\alpha} \gamma^\dag_{\alpha}
,
	\end{align}
and interchanging some dummy indices, gives
	\begin{align}
	j_{ij} 
	&=it_{ij} 
		\sum_{\alpha\beta}
		\left(	
			u_{i\alpha} u_{j\beta} 
		-	v_{j\alpha} v_{i\beta}
		\right)
		\gamma^\dag_{\alpha} \gamma_{\beta}
	,\\
	k_{ij} 
	&=-t_{ij} 
		\sum_{\alpha\beta}
		\left(	
			u_{i\alpha} u_{j\beta} 
		-	v_{j\alpha} v_{i\beta}
		\right)
		\gamma^\dag_{\alpha} \gamma_{\beta}
	.
	\end{align}
Carrying out the rest of the derivation, in analogy with the Anderson case, leads to the same form for the Kubo formula,
	\begin{align}
	\frac{ \Im \Upsilon_{\mu\nu\qqq\omega}}{\pi}
	&= 
		\sum_{\alpha\beta} 
		\Gamma_{\alpha\beta\mu\qqq} \Gamma_{\beta\alpha\nu\bar\qqq}
		(f_\beta - f_\alpha)~
		 \delta(E_\alpha - E_\beta - \omega)
,\\
	\Re \Upsilon_{\mu\nu\qqq\omega}
	&=
		\mean{-k_{\mu\nu}}
+
	\calP \int_{-\infty}^{\infty} \frac{d\omega'}{\pi}~ 
  \frac{ \Im \Upsilon_{\mu\nu\qqq\omega'} }{ \omega - \omega' }
,
	\end{align}
except that $\mean{-k}$ and $\Gamma$ now involve both ``spin components'' of the eigenvectors (\eref{BdGEMQuantities}) and the eigenmode indices $\alpha$ and $\beta$ now run from $1$ to $2N$:
	\begin{align}
	\mean{k_{\mu\nu}}
	&=- \sum_{\alpha ij} r_{ij\mu} r_{ij\nu} t_{ij} 
		\left(	
			u_{i\alpha} u_{j\alpha} 
		-	v_{j\alpha} v_{i\alpha}
		\right)
		f_\alpha
,\nonumber\\
	\Gamma_{\alpha\beta\mu \qqq}
	&=\sum_{ij}  
		e^{-i\qqq\cdot\rrr_i} 
		r_{ij\mu}
		it_{ij}
		\left(	
			u_{i\alpha} u_{j\beta} 
		-	v_{j\alpha} v_{i\beta}
		\right)
.
	\elabel{BdGEMQuantities}
	\end{align}

\chapter{Variational BdG Formalism} \slabel{variationalFormalism}
In this appendix we describe a variational mean-field treatment of the Hubbard model.  This is the only way to decouple the Hubbard interaction in multiple channels without overcounting it.

\section{Variational Method in Statistical Mechanics}
For a system described by Hamiltonian $\hat{H}$ at temperature $T=1/\beta$, the partition function is
	\begin{align}
	Z &= \Tr e^{-\beta\hat{H}}.
	\end{align}
Let $\hat{\rho}$ be an arbitrary density matrix.  Then, formally, we may write
	\begin{align}
	Z 
	&= \Tr \hat{\rho} e^{-\beta\hat{H}-\ln\hat{\rho}}
	= \mean{ e^{-\beta\hat{H}-\ln\hat{\rho}} }_{\hat{\rho}}
	.
	\end{align}
For a classical random variable $X$ and any convex function $f$, Jensen's inequality states that 
$\mean{f(X)} \geq f( \mean{X} )$.  This theorem is easily extended to expectations of a Hermitian operator with respect to a density matrix, i.e., 
$\mean{ f( \hat{X} ) }_{\hat\rho} \geq f( \mean{\hat{X}}_{\hat\rho} )$.
Thus,
	\begin{align}
	Z 
	\geq \exp  \mean{ -\beta\hat{H}-\ln\hat{\rho} }_{\hat{\rho}}
	.
	\end{align}
Taking logs of both sides gives
	\begin{align}
	\Omega = -T \ln Z
	&
	\leq  \mean{ \hat{H} + T \ln\hat{\rho} }_{\hat{\rho}}
	.
	\end{align}
The \emph{variational free energy} $\Omega_\text{var} = \mean{ \hat{H} + T \ln\hat{\rho} }_{\hat{\rho}}$ with respect to \emph{any} density matrix is an upper bound on the true free energy $\Omega$.
This is the variational principle of quantum statistical mechanics.
It is exploited in the variational \emph{method}, in which $\Omega_\text{var}$ is optimized with respect to the trial density matrix $\rho$ to obtain a \emph{least} upper bound to $\Omega$.
\footnote{
One also hopes that the optimal trial density matrix is \emph{close} to the true density matrix,
but this is not guaranteed. 
All variational calculations suffer from the inevitable bias involved in choosing a reasonably simple form for the trial density matrix (or wavefunction).
}

\section{Variational Mean-Field Theory with a Trial Hamiltonian}
Chaikin and Lubensky~\cite{chaikin} describe a general formalism in which the trial density matrix $\hat{\rho}$ is written as the product of arbitrary matrices at each site.  However, in the context of this article, it is most convenient to use the exact density matrix of a solved trial Hamiltonian $\hat{H}_t$,
	\begin{align}
	\hat{\rho}_t
	&= \frac{1}{Z_t} e^{-\beta  \hat{H}_t}
	= e^{\beta \Omega_t - \beta \hat{H}_t},
	\end{align}
where $\Omega_t = -T \ln \Tr e^{-\beta  \hat{H}_t}$ is the trial free energy.
The task is then to optimize the variational free energy
	\begin{align}
	\Omega_\text{var}
	&= \mean{ \hat{H} + T  (\beta \Omega_t - \beta \hat{H}_t  ) }_{\hat{\rho}_t}
	= \mean{ \hat{H}  - \hat{H}_t  }_{\hat{\rho}_t}   +  \Omega_t 
	\elabel{GeneralVariationalFreeEnergy}
	\end{align}
with respect to the trial Hamiltonian $\hat{H}_t$.
To explore the consequences of this decision we must explicitly introduce a model for $\hat{H}_t$.

\section{Hubbard Model: Two-Channel Decoupling}
Let us illustrate the procedure with a clean Hubbard model with an applied chemical potential $\mu^a$,
	\begin{align}
	\hat{H}
	&=
	-	\underbrace{
			\sum_{ij\sigma} t_{ij} \cdag_{i\sigma} \cccc_{j\sigma} 
		}_{\hat{t}}
	-	\underbrace{
			g \sum_{i} \cdag_{i\up} \cdag_{i\dn} \cccc_{i\dn} \cccc_{i\up}
		}_{\hat{g}}
	-	\underbrace{
		 \mu^a \sum_{i\sigma} \cdag_{i\sigma} \cccc_{i\sigma}
		}_{\hat{\mu}^a}
	.
	\elabel{SimpleHubbardHamiltonian}
	\end{align}
We can construct a suitable trial Hamiltonian by decoupling the interaction in terms of a pairing amplitude $\Delta_i$ and a Hartree potential $\mu^H_i$ at every site, which serve as variational parameters:
	\begin{align}
	\hat{H}_t (\{ \Delta_i , \mu^H_i \} )
	&=
	-	\sum_{ij\sigma} t_{ij} \cdag_{i\sigma} \cccc_{j\sigma} 
	-	\underbrace{
			\sum_{i} \Delta_i (\cdag_{i\up} \cdag_{i\dn} + \cccc_{i\dn} \cccc_{i\up})
		}_{\hat{\Delta}}
	-	\underbrace{
		 \sum_i \mu_i (\cdag_{i\up} \cccc_{i\up} + \cdag_{i\dn} \cccc_{i\dn})
		}_{\hat{\mu}}
	,\nonumber
	\end{align}
where $\mu_i = \mu^a + \mu^H_i$.
This bilinear trial Hamiltonian can be solved by diagonalizing it, that is, by 
calculating the eigenvalues $E_\alpha$, eigenvectors $u_\alpha$ and $v_\alpha$, and occupation numbers $f_\alpha$ (see \sref{BdG}).
In the Bogoliubov basis we simply have
$\hat{H}_t = \sum_\alpha E_\alpha (\gammadag_\alpha \gamma_\alpha - \gamma_\alpha \gammadag_\alpha)$.
The resulting trial density matrix,
	\begin{align}
	\hat{\rho}_t
	&=
	\frac{1}{Z_t}
	\exp \left[ 
		-\beta
		 \sum_\alpha E_\alpha (\gammadag_\alpha \gamma_\alpha - \gamma_\alpha \gammadag_\alpha)
	\right]
	,
	\end{align}
is Gaussian, and so the trial free energy can be evaluated straightforwardly,
	\begin{align}
	\Omega_t
	&=
	 -T \ln \Tr e^{-\beta  \hat{H}_t}
	 =
	-T
	\sum_\alpha \ln \left( 2 \cosh \frac{\beta E_\alpha}{2} \right)
	.
	\elabel{SimpleTrialFreeEnergy}
	\end{align}
Now substitue \eref{SimpleHubbardHamiltonian} and \eref{SimpleTrialFreeEnergy} into	
\eref{GeneralVariationalFreeEnergy}.  The first term gives
	\begin{align}
	\mean{ \hat{H}  - \hat{H}_t  }_{\hat{\rho}_t} 
	&=
	\mean{ - \hat{t} - \hat{g} - \hat{\mu}^a - (- \hat{t} - \hat{\Delta} - \hat{\mu})}
	_{\hat{\rho}_t} 
			\nonumber\\
	&=
	\mean{ - \hat{g} + \hat{\Delta} + \hat{\mu}^a - \hat{\mu}}
			\nonumber\\
	&=
		-g \sum_{i} 
			\mean{ \cdag_{i\up} \cdag_{i\dn} \cccc_{i\dn} \cccc_{i\up} }
		+
			\sum_{i} \Delta_i \mean{ \cdag_{i\up} \cdag_{i\dn} + \cccc_{i\dn} \cccc_{i\up} }
		+
			\sum_{i} (\mu_i - \mu^a) \mean{n_i }
			\nonumber
	\end{align}
where all the expectations are taken with respect to the trial density matrix $\hat{\rho}_t$.
Since $\hat{\rho}_t$ is Gaussian, the quartic term reduces to a sum of Wick contractions.
The final expression for the variational free energy is
	\begin{align}
	\Omega_\text{var}  (\{ \Delta_i , \mu^H_i \} )
	&=\Omega_t + \sum_i \Big[ 
			-g ( {F_i}^2 + {n_i}^2 ) + 2 \Delta_i F_i + 2 (\mu_i - \mu^a) n_i )
		\Big]
	\elabel{SimpleVariationalFreeEnergy}
	\end{align}
where $\Omega_t$ is given by \eref{SimpleTrialFreeEnergy} and $F_i$ and $n_i$ are 
	\begin{align}
	F_i 
	&= {1\over 2} \mean{ \cdag_{i\up} \cdag_{i\dn} + \cccc_{i\dn} \cccc_{i\up} }
	= \sum_\alpha (f_\alpha - 1/2) u_{i\alpha} v_{i\alpha}
			,\nonumber\\
	n_i 
	&= {1\over 2} \mean{ \cdag_{i\up} \cccc_{i\up} + \cdag_{i\dn} \cccc_{i\dn} }
	=	{1\over 2} \left [ \sum_\alpha f_\alpha u_{i\alpha}{}^2
		+ \sum_\alpha (1 - f_\alpha) v_{i\alpha}{}^2\right ]
	\end{align}
Here, $n_i \in [0,1]$ is the average density per spin species. It can be verified that, upon minimizing \eref{SimpleVariationalFreeEnergy} with respect to the variational parameters, one recovers the gap and number equations, 
$\Delta_i = g F_i$ and $\mu_i^H = g n_i$, which are familiar from the traditional BdG formalism.

\section{Six-Channel Decoupling}
For completeness we now present the most general mean-field theory for the Hubbard model, in which the Hubbard interaction is decoupled in all six channels.
It is convenient to adopt a $4N\times 4N$ version of the Nambu-Gor'kov matrix formalism and write the Hubbard Hamiltonian as
	\begin{align}
	\hat{H}_\text{true}
	&=
	-	\underbrace{
				\sum_{ijs} \half  t_{ij} \eta_{6ss'} \cdag_{is} \cccc_{js} 
		}_{\hat{t}}
	-	\underbrace{
			\sum_{ilss'} \half  \Sigma^a_{il} \eta_{lss'} \cdag_{is} \cccc_{is'}
		}_{\hat{\Sigma}^a}
	+	\underbrace{
			\sum_{i} U_i x_{i\up} x_{i\dn}
		}_{\hat{U}}
		.
	\end{align}
The notation is as follows:
$t_{ij}$ and $U_i$ are hopping and on-site repulsion,
and $s,s'=1,2,3,4$ are superspin indices that distinguish between the four fermionic degrees of freedom at each site (up-spin and down-spin particles and holes),
	\begin{align}
	\cccc_{is} = \pmat{
		\cccc_{i\up} &
		\cccc_{i\dn} &
		\cdag_{i\up} &
		\cdag_{i\dn} 
	}_s
	.
	\end{align}
The factors of $\half$ in the Hamiltonian compensate for particle-hole doubling.
The operator
$x_{i\sigma} = \half \left( \bar\psi_{i\sigma} \psi_{i\sigma} - \psi_{i\sigma} \bar\psi_{i\sigma} \right)$ is the local density measured with respect to half-filling.  This definition has the advantage of being particle-hole symmetric.
The applied potentials (such as a disorder potential or a Zeeman field) are represented by the self-energy fields $\Sigma^a_{i\chn}$ on sites $i=1,\dotsc,N$ in channels $\chn=1,2,3,4,5,6$.
The six channels correspond to the three components of the Zeeman field $\hhh$, the real and imaginary parts of the pairing potential $\Delta$, and the chemical potential $\mu$:
	\begin{align}
	\Sigma_1 \equiv h_X      &\qquad  \Sigma_2 \equiv h_Y        \qquad \Sigma_3 \equiv h_Z,
	 	\nonumber\\
	\Sigma_4 \equiv \Delta_R &\qquad  \Sigma_5 \equiv \Delta_I   \qquad \Sigma_6 \equiv \mu    
	.
	\end{align}
The basis matrices $\eeeta_\chn$ are
	\begin{align}
	\eeeta_1 = \psmat{0&1&0&0\\1&0&0&0\\0&0&0&-1\\0&0&-1&0}  \qquad
	\eeeta_2 = \psmat{0&-i&0&0\\i&0&0&0\\0&0&0&i\\0&0&-i&0}  \qquad
	\eeeta_3 = \psmat{1&0&0&0\\0&-1&0&0\\0&0&-1&0\\0&0&0&1},
	\nonumber\\
	\eeeta_4 = \psmat{0&0&0&1\\0&0&-1&0\\0&-1&0&0\\1&0&0&0}  \qquad
	\eeeta_5 = \psmat{0&0&0&i\\0&0&-i&0\\0&i&0&0\\-i&0&0&0}  \qquad
	\eeeta_6 = \psmat{1&0&0&0\\0&1&0&0\\0&0&-1&0\\0&0&0&-1}.
	\end{align}
The $\eeeta_\chn$ are Hermitian, particle-hole symmetric, mutually orthogonal, and normalized such that  $\tr \eeeta_\chn^\dag \eeeta_\chn = 4$.
The self-energy term can be written out explicitly as
	\begin{align}
	\hat{\Sigma}^a
	&= 
	\pmat{
		\cccc_{\up} \\[+2pt]
		\cccc_{\dn} \\[+2pt]
		\cdag_{\up} \\[+2pt]
		\cdag_{\dn} 
	}^\dag
	\pmat{
		\mu+h_Z & -h_X+ih_Y & 0 & \Delta_R+i\Delta_I   \\[+3pt]
		-h_X-ih_Y & \mu-h_Z & -\Delta_R-i\Delta_I & 0  \\[+3pt]
		0	& -\Delta_R+i\Delta_I  & -\mu-h_Z & h_X-ih_Y \\[+3pt]
		\Delta_R-i\Delta_I & 0 & h_X+ih_Y & -\mu+h_Z   \\
	}
	\pmat{
		\cccc_{\up} \\[+2pt]
		\cccc_{\dn} \\[+2pt]
		\cdag_{\up} \\[+2pt]
		\cdag_{\dn} 
	}
	\end{align}
(where we have omitted site indices $i$ and superscripts $a$ for clarity).
The self-energy matrix is a Hermitian matrix with particle-hole symmetry; these symmetries constrain the 16 complex matrix elements, so that six real numbers are sufficient to parametrize the self-energy.
Six is the number of generators of the group $SU(2) \times SU(2)$; the six parameters transform into each other under suitable rotations in spin space or particle-hole space.

\subsubsection{Trial Hamiltonian and trial density matrix}
Following the procedure illustrated earlier, we then construct a trial Hamiltonian by decoupling $\hat{U}$ in six channels,
	\begin{align}
	\hat{H}_t
	&=
	-	\sum_{ijs} \half t_{ij} \eta_{6ss'} \cdag_{is} \cccc_{is'}
	-	\underbrace{
			\sum_{ilss'} \half  \Sigma_{il} \eta_{lss'} \cdag_{is} \cccc_{is'}
		}_{\hat{\Sigma}}
	,
	\end{align}
where the total (effective) self-energy $\Sigma_{il} = \Sigma^a_{il} + \Sigma^H_{il}$ is the applied (external) self-energy plus the internal (Hartree/Fock/Bogoliubov) self-energy arising from the decoupling of the interaction.
This bilinear trial Hamiltonian can be constructed explicitly as a $4N\times 4N$ matrix and diagonalized to give eigenvalues $E_\alpha (\alpha=1,\dotsc,4N)$ and eigenvectors $\Phi^{\alpha}_{is}$.
The trial free energy is 
	\begin{align}
	\Omega_t
	&= -T \ln \Tr e^{-\beta\hat{H}_\text{trial}}
	= -\tfrac{T}{2} \sum_\alpha \ln \left( 2 \cosh \tfrac{\beta}{2} E_\alpha \right)
	.
	\end{align}
The variational free energy works out to be
	\begin{align}
	\Omega_\text{var}
	&=
		\Omega_t
	+	\sum_{i} 
			U_i \left(  - m_Z^2 - m_X^2 - m_Y^2 + F_R^2 + F_I^2 + x^2  \right)_i
			\nonumber\\&{}
	+	2 \sum_{i} \big(
			h^H_X m_X + h^H_Y m_Y + h^H_Z m_Z + \Delta^H_R F_R  + \Delta^H_I F_I  + \mu^H x
		\big)_i
	.
	\elabel{FullVariationalFreeEnergy}
	\end{align}
In \eref{FullVariationalFreeEnergy}, there are six densities at every site:
$(m_X, m_Y, m_Z)$ are the three components of magnetization,
$(F_R, F_I)$ are the real and imaginary parts of the anomalous Green function,
and $x$ is the average density per spin species measured with respect to half-filling.
Each of these quantities lies in the interval $[-\half,+\half]$.  Explicitly,
	\begin{align}
	G_1 = m_X = \half\mean{\bar\psi_\up \psi_\dn + \bar\psi_\dn \psi_\up}           \quad
	G_2 = m_Y = \tfrac{1}{2i} \mean{\bar\psi_\up \psi_\dn - \bar\psi_\dn \psi_\up} \quad
	G_3 = m_Z = \half(x_\up - x_\dn)		,\nonumber\\
	G_4 = F_R = \half \mean{\psi_\dn \psi_\up + \bar\psi_\up \bar\psi_\dn}          \quad
	G_5 = F_I = \tfrac{1}{2i} \mean{\psi_\dn \psi_\up - \bar\psi_\up \bar\psi_\dn}  \quad
	~~~~
	G_6 = x   = \half(x_\up + x_\dn)  .
	\nonumber
	\end{align}
These quantities can be calculated from knowledge of the occupation numbers and eigenvectors:
	\begin{align}
	G_{il}
	&=\tfrac{1}{4} \sum_{ss'} \eta_{lss'} 	G_{iss'}
				\nonumber\\
				\text{where}\quad
	G_{iss'}
	&=
		\sum_\alpha
		(f_\alpha - \half)
		\Phi^{\alpha} _ {is}  {}^*
		\Phi^{\alpha} _ {is'}
	.
	\end{align}
Minimizing \eref{FullVariationalFreeEnergy} with respect to 
the $6N$ variational parameters $\Sigma^a_{il}$ 
gives the self-consistency conditions at each site,
	\begin{align}
	\hhh^H_i   = +U_i m_i			,\qquad
	\Delta^H_i = -U_i F_i			,\qquad
	\mu^H_i   = -U_i x_i				.
	\elabel{FullSelfConsistencyConditions}
	\end{align}
This makes physical sense: a repulsive interaction $U>0$ produces positive feedback in the spin channel, which tends to produce spontaneous magnetic order, whereas an attractive interaction $U<0$ produces a tendency towards pairing, charge separation, and charge ordering.

As remarked in \sref{BdG}, the variational mean-field calculation can be performed using Broyden-type methods to solve the self-consistency equations (\eref{FullSelfConsistencyConditions}), while monitoring the variational free energy (\eref{FullVariationalFreeEnergy}) to ensure that the iteration is converging to a minimum of $\Omega_t$ and not a saddle-point or maximum.

\bibliographystyle{forprl}


\end{document}